\documentclass[%
reprint,
 amsmath,amssymb,
 aps,
rmp,
floatfix,
]{revtex4-2}
\usepackage{graphicx}% Include figure files
\usepackage{dcolumn}% Align table columns on decimal point
\makeatletter
\newcommand\definecommand{%
  \@ifstar{}{\define@command@nostar}}
\newcommand\define@command@nostar[1]{%
  \@ifundefined{}{\newcommand{#1}}
}
\definecommand\@reinserts{}
\makeatother
\usepackage{manyfoot}
\DeclareNewFootnote{B}
\renewcommand*{\thefootnoteB}{\alph{footnoteB}}
\renewcommand*{\footnotetextB}[2][]{{\stepcounter{footnoteB}\begin{NoHyper}\Footnotemark\thefootnoteB\,\end{NoHyper}}\footnotesize{#2}}
\DeclareNewFootnote{S}
\renewcommand*{\thefootnoteS}{\fnsymbol{footnoteS}}
\renewcommand*{\footnotetextS}[2][]{{\stepcounter{footnoteS}\begin{NoHyper}\Footnotemark\thefootnoteS\,\end{NoHyper}}\footnotesize{#2}}
\usepackage{bm}% bold math
\usepackage{xcolor}
\usepackage{longtable}
\usepackage{makecell}
\usepackage{multirow}
\usepackage{array}
\usepackage[version=3]{mhchem} % Formula subscripts using \ce{}
\usepackage{textcomp}
\usepackage{dcolumn}
\usepackage{gensymb}
\usepackage{amsmath}
\usepackage{float}
\usepackage{hyperref}% add hypertext capabilities
\hypersetup{
    colorlinks = true,
    allcolors = [rgb]{0,0,0.5},
}

\begin{document}

\preprint{APS/123-QED}

\title{Respiratory aerosols and droplets in the transmission of infectious diseases}
% Multimodal size distributions of respiratory particles in airborne disease transmission

\author{Mira L. P\"ohlker}
 %\altaffiliation[]{}%Lines break automatically or can be forced with \\
 \email{m.pohlker@mpic.de}
 
\author{Ovid O. Kr\"uger}
 %\email{o.krueger@mpic.de}
 
\author{Jan-David F\"orster}

\author{Thomas Berkemeier}

\author{Wolfgang Elbert}

\author{Janine Fr\"ohlich-Nowoisky}

\author{Ulrich P\"oschl}

\author{Christopher P\"ohlker}
\email{c.pohlker@mpic.de}

\affiliation{Multiphase Chemistry Department, Max Planck Institute for Chemistry, Mainz, Germany.}

\author{Gholamhossein Bagheri}

\author{Eberhard Bodenschatz}

\affiliation{Laboratory for Fluid Physics, Pattern Formation and Biocomplexity, Max Planck Institute for Dynamics and Self-Organization, G\"ottingen, Germany.}%

\author{J. Alex Huffman}
\affiliation{Department of Chemistry \& Biochemistry, University of Denver, Denver, USA.}

\author{Simone Scheithauer}
\affiliation{Institute of Infection Control and Infectious Diseases, University Medical Center, Georg August University, G\"ottingen, Germany.}

\author{Eugene Mikhailov}
\affiliation{St. Petersburg State University St. Petersburg, Russia}
\affiliation{Multiphase Chemistry Department, Max Planck Institute for Chemistry, Mainz, Germany}

\date{\today}

\begin{abstract} 
\vspace{5mm}
\noindent Knowing the physicochemical properties of exhaled droplets and aerosol particles is a prerequisite for a detailed mechanistic understanding and effective prevention of the airborne transmission of infectious human diseases. This article provides a critical review and synthesis of scientific knowledge on the number concentrations, size distributions, composition, mixing state, and related properties of respiratory particles emitted upon breathing, speaking, singing, coughing, and sneezing. We derive and present a parameterization of respiratory particle size distributions based on five lognormal modes related to different origins in the respiratory tract, which can be used to trace and localize the sources of infectious particles. This approach may support the medical treatment as well as the risk assessment for aerosol and droplet transmission of infectious diseases. It was applied to analyze which respiratory activities may drive the spread of specific pathogens, such as \textit{Mycobacterium tuberculosis}, influenza viruses, and SARS-CoV-2 viruses. The results confirm the high relevance of vocalization for the transmission of SARS-CoV-2 as well as the usefulness of physical distancing, face masks, room ventilation, and air filtration as preventive measures against COVID-19 and other airborne infectious diseases.\par

\clearpage
\end{abstract}

\maketitle
\tableofcontents

\section{\label{sec:level1}Introduction}

Diseases that spread via the respiratory tract, such as measles, tuberculosis, influenza, and the coronavirus disease 2019 (COVID-19) have played dramatic and important roles in global public health \cite[e.g.,][]{RN1915,RN1911,RN1661}. Disease outbreaks can be driven, in whole or large part, by emissions of pathogen-laden particles that can infect nearby persons, such as those expelled in a spray of droplets or clouds of small airborne aerosols \cite[e.g.,][]{RN1908,RN1877,RN1661,RN2019,RN2075}. Knowledge on the relative importance of transmission pathways and the mechanisms of host-to-host transmission, in relation to pathogen-host interactions and environmental factors, remain critical gaps of knowledge \cite[][and references therein]{RN1907,RN2010}. A detailed understanding of pathogen transmission dynamics in space and time is needed to improve public health interventions, which are a cornerstone in pandemic control besides vaccination strategies, to control outbreaks and reduce infection rates \cite{RN1606,RN1877}.\par

Several transmission routes for pathogens passable between humans are known \cite{RN1768,RN1717,RN1881,RN2076,RN2092}. The most widely occurring and relevant routes can be grouped into contact vs contact-free. Contact transmission from an infected individual to a susceptible recipient can occur through either direct person-to-person contact (e.g., a handshake) or indirect contact via contaminated objects or surfaces (fomites), followed by a hand-to-face transport of pathogens (i.e., self-inoculation of eyes, nose or mouth) \cite{RN1688,RN1916}. Transmission without physical contact, through the air, can occur via both near- and far-field transmission from either small or large particles being emitted by the mouth or nose of infected individuals via coughing, sneezing, talking, or even breathing \cite[e.g.,][]{RN1841,RN1871,RN1916,RN1783,RN2019,RN2021,RN2075,RN2076}. For this pathway, the terms contact-free and airborne are commonly used. Particles containing viable pathogens can then land directly on the mucosal surfaces of the recipient or can be inhaled. This review focuses only on emissions from the respiratory tract, defined here as the path of respiratory air from deep lungs through the mouth and nose. Nevertheless, other emissions of potentially pathogen-laden particles, e.g. from the human body (i.e., the "personal cloud" effect of skin, clothing emissions) and of aerosolized fecal material can also play critical roles in disease transmission, if deposited in the respiratory tract of a recipient \cite[e.g.,][]{RN1990,RN1992,RN1994,RN1995,RN1791,RN2009,RN2061,RN2077}.

\begin{figure*}
\includegraphics[width=16cm]{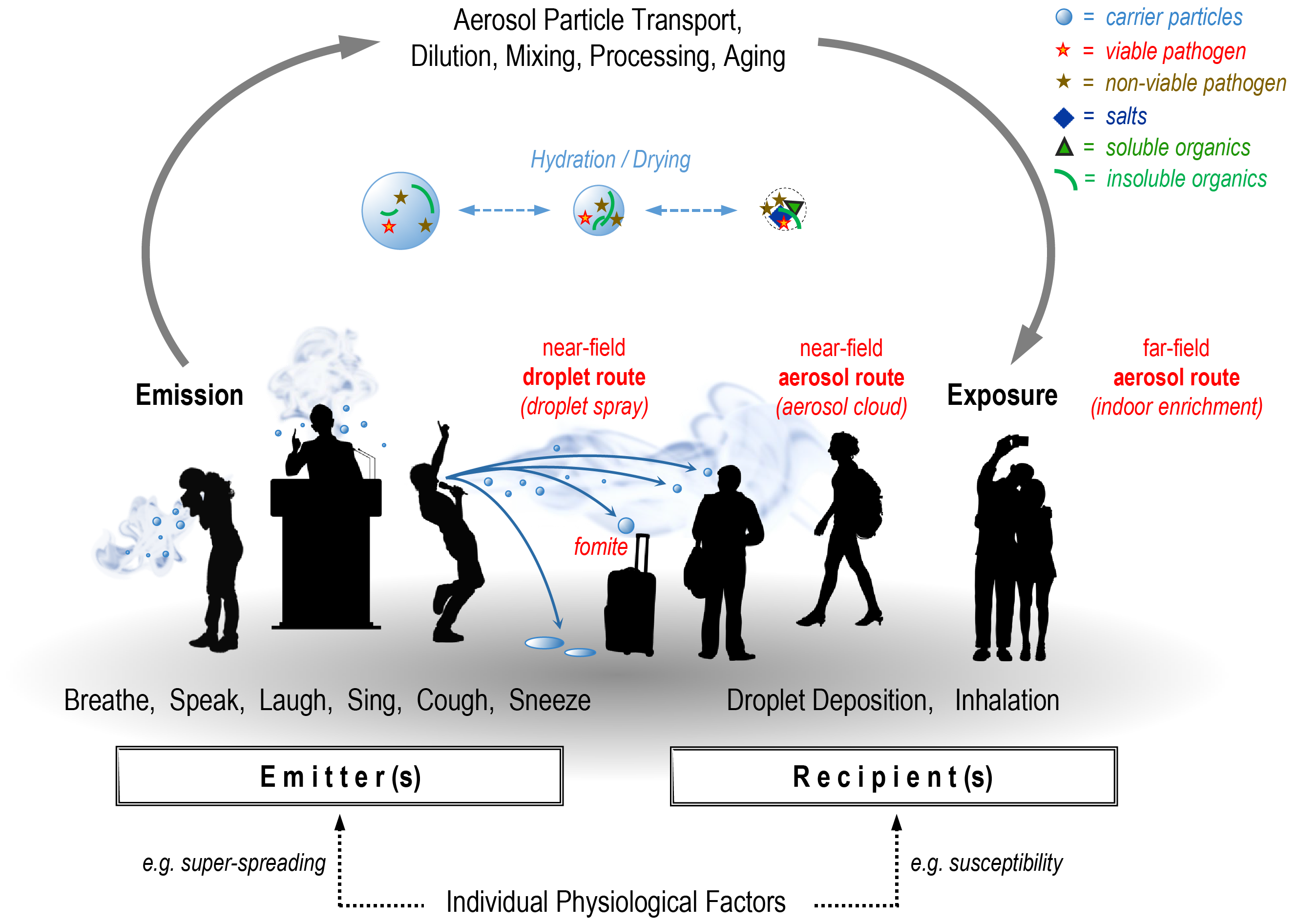} 
\caption{Conceptual scheme of the aerosol and droplet pathogen transmission routes along with relevant physicochemical properties of respiratory particles.}
\label{fig:scheme}
\end{figure*}

Particles with diameters larger than $\sim$100\,\micro m are often called droplets in the context of respiratory emissions \cite{RN1875}. After being emitted, droplets follow (semi-)ballistic trajectories to deposit directly onto objects or mucosal surfaces, i.e., in the nose, mouth, or eyes. Due to the large mass they are only weakly influenced by the airflow and inhalation is relatively unlikely \cite{RN1691,RN2024,RN1902}. Droplets also have too much mass to stay suspended for longer than a few seconds and so fall to the ground rapidly within about one to two meters of the emitter \cite{RN1871,RN1875}. The near-field deposition of droplets largely depends on particle size and emission velocity, and thus extends increasingly further from the source via singing or shouting as well as coughing or sneezing \cite[e.g.,][]{RN1686,RN1907,RN1720,RN1677}. Particles with diameters smaller than $\sim$100\,\micro m are often called aerosols or droplet nuclei in the context of respiratory emissions and have sufficiently small mass and momentum to be inhalable via airflow into the respiratory tract \cite[e.g.,][]{RN2024}. Respiratory aerosols are generally present in the highest concentration in the near-field in a plume closest to the emitter, but can also mix readily into a volume of air in the far-field \cite[e.g.,][]{RN1871,RN1978,RN1986,RN2019}. They can remain suspended for minutes or even hours, depending on particle size and the free air turbulence or flow dynamics \cite{RN1907}. Thus, it can be possible for aerosols to be a dominant mode of transmission in both the near-field (1-2\,m) and far-field (e.g., mixed into room air) \cite[e.g.,][]{RN1893,RN1935,RN1871,RN2076}, though mitigation strategies may differ within the two categories of distance.\par

Disease transmission through exhaled particles can tend predominantly toward the aerosol route if mostly small and readily airborne particles containing viable pathogens are emitted, or predominantly towards the droplet route if mostly large and (semi-)ballistically distributed droplets are involved. This distinction is largely a function of pathogen-host interaction and physiological factors. The boundary between aerosols and droplets is fraught, because, from a physical perspective, they each lie at different ends of a size continuum \cite{RN2012}. Historically, a dividing size of 5\,\micro m has been utilized, but this has relatively little support from either physical or physiological reasoning \cite{RN2019,RN2076}. For example, the settling velocity of a 5\,\micro m leads to a residence time of ~25 minutes. Upward convection from body heat and other turbulent dynamics make the estimation of residence time more complicated, but even particles smaller than this will mix into the majority of the room volume and are thus relevant for inhalation in the far-field. More recently a dividing size of 100\,\micro m has become more commonly accepted, largely because this is approximately the size above which particles are unlikely to be inhaled, based on the physics of airflow into the respiratory tract \cite{RN1765,RN1767,RN1686,RN1875}. It should also be noted that, while many communities differentiate between aerosol and droplet size regimes, the specific cut-point definition between the regimes can be different. In the near-field, infection can be caused by ballistic droplet spray and contact-based mechanisms, as well as by inhalation of concentrated aerosol clouds, which means that near-field aerosol infection can mimic epidemiological patterns of large-droplet spray or contact infections \cite{RN1878,RN1907,RN1871}. Accordingly, it can be very complicated to separate these processes to identify the most significant route for a given pathogen \cite{RN1783,RN1661,RN1704,RN1879}. The size range of deposited pathogen-containing particles can also play an important role in the development and severity of the disease \cite[e.g.,][]{RN1930,RN1829}.\par
 
The 'classical' examples of pathogens that spread predominantly via aerosols are \textit{Mycobacterium tuberculosis} causing tuberculosis (TB), the measles morbillivirus (MeV) causing measles, and the varicella-zoster virus (VZV) causing chickenpox \cite[e.g.,][]{RN1661,RN1885,RN1676,RN1673,RN1674,RN1943,RN2092}. More 'exotic' pathogens that can spread via aerosols are the bacteria \textit{Coxiella burnetii} causing Q fever and the spores of \textit{Bacillus anthracis} causing anthrax \cite[e.g.,][]{RN1940,RN2011}. For the COVID-19 pandemic, caused by the readily transmissible severe acute respiratory syndrome coronavirus 2 (SARS-CoV-2) \cite{RN1715,RN1829}, the preponderance of evidence suggests that SARS-CoV-2 is aerosol-transmissible with virus-laden respiratory particles transmitted through the air being (key) drivers of infection \cite[e.g.,][]{RN1685,RN1688,RN1657,RN1704,RN1743,RN1841,RN1875,RN1651,RN1595,RN1652,RN1740,RN1815,RN1933,RN1934,RN1935,RN1937,RN1741,RN2019,RN2022,RN2021,RN2025,RN2078}. This is in line with studies suggesting that the closely related coronaviruses SARS-CoV-1 and MERS-CoV (Middle-East Respiratory Syndrome coronavirus) also spreads through aerosols as well \cite[e.g.,][]{RN1670,RN1669,RN1671,RN1672,RN1884}.\par

\autoref{fig:scheme} illustrates the life cycle of respiratory particles in contact-free pathogen transmission, spanning from exhalation over airborne transport to the potential infection of a recipient \cite{RN1978}. The emission of particles in relation to respiratory activities, such as breathing, speaking, singing, coughing, and sneezing has been analyzed in numerous studies, as summarized, e.g., in reviews by \citet{RN1683}, \citet{RN1714}, and \citet{RN1684}. The fluid dynamics involved in the spread of the exhaled multiphase cloud of potentially pathogen-laden particles was summarized, e.g., by \citet{RN1907}, \citet{RN1908}, and \citet{RN1677}. The transmitted pathogen dose response of the recipient is determined by multiple factors, such as the number, size distribution, and physicochemical properties of the pathogen bearing particles being inhaled or deposited on mucosal surfaces during a given exposure time \cite{RN1887}. The droplet route is exclusively relevant in the near-field through droplet spray deposition on persons or objects (creating fomites). The aerosol routes can be relevant in the near-field upon inhalation of concentrated clouds of small particles near the emitter as well as in the far-field when small particles accumulate in indoor environments (e.g., in schools, restaurants, public transport) or are distributed via directed air flows (e.g., air conditioning) before significant dilution occurs \cite[e.g.,][]{RN1884,RN1871,RN1893,RN1899,RN1765,RN1720,RN1658,RN1804,RN1807,RN1935,RN2078}. \par

The aerosol infection pathway is largely influenced by individual physiological factors of both the emitter (e.g., high individual variation in infectiousness with superspreading individuals) and the recipient (enhanced susceptibility due to pre-existing conditions, co-infections, etc.) \cite[e.g.,][]{RN1740,RN1709,RN2037,RN2092,RN1771}. Important is also the timing of maximum pathogen replication with associated exhalation and the encounter of emitter and recipient. Current evidence on the replication and emission of wild-type SARS-CoV-2, for instance, peaks two to three days prior to and on the first days of symptom onset \cite{RN1931,RN1932,RN1912}. A mechanistic understanding of essential processes in Fig.\,\ref{fig:scheme} requires transdisciplinary bridges between infection epidemiology, virology, pulmonology, and immunology as well as aerosol physics and chemistry, fluid dynamics, and related fields \cite{RN1877,RN1581,RN2075}.\par

This review article addresses the microphysical particle properties involved in contact-free diseases transmission. It focuses particularly on the concentrations and size distributions of respiratory particles, as well as the distribution of pathogens within these carrier particle populations. \autoref{sec:ReviewSec} provides a general and concise summary of definitions, nomenclature, and key parameters as well as mechanistically relevant information on droplet formation in the respiratory tract, followed by rapid droplet desiccation after emission. \autoref{sec:MainSec} follows with an in-depth review and synthesis of the scientific literature on respiratory particle size distributions (PSDs) from breathing, speaking and singing. A generalized multimodal, lognormal parameterization of exhaled PSDs is introduced here, which is based on previous observations and parameterization approaches \cite{RN1608,RN1611,RN1714,RN1707,RN1663}. The new parameterization (i) covers the size range from $<$10\,nm to $>1000$\,\micro m, (ii) is based on the smallest number of modes needed to adequately represent the available experimental PSDs weighted both by particle number and volume, (iii) is widely representative of the existing literature, and (iv) is readily applicable to modelling studies. The parameterization has further been related to the emission mechanisms and sites in the respiratory tract as well as to mode-specific particle number concentrations in relation to different respiratory activities. Finally, using the multimodal parameterization, emission mechanisms and particle size modes are identified, which are most closely associated with the spread of common pathogens. The literature summary and consistent parameterization of the exhaled PSDs can be used as a framework for improved understanding, control, and prevention of infectious disease transmission, such as for COVID-19 and other present and future diseases, which spread via the respiratory tract.\par
% , as well as coughing and sneezing. The published PSDs show significant inconsistencies, likely due to experimental challenges and limitations in characterizing this highly dynamic particle population.

\section{\label{sec:ReviewSec}Definitions, nomenclature, and key parameters in aerosol and droplet pathogen transmission}

\subsection{\label{sec:Def}Definitions and nomenclature on respiratory aerosol and droplets}

A major challenge in the multidisciplinary field of airborne disease transmission is that the different scientific communities involved often "don't speak the same language" \cite{RN1877}. Moreover, the terminology is not always clearly defined or consistently used and, thus can promote misunderstandings \cite{RN2019,RN1878}. The fundamental terms aerosol and droplet are used especially inconsistently by different scientific communities and with different meanings and implications.\par 

The term droplet can convey two broadly different concepts: (i) it is often used as a counterpart for aerosol in a dichotomous classification of airborne vs (semi)ballistic transmission routes (see details below). In this sense, the term droplet is meant to differentiate a particular (large) particle size regime marked by aerodynamic behavior separate from the smaller range of the size continuum of respiratory particles. (ii) Moreover, the term droplet is also frequently used to mean respiratory particles that contain water. Note, however, that all respiratory particles comprise significant amounts of water upon exhalation and, thus can be considered as droplets, regardless of size \cite{RN1907,RN2019}. Further, the term droplet nuclei is commonly used, within broadly medically-oriented communities, for the remaining residues after evaporation of respiratory droplets \cite{RN1720,RN1611,RN1817}. This use of the terms nucleus or nuclei is somewhat inconsistent from a process perspective. "Nuclei", as defined within the scope of atmospheric physics, are those particles involved in a nucleation process that lead to particle growth, e.g., a physical surface on which cloud droplets or ice crystals initiate growth. The term droplet nuclei, however, here refers to the end point of droplet desiccation and, thus, shrinkage of a droplet rather than growth from a seed nucleus.\par

The term aerosol is used in some fields as a synonym for droplet nucleus, whereas communities of physical science typically use the term aerosol to refer to particles small enough to stay suspended in air for some period of time (as defined more specifically below). Use of the term airborne is especially contentious and inconsistently used within the context of respiratory disease \cite{RN1878,RN1657,RN2019,RN1661,RN2076}. In most sub-fields of the physical sciences the term airborne is associated broadly with particles that are suspended in or transported through the air. In this context, both smaller aerosol particles and droplets can be considered airborne, although the residence time in air and ability to be inhaled varies strongly with particle size \cite{RN1978,RN1875,RN2053,RN1691}. For these reasons, the majority of discussion associated with the term airborne among physical scientists is associated with aerosols \cite{RN1871}. In stark contrast to this perspective, non-physical sub-fields of science and medicine have historically applied additional limitations to the use of the term airborne, e.g. to aerosol-based diseases with demonstrably high basic reproduction numbers ($R_0$) such as measles, but without specific physical or mechanistic reasoning \cite{RN1877,RN2019,RN2076,RN2092}. A broader discussion of the differences in the way different communities define this term will not be addressed in detail here. For the purposes of this discussion, we will adopt a physical perspective of the term airborne, comprising both smaller aerosols and larger droplets.\par

For clarity, we define here the terminology used throughout this manuscript:
\begin{itemize}
    \item \textbf{Aerosol}: According to the established text book definition, an aerosol is defined as a suspension of liquid or solid particles in a gas, with particle diameters ranging from few nm up to about 100\,\micro m \cite{RN433,RN8,RN1691,RN999}. Note that sometimes the term aerosols (with plural s) is used to refer specifically to the suspended particles, which frequently causes confusion in relation to the aforementioned rigorous definition of aerosol, which is already plural with respect to the particles involved. The term droplet nuclei is frequently used within medical communities to be synonymous with aerosol.
    
    \item \textbf{Droplets}: Generally speaking, droplets are liquid particles. In the medical and epidemiological literature the term droplet is frequently used for aqueous liquid particles larger than 5\,\micro m in diameter. Here, the term droplet has been used cautiously to avoid misunderstandings. We used it only for aqueous droplets (of all sizes) directly upon emission from the respiratory tract, regardless of particle size. After the onset of drying (which happens quickly), the terms residue or dried particle are used for the partially or fully dried droplets. Note that we avoid using the term droplet nuclei, because the term nucleus has the aforementioned different meaning in aerosol physics.
    
    \item \textbf{Particles}: The term particle refers here to the entire population of liquid or solid particles encompassing the full spectrum of possible sizes, as well as physical and chemical states. 
\end{itemize}\par

Throughout the text, we predominantly use the terms particle and aerosol, and to only a minor degree the term droplet. Where not clear from the context, we further specified these terms with attributes defined by physical properties such as aerodynamic behavior, i.e., airborne vs (semi-)ballistic, or water content, i.e., wet vs dry/dried. Often, a particular particle size range is of relevance -- in these cases we have numerically specified the diameter thresholds (e.g., $>1$\,\micro m or from 0.1 to 10\,\micro m). To broadly subdivide the modes of the particle size distribution into two groups according to the commonly applied set of measurement instrumentation (see Tab. \ref{tab:LitSyn}, Sect. \ref{sec:SpeakSingResults}, and Sect. \ref{sec:CoughSneezeResults}), we have used the relative terms small for all modes centered below 5\,\micro m and large for all modes centered above 5\,\micro m. Note that this 5\,\micro m threshold is not related to the traditionally used 5\,\micro m threshold in a dichotomous classification of aerosol vs droplet infection routes (see details below). All particle sizes in this manuscript refer to diameter and never to radius. 

Neither the lower nor upper aerosol size ranges have rigorously defined physical limits: The lower limit is marked by a gradual transition from gas molecules and larger molecular clusters to nanometer-sized particles. The upper limit is given by the gradually changing aerodynamic properties of particles that vary as a function of size when moving in a gas, the most important of which is the increased sedimentation rate of large particles due to gravity. An alternative, but parallel transition at the upper size limit is the point at which particles become too big to be efficiently inhalable \cite[e.g.,][]{RN2024,RN1691}. Here, the particle's Reynolds number ($Re$), which is the ratio of the resisting force of the viscous gas to the inertial force of the moving particle, separates the \textit{Stokes regime} (with $\text{viscous forces}>>\text{inertial forces}$ and $Re<$\,1) from the purely ballistic \textit{Newton regime} ($\text{with viscous forces}<<\text{inertial forces}$ and $Re>$\,1000) with a semi-ballistic \textit{transition regime} in between (1\,$<Re<$1000) \citep{RN1691}. Thus, $Re$ quantifies to which degree the particles are prone to follow the air streams patterns, e.g., upon inhalation. $Re$ is calculated through 
\begin{equation} \label{eq:1}
    Re\,=\,\frac{\rho_\text{g}\,v_\text{s}\,D}{\eta}
\end{equation}

\noindent with the density of the gas ($\rho_\text{g}$), which is the density of air here, the relative velocity between air and particle ($v_\text{s}$), which is here the settling velocity, the particle diameter ($D$), and the gas dynamic viscosity ($\eta$), also for air here. For a spherical particle in the Stokes regime, $v_\text{s}$ is calculated through 
\begin{align} 
v_\text{s}\,& =\,\frac{\rho_\text{p}\,D^2\,g\,C_\text{c}}{18 \eta} \label{eq:2}\\
\text{with}\quad C_\text{c}\,& =\,1+\frac{\lambda}{D}\left[2.34 +1.05 \exp\left(-0.39\,\frac{D}{\lambda}\right)\right]\nonumber\\
\text{for}\quad(Re\,&<\,1)\nonumber
\end{align}

\noindent with the particle density ($\rho_\text{p}$), the gravitational acceleration ($g$), the Cunningham slip correction ($C_\text{c}$), and $\lambda$ as the gas mean-free-path \citep{RN1691}. $C_\text{c}$ matters primarily for particles with $D<1$\,\micro m and converges to unity for $D>1$\,\micro m. For a spherical particle in the transition regime, $v_\text{s}$ is calculated through  
\begin{align} 
v_{s}\,& =\,\left(\frac{4\,\rho_\text{p}\,D\,g}{3\,C_\text{D}\,\rho_\text{g}}\right)^{\frac{1}{2}}\label{eq:3} \\
\text{with}\quad C_\text{D}\,& =\,\frac{24}{Re} \left(1+0.15\,Re^{0.687} \right)\nonumber\\
\text{for}\quad (Re\,&>\,1)\nonumber
\end{align}

\noindent with the drag coefficient ($C_\text{D}$). $Re$ for water drops with $D<80$\,\micro m in room conditions is $<1$ and so for this size range, using the Stokes drag would not result in significant errors. Note that the drag coefficient ($C_\text{D}$) in Eq.\,\ref{eq:3} is an empirical correction, which depends on $v_\text{s}$.

\begin{figure}[t]
  \includegraphics[width=\columnwidth]{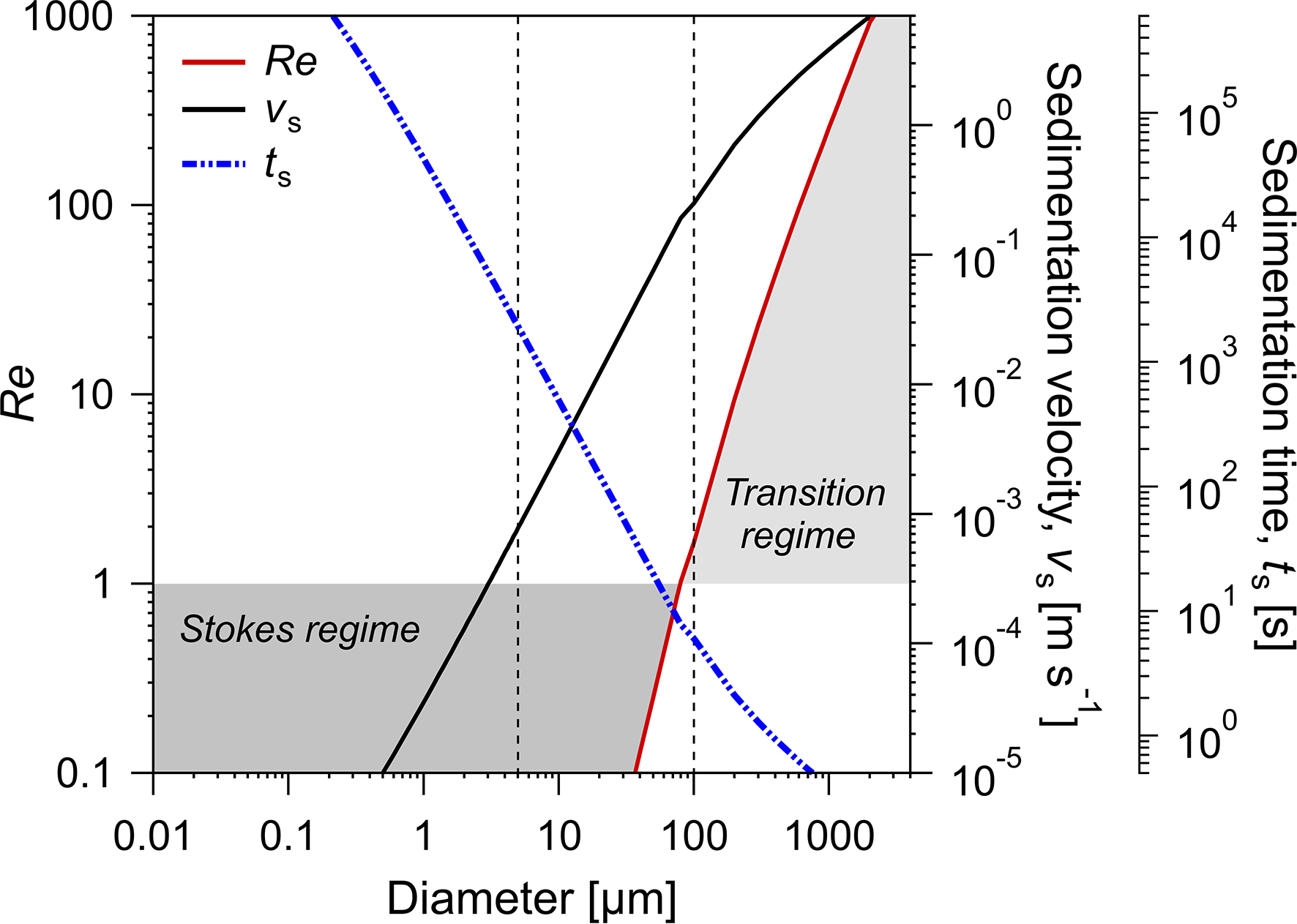}
  \caption{\label{fig:Re}Size dependence of the Reynolds number ($Re$), settling velocity ($v_\text{s}$), and sedimentation time from 1.5\,m height ($t_\text{s}$) for spherical particles in still air. Gray shadings highlights Stokes ($Re<1$) vs transition regimes ($1<Re<1000$). Vertical dashed lines mark frequently used threshold at 5\,\micro m to separate large vs small particles as well as threshold at 100\,\micro m as proposed by \citet{RN1875}.}
\end{figure}

Within the Stokes regime and in quiescent/still air, $v_\text{s}$ scales with $D^2$, spanning from essentially infinite airborne residence times for smaller particles (i.e., \mbox{$D<\,$1\,\micro m)} to finite settling velocities for large particles (Fig. \ref{fig:Re}). For example, for an exhalation at 1.5\,m height, a 5\,\micro m particle settles out within $\sim$30\,min, a 8\,\micro m particle settles out within $\sim$10\,min, whereas a 20\,\micro m particle settles out within $\sim$2\,min. In real-world settings, however, the air is typically not still, but rather influenced by air movements on different scales, including upward convection due to heating from bodies, and so the particle residence time in the air of occupied rooms is usually much longer than in still air. The drag of the ambient air can extend particle residence times significantly \cite{RN1877,RN2010}. For non-spherical particles, Eq.\,\ref{eq:2} should be modified to take particle shape into account. If particle shape does not deviate clearly from a sphere, the errors from using Eq.\,\ref{eq:2} are not significant \cite{RN1891}. 

Figure\,\ref{fig:Re} shows that aerosol and droplet transmission routes cannot be strictly separated, but exist in different size regimes of a particle size and fluid mechanical continuum. The terminology used to separate the concepts of aerosol and droplet transmission often causes more confusion than clarity by introducing the false notion that droplets and aerosols are somehow fundamentally different \cite{RN1767,RN1686}. Further, historically inaccurate understandings of the physics of airborne particles persists in many scientific disciplines today, and deeply held disciplinary beliefs and nomenclature add to the confusion \cite[e.g.,][]{RN1878,RN1978,RN2019,RN2076,RN1713,RN1716,RN1683,RN1744,RN1767,RN1877}. \citet{RN1686}, for instance, criticized the widely used and "overly simplified [...] dichotomous classification between large vs small droplets [with] various arbitrary droplet diameter cutoffs, from 5 to 10\,\micro m", which still underlies the current risk management, recommendations, and infection control. \citet{RN1686} further proposes to consider aerosol and droplet spray transmission as a multiphase turbulent cloud of buoyant hot and moist air that contains a continuous range of particle sizes \cite{RN1720,RN1686,RN2027}. In this context, \citet{RN1875} advocated that 100\,\micro m would be a more appropriate threshold than the historically used 5 or 10\,\micro m. All such clear-cut boundaries are a simplifications of the physics involved. Nevertheless, if a threshold has to be defined, 100\,\micro m appears as the most appropriate choice, as it is roughly the diameter at which $Re\approx1$ and beyond which the drag force acting on the particle no longer varies linearly with the $Re$ (Fig.\ref{fig:Re}). It further corresponds to the largest particle size that is typically inhalable, and so provides compelling practical benefit with respect to transmission pathways and mitigation strategies.\par

Concerning the composition of respiratory particles, we use the following terminology throughout the text: Mucosal fluids comprise a large group of liquid surface films (e.g., tear fluid, nasal mucus, bronchial mucus, gastric mucus, sweat, etc.) that cover different parts of the body or organ surfaces exposed to the external environment \cite{RN1952,RN1909}. These fluids typically have site-specific composition and fulfill a variety of specific functions. The following two mucosal fluids play an essential role in respiratory particle emission:
\begin{itemize}
    \item \textbf{Saliva} is present in the oral cavity, where it is produced by different salivary glands \cite{RN1952,RN1951}.
    \item \textbf{Epithelial lining fluid (ELF)} cover the air-facing surfaces of the lower respiratory tract (LRT). Note that in addition to ELF, the terms respiratory tract lining fluid (RTLF), airways surface liquid (ASL), and mucus are also widely used in the literature \cite{RN1909}. Here we restrict use to the term ELF. 
\end{itemize}
  
\noindent Further, we use the terms mucosalivary or mucosal films and fluids to refer to both ELF and saliva. An overview of the complex and variable composition of saliva and ELF can be found in Sect.\,\ref{sec:Hygros}. The term respiratory particles here refers to emissions from both saliva and ELF.

\subsection{\label{sec:Props} Properties of respiratory particles}

The physicochemical aerosol properties that matter most in the spread of diseases are 
\begin{itemize}
    \item particle number and volume concentration (Sect.\,\ref{sec:Conc_Review} and Sect.\,\ref{sec:ConcRates})
    \item particle number and volume size distribution (Sect.\,\ref{sec:PSD_Review} and Sect.\,\ref{sec:MainSec})
    \item size-dependent distribution of pathogens in the respiratory carrier particle population (Sect.\,\ref{sec:Mixing} and Sect.\,\ref{sec:Modality})
    \item composition and hygroscopicity of the mucosalivary particles along with their desiccation and (re)humidification properties (Sect.\,\ref{sec:Hygros})
\end{itemize}
\noindent Further, physical and biochemical properties of the pathogens themselves play important roles, which are addressed only briefly in Sect.\,\ref{sec:Survival}. For further details, refer to the cited literature.

\begin{table*}
\caption{\label{tab:RespParam}Parameters of different respiratory activities/events, summarized from previous studies. The following events are specified: breathe (event\,=\,one exhalation), speak (event\,=\,one spoken word of average length), cough (event\,=\,one cough), and sneeze (event\,=\,one sneeze). Relevant event-specific parameters are the emitted air volume ($V_{\text{air}}$), duration ($\Delta t$), and peak flow rate ($q$) per event. Relevant time-averaged parameters are the event rate ($f$), which is the number of event repetitions per hour, and the average air emission rate ($\dot{V}$), obtained through $\dot{V}\,=\,V_{\text{air}}\cdot f$ \cite{RN1847}. For speaking, two (short) words spoken per second were assumed to obtain a speaking-related $f$, according to \citet{RN1663}. Several parameters show an inherently high inter- and intrasubject variability, which is reflected in the table as typical parameter ranges. The values in brackets show the characteristic values used in calculations in this review article (e.g., Sect.\,\ref{sec:ConcRates} and Sect.\,\ref{sec:Modality}). Values represent the average of male and female adults. Values mostly represent healthy subjects -- only for the cough rate, healthy and diseased subjects are distinguished.}
\begin{ruledtabular}
\begin{tabular}{llllll}
 & \multicolumn{3}{c}{Properties per respiratory event} & \multicolumn{2}{c}{Time-averaged properties}\\
 & 
{Exhaled volume} & 
{Duration} & 
{Peak flow rate} & 
{Event rate} & 
{Air emission rate}\\
 & 
$V_{\text{air}}$ {[$\text{L}$]} & 
$\Delta t$ {[s]} & 
{$q$ [$\text{L}\,\text{s}^{-1}$]} & 
$f$ {[$\text{h}^{-1}$]} & 
{$\dot{V}$ [$\text{L}\,\text{h}^{-1}$]} \\[2pt]
\hline & \\[-1.7ex]
{Tidal breath\vspace{3mm}} & 
{\makecell[tl]{{$0.4-1.7$\,\footnotemarkB[2]\textsuperscript{,}\footnotemarkB[8]\textsuperscript{,}\footnotemarkB[11]\textsuperscript{,}\footnotemarkB[14]}\\{[0.5]\,\footnotemarkB[3]\textsuperscript{,}\footnotemarkB[7]\textsuperscript{,}\footnotemarkB[16]}}} & 
{\makecell[tl]{{$1.5-2.5$}\,\footnotemarkB[6]\textsuperscript{,}\footnotemarkB[10]\\{[2]\,\footnotemarkB[3]}}} & 
{$0.2-0.7$\,\footnotemarkB[9]\textsuperscript{,}\footnotemarkB[10]\textsuperscript{,}\footnotemarkB[11]} & 
{\makecell[tl]{{$600-1200$\,\footnotemarkB[2]\textsuperscript{,}\footnotemarkB[8]\textsuperscript{,}\footnotemarkB[14]}\\{[720]\,\footnotemarkB[3]\textsuperscript{,}\footnotemarkB[7]\textsuperscript{,}\footnotemarkB[16]}}} &
{\makecell[tl]{{$360-800$}\,\footnotemarkB[7]\textsuperscript{,}\footnotemarkB[10]\textsuperscript{,}\footnotemarkB[12]\\{[360]\,\footnotemarkS[1]}}} \\

%\makecell[tl]{Breath with\\airway closure\vspace{3mm}} & 
%\makecell[tl]{$1.2-5.0$\,\footnotemarkB[3]\textsuperscript{,}\footnotemarkB[4]\textsuperscript{,}\footnotemarkB[13]\\{[1.5]\footnotemarkB[22]}} & 
%{2\,\footnotemarkB[20]} 
%& {$0.1-1.2$\,\footnotemarkB[3]} & 
%{[720]\,\footnotemarkB[20]} & 
%{[1080]\,\footnotemarkB[21]}
\\

\makecell[tl]{Spoken word\\\vspace{3mm}} & % \\shouting 
{\makecell[tl]{{[0.1]\,\footnotemarkS[1]}}} &

{\makecell[tl]{$0.5$\,\footnotemarkB[12]}}  & 

{\makecell[tl]{{$0.3-1.6$\,\footnotemarkB[1]\textsuperscript{,}\footnotemarkB[5]\textsuperscript{,}\footnotemarkB[10]}}} & 

{[7200]\,\footnotemarkB[12]} & 

{\makecell[tl]{{$450-700$\,\footnotemarkB[10]\textsuperscript{,}\footnotemarkB[12]}\\{[700]\,\footnotemarkB[10]}}}  
\\

{Cough\vspace{3mm}} & 

{\makecell[tl]{{$0.3-4$\,\footnotemarkB[2]\textsuperscript{,}\footnotemarkB[3]\textsuperscript{,}\footnotemarkB[4]\textsuperscript{,}\footnotemarkB[9]\textsuperscript{,}\footnotemarkB[12]\textsuperscript{,}\footnotemarkB[13]\textsuperscript{,}\footnotemarkB[16]\textsuperscript{,}\footnotemarkB[22]}\\
{[1.5]\,\footnotemarkB[18]}}} & 

{$0.2-1$\,\footnotemarkB[2]\textsuperscript{,}\footnotemarkB[3]\textsuperscript{,}\footnotemarkB[9]\textsuperscript{,}\footnotemarkB[17]\textsuperscript{,}\footnotemarkB[22]} & 

{$0.2-15$\,\footnotemarkB[2]\textsuperscript{,}\footnotemarkB[9]\textsuperscript{,}\footnotemarkB[10]\textsuperscript{,}\footnotemarkB[13]\textsuperscript{,}\footnotemarkB[18]\textsuperscript{,}\footnotemarkB[23]} & 

\makecell[tl]{{healthy: $0-4$\,\footnotemarkB[19]\textsuperscript{,}\footnotemarkB[20]\textsuperscript{,}\footnotemarkB[24]}\\{healthy smoker: $0-8$\,\footnotemarkB[20]\textsuperscript{,}\footnotemarkB[24]}\\{diseased: $0-140$\,\footnotemarkB[15]\textsuperscript{,}\footnotemarkB[16]\textsuperscript{,}\footnotemarkB[19]\textsuperscript{,}\footnotemarkB[20]\textsuperscript{,}\footnotemarkB[21]\textsuperscript{,}\footnotemarkB[24]}\\{[10]\,\footnotemarkB[3]}} &

{[15]\,\footnotemarkS[1]}  
\\

Sneeze & 
{\makecell[tl]{{$1-4$}\,\footnotemarkB[3]\\{[2]\,\footnotemarkB[16]}}} & 

{$0.1-0.2$\,\footnotemarkB[3]} & 

{$10-20$\,\footnotemarkS[1]} & 

{\makecell[tl]{{$5-30$}\,\footnotemarkB[3]\\{[10]\,\footnotemarkS[2]}}} & 

{[20]\,\footnotemarkS[1]} 
\\

\end{tabular}
\end{ruledtabular}
\raggedright{\par
\footnotetextB[1]{\citet{RN1895}} % Abkarian  1
\footnotetextB[2]{\citet{RN1609}}; % Ai & Melikov 2017  2
\footnotetextB[3]{\citet{RN1907}}; % Bourouiba 2020  3
\footnotetextB[4]{\citet{RN1611}}; % Chao 2009 0.4 speak vol. from tidal volume one:0.03  4
\footnotetextB[5]{\citet{RN2064}}; % Chi  5
\footnotetextB[6]{\citet{RN2062}} % Conrad & Schönle 79  6
\footnotetextB[7]{\citet{RN1847}}; % Derrickson 2017 Book  7
\footnotetextB[8]{\citet{RN1873}}; % Gao 2018
\footnotetextB[9]{\citet{RN1858}}; % Gupta 2009
\footnotetextB[10]{\citet{RN1857}}; % Gupta 2011
\footnotetextB[11]{\citet{RN1613}}; % Holmgren 2013
\footnotetextB[12]{\citet{RN1663}}; % Johnson
\footnotetextB[13]{\citet{RN1848}}; % Lee
\footnotetextB[14]{\citet{RN1859}}; % Levitzky
\footnotetextB[15]{\citet{RN2018}}; % Patterson 2018
\footnotetextB[16]{\citet{RN2017}}; % Patterson 2019
\footnotetextB[17]{\citet{RN1861}}; % Ran
\footnotetextB[18]{\citet{RN1861}}; % Ren 2020
\footnotetextB[19]{\citet{RN2015}}; % Sinha 16
\footnotetextB[20]{\citet{RN2014}} % Sumner 13
\footnotetextB[21]{\citet{RN1862}}; % Sunger 2012
\footnotetextB[22]{\citet{RN1863}}; % Wei
\footnotetextB[23]{\citet{RN1607}}; % Yan
\footnotetextB[24]{\citet{RN2029}}; % Yousaf 13
\footnotetextS[1]{Calculated through $\dot{V}\,=\,V_{\text{air}}\cdot f$};
\footnotetextS[2]{Sparse literature, values adopted from coughing}.
\setcounter{footnoteS}{0}
% \footnotetextB[26]{Calculated as tidal volume plus expiratory reserve volume (average of male \& female values), according to \citet{RN1847}}; % Derrickson 2017 Book
}
\end{table*}

\subsubsection{\label{sec:Conc_Review}Particle number and volume concentrations and emission rates}

The particle number ($N$) or volume ($V_\text{p}$) concentration ($C$) in an air volume ($V_\text{air}$) defines the overall abundance of exhaled (and potentially pathogen bearing) particles according to
\begin{equation} \label{eq:C_NV}
    C_\text{N}\,=\,\frac{N}{V_\text{air}}\quad\text{and}\quad C_\text{V}\,=\,\frac{V_\text{p}}{V_\text{air}}
\end{equation}

\noindent The particle emission rates ($Q$) are derived through
\begin{equation} \label{eq:Q_NV}
    Q_\text{N}\,=\,N\,f\quad\text{and}\quad Q_\text{V}\,=\,V_\text{p}\,f\\
\end{equation}
\noindent as well as
\begin{equation} \label{eq:Q_NV_2}
    Q_\text{N}\,=\,C_\text{N}\,\dot{V}\quad\text{and}\quad Q_\text{V}\,=\,C_\text{V}\,\dot{V}
\end{equation}

\noindent with a given $N$ and $V_\text{p}$, the rate ($f$) of a given respiratory event, as well as the air emission rate ($\dot{V}$). Table\,\ref{tab:RespParam} summarizes average parameters for respiratory events and specifies those values used in the calculations in this work. The particle source is highly variable in terms of strength and frequency. It spans from semi-continuous tidal breathing to short and intense events such as sneezing with a duration $\Delta t<$\,1\,s (Table\,\ref{tab:RespParam}). The inter- and intrasubject variability in $C_\text{N}$ as a function of respiratory activity and physiological factors is remarkably high, spanning two to three orders of magnitude from $\sim$0.1 up to $\sim$100\,$\text{cm}^{-3}$ \cite[e.g.,][]{RN1701,RN1707,RN1613,RN1608,RN1708,RN1750,RN1734,RN1830}. The $C_\text{N}$ and $Q_\text{N}$ can be converted in $C_\text{V}$ and $Q_\text{V}$ through 
\begin{equation} \label{eq:NtoV}
   C_V\,=\,\frac{\pi}{6}\,D^3\,C_N\,\,\,\,\,\,\text{and}\,\,\,\,\,\,Q_V\,=\,\frac{\pi}{6}\,D^3\,Q_N
\end{equation}

\noindent under the assumption that the particles have a (nearly) spherical shape. 

Exhaled puffs of air are discontinuous, turbulent fluid volumes emitted from a point source (mouth or nose) and driven at first by momentum and subsequently by buoyancy. The puffs are spatially heterogeneous due to mixing and dilution in turbulent eddies and variable water vapor and temperature fields, which results in a decrease in $C_\text{N}$ with distance from the source and, therefore, makes analysis of measurements complicated to interpret \cite{RN1708,RN1907,RN2027}. Note further that the $C_\text{N}$ levels are typically much lower than ambient aerosol concentrations (both indoor and outdoor), which typically range from few hundreds to few thousands particles per $\text{cm}^3$ \cite[e.g.,][]{RN633,RN1752}. This imposes further experimental challenges since the ambient background aerosol must either be removed (i.e., filtration) or carefully characterized (i.e., background subtraction). Even in a clean-room environment, however, it is very challenging to detect the influence of respiratory aerosols on top of existing particle concentrations unless particles are selectively detected, e.g., based on differences in composition. For a comparison of respiration $C_\text{N}$ levels from different studies, it is essential to specify the measurement size range as most instruments or techniques cover only a limited band of the overall relevant size distribution (see Table\,\ref{tab:LitSyn}). Thus, most reported number concentrations do not account for the total $C_\text{N}$, but rather a subset within a certain size range determined by the instruments' specifications. As one result of the parameterization of particle size distributions presented in this study, Sect.\,\ref{sec:ConcRates} presents a statistical summary of measured particle number and volume concentrations as well as emission rates in relation to respiratory activities. 
% Alex on clean room: The line before strongly implied that filtering the air is sufficient to just see the respiratory aerosol on top. I have had this discussion several times with various people doing these experiments, and everyone seems to agree it is essentially impossible to do this, unless you somehow can sub-select the respiratory particles (by yet undeveloped techniques for real-time analysis; or using long-integration time based on culturable methods).

\subsubsection{\label{sec:PSD_Review}Particle number and volume size distributions}

The initial particle size distribution (PSD) right immediately following emission depends on the formation mechanisms and sites within the respiratory tract (Sect.\,\ref{sec:Emission}) \cite[e.g.,][]{RN1707,RN1608,RN1617,RN1663}. After emission and rapid evaporation (Sect.\,\ref{sec:Hygros}), the PSD determines the aerosol residence time and mobility in the air \cite[e.g.,][]{RN2027}. The particle movement and transport through air is driven by multiple forces, such as drag, inertial, electrostatic, radiative, gravitational, and thermophoretic forces, as well as Brownian motion and turbulent diffusion \cite{RN1765,RN1761,RN1691}. The influence of these forces strongly depends on the PSD and, thus, affects transport over distances as well as either dilution or potential enrichment under given air conditions \cite{RN1708,RN1986,RN1691}. Moreover, the PSD defines the filtration efficiency of face masks (upon both in- and exhalation) as well as the deposition sites of particles in the upper respiratory tract (URT) and lower respiratory tract (LRT) (see also Fig.\,\ref{fig:RT}) \cite{RN1702,RN1785}. 

The respiration PSDs have a characteristic multimodal shape \cite{RN1608,RN1611,RN1714,RN1707,RN1663,RN1728}. They can be described well by a multimode lognormal fit function with $n$ individual modes ($i$) according to
\begin{equation} \label{eq:f_NSD}
    f_N\left(D\right)\,=\,\sum_{i=1}^{n}A_i\exp\left\{-\left[ \frac{\ln\left(\frac{D}{D_{i}}\right)}{\sigma_i}\right]^2\right\}
\end{equation}

\noindent with $D$ as the particle diameter, $D_i$ as the mode mean geometric diameter, $A_i$ as the number concentration at $D_i$, and $\sigma_i$ the modal geometric standard deviation which defines the mode width \cite{RN433}. In this study, $f_N\left(D\right)$ is given as d$N$/d$\log D$, which is broadly established in aerosol science. Thus, Equation \ref{eq:f_NSD} uses both, $log$ and $ln$. To convert uniformly to $log$ (if preferred), $\ln(D/D_i)$ can be replaced by $2.303\cdot \log(D/D_i)$ to make the equation uniform. For the multimodal lognormal fitting, the smallest possible number of modes yielding a good representation of the experimental data is preferred. Ideally, the individual modes can be associated with the mechanisms and sites of specific emission processes in the respiratory tract. Equation\,\ref{eq:f_NSD} has been used throughout this work and in the parameterization present here. Note, that different versions of lognormal fit functions have been broadly used in aerosol studies and in the field of respiratory aerosols in particular. As an alternative example, the following Eq.\,\ref{eq:HZ} from \citet{RN1905} has been used widely \cite[e.g.,][]{RN1608,RN1611,RN1714,RN1707,RN1663} 
\begin{equation} \label{eq:HZ}
  f_H\left(D\right)\,=\,\sum_{i=1}^{n} \frac{C_i}{\sqrt{2\pi}\ln(\sigma_i)}\exp\left\{-\left[\frac{\ln\left(\frac{D}{D_{i}}\right)}{\sqrt{2}\ln(\sigma_i)}\right]^2\right\}
\end{equation}

\noindent with $D_i$ as the mode mean geometric diameter as in Eq.\,(\ref{eq:f_NSD}), $C_i$ as the integral particle number concentration of the mode, and $\sigma_i$ as the modal geometric standard deviation. Importantly, both functions Eq.\,(\ref{eq:f_NSD}) and Eq.\,(\ref{eq:HZ}) yield the same fitting results. They only differ in the definition or meaning of the fit parameters. 

Essentially all measurements of respiratory aerosols yield particle number size distributions (NSDs) as primary data (i.e., number of counted particles in a given sequence of size bins). The NSDs can be converted in size distributions of the number emission rates ($Q_\text{N}$) according to Eq.\,\ref{eq:Q_NV} and $f$ from Table\,\ref{tab:RespParam}. Further, the NSDs can be converted in particle volume size distributions (VSDs) according to    
\begin{gather} \label{eq:f_VSD}
     f_V\left(D\right)\,=\,\frac{\pi}{6}D^3f_N(D)\\
     =\,\frac{\pi}{6}D^3\cdot \sum_{i=1}^{n}A_i\exp\left\{-\left[ \frac{\ln\left(\frac{D}{D_{i}}\right)}{\sigma_i}\right]^2\right\}\nonumber
\end{gather}

\noindent assuming a (nearly) spherical particle shape.

Inconsistencies and deviations in the PSDs reported in published literature can presumably be explained by the highly dynamic properties of the respiratory particle population, along with experimental challenges in its characterization \cite[e.g.,][]{RN1750,RN1663,RN1600}. Moreover, many different measurement techniques have been used, each of which can detect aerosol particles over a narrow band of the overall PSD. To construct a PSD over the full range of particle sizes emitted by human respiration ($<10$\,nm to $>1000$\,\micro m) requires data from a set of instruments to be stitched together, each of which may use different physical parameters for detection (i.e., optical, aerodynamic, physical, or electric mobility sizing). The matching of PSDs using these different techniques introduces additional uncertainties to be considered in the interpretation of those measurements \cite{RN433}. Section\,\ref{sec:MainSec} follows up on this general overview of respiratory particle NSDs and VSDs with a comprehensive summary of the available literature and the development of an efficient parameterization scheme, representing their characteristic multimodal shape.

\begin{figure}[b]
  \includegraphics[width=\columnwidth]{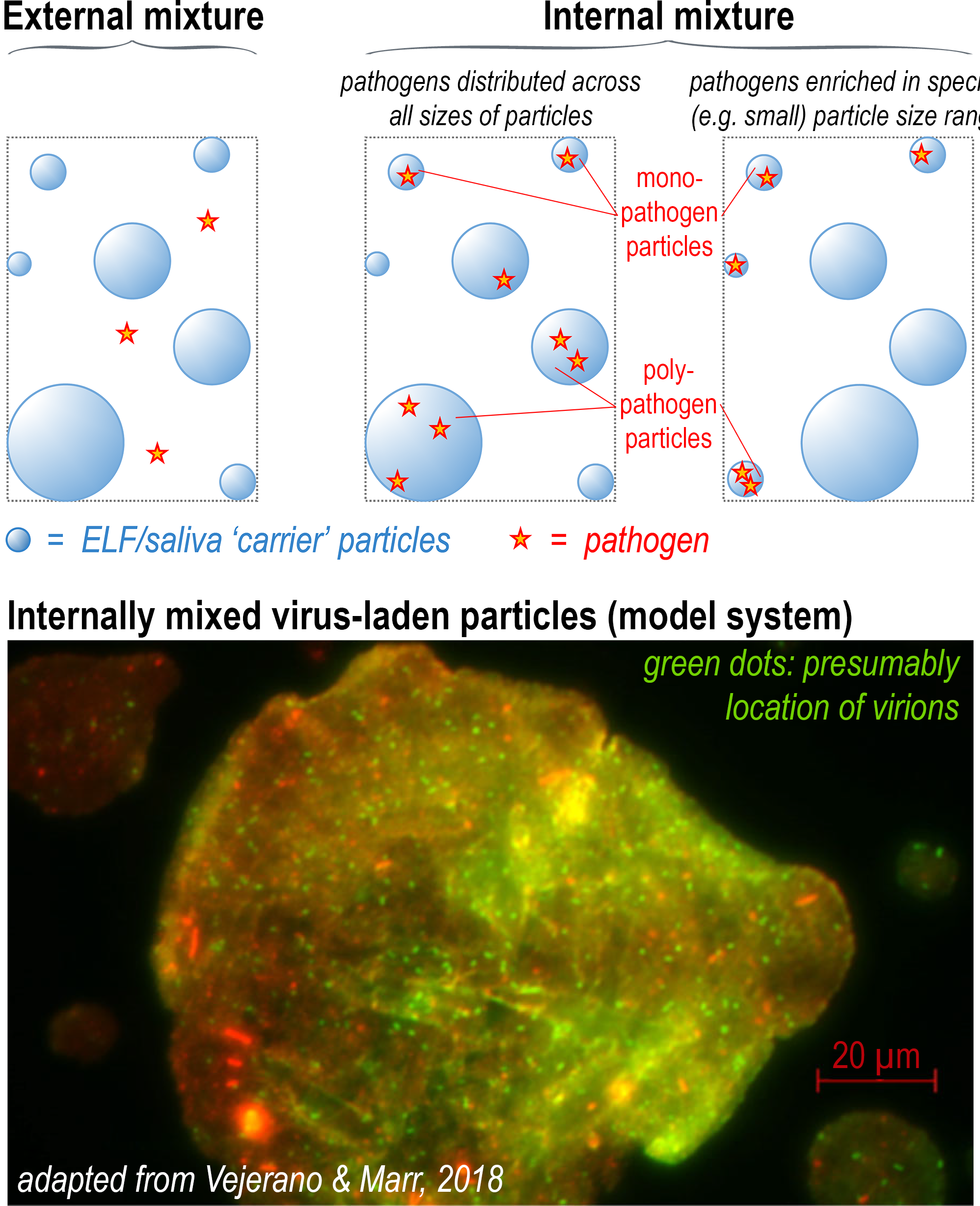}
  \caption{\label{fig:Mixing}Figure illustrates concepts of internal vs external mixing states in relation to pathogen distribution in mucusalivary particle populations. It further emphasizes the difference between mono- and poly-pathogen particles \cite[e.g.,][]{RN1894}. Microscopy images adapted from \citet{RN1610} shows laboratory-generated virus-laden and internally mixed particles as surrogates for authentic respiratory particles. Here, particles comprise salt, glycoprotein, surfactant, and the well-studied \textit{Pseudomonas} $\phi$6 virions. Green fluorescence is associated with surfactants, which are assumed to partition to the $\phi$6 virions. Adapted with permission under CC Attribution License 4.0.}
\end{figure}

\begin{table*}[ht!]
\caption{\label{tab:HalfLife}%
Summary of experimentally determined half-lives ($T_{1/2}$) in exponential decay of pathogen viability in aerosol. This table illustrates the wide range of 'typical' time scales involved in pathogen viability and their dependence on environmental factors by means of selected studies. It does not provide a comprehensive overview of the published literature. Temperature ($\theta$), relative humidity (RH), and irradiance ($E$, integrated UVB) are specified as important environmental factors. $T_{1/2}$ values in parentheses represent 95\,\% confidence interval. Where $E$ is not specified in the table, experiments were conducted in the dark.
}
\begin{ruledtabular}
\begin{tabular}{llllll}
\makecell[tl]{Pathogen} & {$T_{1/2}$ {[$\text{min}$]}} & \multicolumn{3}{c}{Conditions} & \makecell[tl]{Reference}\\
 &
 &
\makecell[tc]{$\theta$ [\degree\,C]}&
\makecell[tc]{RH [\%]}&
\makecell[tc]{$E$ [$\text{Wm}^{-2}$]}&
\\[2pt]
\hline & \\[-1.7ex]

\makecell[tl]{\textit{M. tuberculosis}, \textit{M. avium}, \textit{M. intracellulare}} & \makecell[tc]{7.5 ($6-9$)\footnote{All data points from this study were fitted jointly with exponential decay function to obtain average half-life (with values for 95\% confidence interval) shown here.}} & \makecell[tc]{$24\pm2$} & \makecell[tc]{$74\pm3$} & & \cite{RN1944}\vspace{2mm}\\

\makecell[tl]{\textit{Mycobacterium tuberculosis}} & \makecell[tc]{90} & \makecell[tc]{20-25} & \makecell[tc]{60-70} & \makecell[tc]{--} & \cite{RN1917}\vspace{2mm}\\

\makecell[tl]{\textit{Pseudomonas aeruginosa}} & \makecell[tc]{50 ($30-151$)} & \makecell[tc]{n.a.} & \makecell[tc]{n.a.} & \makecell[tc]{--} & \cite{RN1942}\vspace{2mm}\\

\makecell[tl]{Influenza A (i.e., H1N1, H5N1, H3N2)} & \makecell[tc]{$15-90$} & \makecell[tc]{25} & \makecell[tc]{55} & \makecell[tc]{--} & \cite{RN1946}\vspace{2mm}\\

\makecell[tl]{Influenza A (i.e., H1N1)} & \makecell[tc]{$\sim32$} & \makecell[tc]{20} & \makecell[tc]{20-70} & \makecell[tc]{0} & \cite{RN1770}\\

 & \makecell[tc]{$\sim2$} & \makecell[tc]{20} & \makecell[tc]{20-70} & \makecell[tc]{1.44} & \vspace{2mm}\\

\makecell[tl]{MERS-CoV} & \makecell[tc]{$>$70} & \makecell[tc]{25} & \makecell[tc]{79} & \makecell[tc]{--} & \cite{RN1945}\\
 & \makecell[tc]{$\sim$24} & \makecell[tc]{38} & \makecell[tc]{24} & \makecell[tc]{--} & \vspace{2mm}\\

\makecell[tl]{SARS-CoV-1\vspace{2mm}} & \makecell[tc]{72 ($47-146$)} & \makecell[tc]{23} & \makecell[tc]{40} & \makecell[tc]{--} & \cite{RN1595}\vspace{2mm}\\

\makecell[tl]{SARS-CoV-2\vspace{2mm}} & \makecell[tc]{66 ($38-158$)} & \makecell[tc]{23} & \makecell[tc]{40} & \makecell[tc]{--} & \cite{RN1595}\vspace{2mm}\\

\makecell[tl]{SARS-CoV-2\vspace{2mm}} & \makecell[tc]{$>$960\footnote{This value is based on one measurement only.}} & \makecell[tc]{$25\pm2$} & \makecell[tc]{$53\pm11$} & \makecell[tc]{--} & \cite{RN1742}\vspace{2mm}\\

\makecell[tl]{SARS-CoV-2\vspace{2mm}} & \makecell[tc]{$30-177$} & \makecell[tc]{$19-22$} & \makecell[tc]{$40-88$} & \makecell[tc]{--} & \cite{RN1947}\vspace{2mm}\\

\makecell[tl]{SARS-CoV-2} & \makecell[tc]{$10\,\,- >$100} & \makecell[tc]{$10-40$} & \makecell[tc]{$20-70$} & \makecell[tc]{0} & \cite{RN1938}\\
 & \makecell[tc]{$3-5$} & \makecell[tc]{$10-30$} & \makecell[tc]{45} & \makecell[tc]{0.9} & \\
 & \makecell[tc]{$1-3$} & \makecell[tc]{$10-40$} & \makecell[tc]{$20-70$} & \makecell[tc]{1.9} & \\
\end{tabular}
\end{ruledtabular}
\end{table*}

\subsubsection{\label{sec:Mixing}Size distributions of pathogens within respiratory particle populations}  

The pathogen mixing state -- which can be defined in relation to \citet{RN1780} as the distribution of the pathogens across the carrier particle population -- is of prime relevance for a mechanistic understanding of aerosolization, transport, and deposition \cite{RN1683,RN1783,RN1829}. Small pathogens such as viruses are generally not emitted alone, but are embedded within (much) larger mucusalivary particles (Fig.\,\ref{fig:Mixing}) \cite{RN2049}. Accordingly, the pathogen-mucusalivary aerosol can be generally considered as an internal mixture \cite{RN1780,RN1652,RN1765,RN1610}. Note, however, that published data on the pathogen mixing state is sparse and microphysical details are widely unknown. For example, it is unknown whether small pathogens, such as virions, are located within the carrier particles or on their surface, which is relevant for their exposure to environmental conditions and, thus, their viability \cite[e.g.,][]{RN1610,RN1919}. It is further unknown, whether respiratory particles hold one or more pathogen copies (mono- vs poly-pathogen particles), which has implications for the assessment of infection risks \cite{RN1894}.   

If a given type of pathogen were embedded in emitted particles at a consistent concentration, one would expect the number of, e.g., virions to increase with particle size, and thus the dose response might scale with the cumulative particle volume (Fig.\,\ref{fig:Mixing}). If, instead, the pathogen were embedded with a relatively constant number of virions in a particular (e.g., smaller) particle size fraction, the dose response would likely rather scale with the number of inhaled particles in this specific size range (e.g., a specific mode of the PSD). Size-resolved sampling and detection of airborne pathogens from infected individuals, which allows important conclusions on the mixing state, is however experimentally demanding and corresponding studies are therefore rare \cite{RN1878}. The answer may also be more complicated still, e.g. as a mixture between these endpoints, where virions may be enriched on particle surfaces or in the bulk, but not completely devoid in the contrasting phase. The existing studies -- especially on \textit{M. tuberculosis} \cite{RN1695,RN1856,RN2018}, influenza virus \cite{RN1599,RN1827,RN1869,RN1605,RN2099,RN1783}, and SARS-CoV-2 \cite{RN1705,RN1652,RN1651,RN1937} -- suggest that pathogens are typically enriched in specific size ranges of the carrier PSD, likely defined by the particle formation mechanisms and sites, in relation to the site of infection (see Sect.\,\ref{sec:Emission}). Furthermore, PSDs combined with the pathogen number density in mucusalivary particles determines the infection risk associated with multi-pathogen aerosols, which has been recently addressed in a new dose-response model \cite{RN1894}.

Beyond the actually inhaled dose, the mixing state is further important for the dose-response relationship of a given pathogen. The size range of the carrier PSD in relation to its particle deposition properties defines which regions of the URT and LRT are reached. As different deposition sites in the respiratory tract can show different susceptibilities to a given pathogen, the dose-response can be size-dependent in terms of the inhaled PSD \cite[e.g.,][]{RN1783,RN1889,RN1978,RN2094}. Previous studies have shown clearly size-dependent infectivities for airborne influenza virions \cite{RN1803} as well as \textit{M. tuberculosis} \cite{RN1797,RN1798}, both being significantly more infectious if carried in smaller particles. Further, pathogen deposition in the LRT relative to the URT is often associated with higher severity, morbidity and fatality \cite[e.g.,][]{RN1683,RN1795,RN1912,RN1930,RN1829,RN2094}. \citet{RN1879} introduced the concept of \textit{anisotropic infection}, emphasizing that the clinical severity of certain disease depend on the mode of acquisition. This means that differences in pathogen deposition site in the respiratory tract can be associated with dramatic differences in disease development, ranging from an acute disease response to a much milder course.

\subsubsection{\label{sec:Survival}Decay of pathogen viability in aerosol}  

Another essential parameter in disease transmission through the aerosol route is the survival of the corresponding pathogens during airborne transport. The decline in the pathogen's viability defines the lifetime of its infectious potential upon airborne transport and must be considered in infection control measures and risk assessments. The inactivation typically follows a first-order exponential decay function 
\begin{equation} \label{eq:Decay}
    N_{p}(t)\,=\,N_{p,0}\,\exp\left(-k\,t\right)
\end{equation}

\noindent with $N_{p}(t)$ as the number of airborne and viable pathogens at a given time $t$, $N_{p,0}$ as the initial number of viable pathogens at $t=0$, and $k$ as the inactivation rate, which further gives the half-life as $T_{1/2}=\ln2/k$ \cite{RN1942,RN1949}. Various environmental factors, such as temperature, desiccation at low humidity, radiation, i.e., ultraviolet light, and reactive gaseous species can damage the pathogens' lipids, proteins, or nucleic acids and, thus inactivate them \cite[e.g.,][]{RN1783,RN1817,RN1938,RN2089,RN2092}. An overview of the effects of temperature and humidity on pathogen viability is provided by \citet{RN1950}. The pathogen viability further depends on the chemical microenvironment and, thus on ELF and saliva composition as well as water content \cite{RN1610,RN1817}. For example, ELF and saliva can act as an organic barrier against environmental exposure after droplet desiccation. Typical half-lives ($T_{1/2}$) for selected pathogens in aerosol are summarized in Table\,\ref{tab:HalfLife}. The variability in $T_{1/2}$ is large, and strongly enhanced decay rates were found, e.g., for increased temperature and irradiance \cite{RN1770,RN1938}.

\subsection{\label{sec:Hygros}Hygroscopic growth and shrinkage of respiratory particles in relation to their chemical composition and hygroscopicity}
 
The composition of the exhaled mucosal fluid droplets is complex, highly variable, and not well characterized \cite{RN1817,sarkar2019}. In the respiratory tract and, thus also in freshly emitted droplets, the fluids contain large mass fractions of water with $>$99\,\% in saliva \cite{RN1951} and $\sim$95\,\% in ELF \cite{RN1909,RN2040}. Dissolved in the water are various salts acting as electrolytes and buffers with sodium (\ce{Na+}), potassium (\ce{K+}), and calcium (\ce{Ca^2+}) as main cations as well as chloride (\ce{Cl-}), hydrogen carbonate (\ce{HCO3-}), and phosphates (mainly \ce{H2PO4-} and \ce{HPO4^2-}) as major anions (see Table\,\ref{tab:Composition}). Further, a broad variety of organic constituents is either dissolved or suspended in the water. This includes proteins for defensive purposes (e.g., lysozyme, immunoglobulins), glycoproteins responsible for the viscous and elastic gel-like properties of the fluids (i.e., mucins), as well as further compounds such as lipids, surfactants (i.e., phospholipids), cholesterol, and urea \cite{RN1909}. For instance, \citet{bredberg2012} detected more than 100 different proteins in exhaled particle samples. Further, the ELF composition changes from the upper conducting airways towards the alveolar region \cite{RN1737,RN1792,RN1793,RN1794,RN1817}. Table\,\ref{tab:Composition} provides a general overview of the main constituents and specifies typical concentration ranges, however, does not seek to resolve the entire chemical complexity or variability of the fluids.\par 

The life cycle of the respiratory particles is very dynamic \cite{RN704}. At the moment of exhalation, the particles are in a liquid state and can be regarded as droplets of mucosal fluids. The warm ($\sim$37\,\celsius), water-saturated ($\sim$100\,\%\,RH), and particle-laden air leaves the respiratory tract and typically experiences sudden changes in temperature and RH \cite[e.g.,][]{RN2028,RN2026,RN2027}. A frequent scenario is the exhalation into subsaturated water vapor conditions (i.e., $\text{RH}<100\,\%$), such as room air at typically 40 to 80\,\% RH \cite[e.g.,][]{RN1707,RN1611,RN1775,RN2063}. Under these conditions, the particles leave the respiratory tract at $\text{RH}\approx100\,\%$ with an initial diameter ($D_\text{exh}$). In relation to the particle composition and the associated hygroscopicity, evaporation of water occurs fast (i.e.,\,$<1$\,s) and causes a substantial decrease in $D$ \cite{RN1822,RN1823,RN2063,RN2079}. For complete evaporation, the droplets shrink completely to their dry diameter ($D_\text{dry}$) defined by the remaining nonvolatile solutes, whereas at typical intermediate RH levels the droplets shrink to an equilibrium (wet) diameter ($D_\text{wet}$) \cite{RN1707,RN1600,RN1737}. The ratio of the wet diameter as a function of RH ($D_\text{wet}\text{(RH)}$) and the dry diameter ($D_\text{dry}$) defines the particles' growth factor ($g_\text{d}$) as
\begin{equation}
    g_d=\frac{D_\text{wet}\text{(RH)}}{D_\text{dry}}\,\,\,.
\end{equation}

\noindent The cube of $g_\text{d}$ gives the volume growth factor ($g_\text{V}$) through $g_\text{V}=g_\text{d}^3$. Under certain conditions, such as during mixing of the respiratory puff with cold and humid outdoor air, also supersaturated water vapor conditions (i.e., $\text{RH}>100\,\%$) can occur \cite{RN2026,RN2027}. Supersaturation can cause an initial and significant growth of the droplets, as outlined in more detail below.\par
\pagebreak
\begin{table}[ht!]
\caption{\label{tab:Composition}%
Overview of main constituents in saliva and ELF. Table specifies the range of typical mass concentrations $\beta$ in mg per 100\,mL, reported in the literature. For saliva, the cited studies often compared concentrations under resting vs stimulated conditions, which were both implemented here. The values in brackets represent best guess values based on the (in the authors' view) most robust studies or means for values with low data coverage.% not aerosol ANA!
%adapted from Bicer, revised and updated.
}
%%%NEW TABLE%%%
\begin{ruledtabular}
\begin{tabular}{lrlrl}
\setcounter{footnoteB}{0}
Constituent          &\multicolumn{2}{c}{Saliva} & \multicolumn{2}{c}{ELF} \\
  &\multicolumn{2}{c}{$\beta$ [mg\,dL$^{-1}$]}   & \multicolumn{2}{c}{$\beta$ [mg\,dL$^{-1}$]}  \\[2pt]
\hline & \\[-1.7ex]
\textbf{Inorganic}             & & &                        &              \\
Ca                   & 2--11 [6]&\footnotemarkB[1]\textsuperscript{,}\footnotemarkB[9]\textsuperscript{,}\footnotemarkB[10]\textsuperscript{,}\footnotemarkB[11]\textsuperscript{,}\footnotemarkB[20]\textsuperscript{,}\footnotemarkB[26]        %sarkar2019,kumar2017a,Dawes1995,dawes1974,dawes1969,ben-aryeh1985
&&              \\
Cl                   & 30--130 [70]& \footnotemarkB[9]\textsuperscript{,}\footnotemarkB[10]\textsuperscript{,}\footnotemarkB[11] %dawes1974,dawes1969,Dawes1995
& 250--300 & \footnotemarkB[19]%\knowles1997
              \\
HCO\textsubscript{3}\textsuperscript{-}                  & 6--220 [85]&%dawes1974,dawes1969,Dawes1995
\footnotemarkB[9]\textsuperscript{,}\footnotemarkB[10]\textsuperscript{,}\footnotemarkB[11]
& 150--230 & \footnotemarkB[19]%\knowles1997     
\\
K                     & 51--130 [66]&\footnotemarkB[2]\textsuperscript{,}\footnotemarkB[9]\textsuperscript{,}\footnotemarkB[10]\textsuperscript{,}\footnotemarkB[11]\textsuperscript{,}\footnotemarkB[20]\textsuperscript{,}\footnotemarkB[26] %kumar2017a,dawes1974,dawes1969,Dawes1995,Sarkar2019,Ben-Aryeh1990
& 51--94&\footnotemarkB[19]%knowles1997
\\
Mg                     & 0.1--1.2&\footnotemarkB[1]\textsuperscript{,}\footnotemarkB[10]\textsuperscript{,}\footnotemarkB[24]%renke2016,dawes1974,ben-aryeh1985
&                        &              \\
Na                     & 0--130 [39]&\footnotemarkB[2]\textsuperscript{,}\footnotemarkB[9]\textsuperscript{,}\footnotemarkB[10]\textsuperscript{,}\footnotemarkB[11]\textsuperscript{,}\footnotemarkB[20]\textsuperscript{,}\footnotemarkB[26]%kumar2017a,dawes1974,dawes1969,Dawes1995,Sarkar2019,Ben-Aryeh1990
& 190--230&\footnotemarkB[19]%knowles1997
\\
Phosphates             & 20--220 [34]&\footnotemarkB[9]\textsuperscript{,}\footnotemarkB[10]\textsuperscript{,}\footnotemarkB[11]\textsuperscript{,}\footnotemarkB[24] %dawes1969,dawes1974,Dawes1995,renke2016
& &              \\[2pt]
%\quad Ammonium                   &20--130&\footnotemarkB[8]%renke2016&                        &              \\
%(F-)                   & & &                        &              \\
%(SN-)                  & & &                        &              \\
\hline & \\[-1.7ex]
\textbf{Organic}               &  & &                       &              \\
Glutathion             & $\leq$0.1 &\footnotemarkB[8]\textsuperscript{,}\footnotemarkB[17]%iwasaki2006,dauletbaev2001
&                        3--5&\footnotemarkB[3]\textsuperscript{,}\footnotemarkB[13]\textsuperscript{,}\footnotemarkB[32]%bicer2014,vandervliet1999,hatch1992
\\
Lactate                & 1--50 [12]&\footnotemarkB[11]%Dawes1995 
&                      &              \\
Lipids                 & 1.4--3 &\footnotemarkB[21]\textsuperscript{,}\footnotemarkB[24]%larsson1996,renke2016
&489--1204&\footnotemarkS[1]      \\% footnote the sum of subsequent entries!
&&& [762] &\footnotemarkS[1]\\
\quad Cholesterol            &0.13--50 [8]%larsson1996 (0.13) ,kumar2017a
&\footnotemarkB[20]\textsuperscript{,}\footnotemarkB[21] &    9--14 [12]%Bicer2014      
&\footnotemarkB[3] \\
\quad Phospholipids          & $<$0.1& \footnotemarkB[21]%larsson1996
&                    480--1190&\footnotemarkB[3]\textsuperscript{,}\footnotemarkB[16]\textsuperscript{,}\footnotemarkB[33]%bicer2014,hull1997,(veldhuizen1998)
\\ 
&&&[750]&\\% footnote! from total lipids 80\% are phosphatidylcholine half of which is disaturated most in the form of DPPC, see Veldhuizen1998, in agreement with hull %phospholipids account for approximately 90\% mass(?) of RTLF, rest 10% (goerke1974) proteins according to Bredberg2012)
\quad \quad DPPC &&& 330&\footnotemarkB[16]\textsuperscript{,}\footnotemarkB[33]%hull1997,(Veldhuizen1998)
\\[4pt]
Proteins               & 128--640 &
\footnotemarkB[2]\textsuperscript{,}\footnotemarkB[6]\textsuperscript{,}\footnotemarkB[9]\textsuperscript{,}\footnotemarkB[12]\textsuperscript{,}
%Dawes1969 (110-390), rantonen2000 (120),cheaib (157), kumar2017a (140-640), sarkar(126-142),dwyer(140),suh2009 (186),ben-aryeh1990 (128)
&                   470--1290 &\footnotemarkB[7]\textsuperscript{,}\footnotemarkB[30]\textsuperscript{,}\footnotemarkB[32]%dargaville1999 (4700) ,sutinen1995 (10700) ,vandervliet1999 (9500)
\\
&[220]&\footnotemarkB[20]\textsuperscript{,}\footnotemarkB[23]\textsuperscript{,}\footnotemarkB[26]\textsuperscript{,}\footnotemarkB[29]&[1000]&
\\
\quad Albumin               & 8--50 [20]%meurman2002 (40,50,20) ,rantonen2000 (17), cheaib2013 (9.5) ,henskens1992 (8)
&\footnotemarkB[6]\textsuperscript{,}\footnotemarkB[15]\textsuperscript{,}\footnotemarkB[22]\textsuperscript{,}\footnotemarkB[23]  &  290--730& \footnotemarkB[19]\textsuperscript{,}\footnotemarkB[30]%sutinen1995 (290-470), knowles1997 (730)
\\ %albumin (26% of the total protein content in RTLF Bredberg2012)
\quad Amylase                &5--121 [38]    &\footnotemarkB[6]\textsuperscript{,}\footnotemarkB[20]%kumar2017a (38) ,cheaib2013
&                     0 & \footnotemarkB[5]%bredberg2012             
\\ %amylase 60% of total protein in saliva vitorino2004
\quad IgA                    & 3--19%,cheaib2013, rantonen2000,brandtzaeg1998,ben-aryeh1990
&\footnotemarkB[2]\textsuperscript{,}\footnotemarkB[4]\textsuperscript{,}\footnotemarkB[6]\textsuperscript{,}\footnotemarkB[23] & 4--140& \footnotemarkB[3]\textsuperscript{,}\footnotemarkB[13]\textsuperscript{,}\footnotemarkB[30]
%bicer2014,hatch1992,sutinen1995  
\\
\quad IgG                    & 1.1--1.4%rantonen2000,brandtzaeg1998
&\footnotemarkB[4]\textsuperscript{,}\footnotemarkB[23]& 56--260&
\footnotemarkB[3]\textsuperscript{,}\footnotemarkB[13]\textsuperscript{,}\footnotemarkB[25]\textsuperscript{,}\footnotemarkB[30]%bicer2014,hatch1992,sutinen1995,rennard1990 
\\
\quad IgM                    & 0.21--0.48%brandtzaeg1998,rantonen2000
&\footnotemarkB[4]\textsuperscript{,}\footnotemarkB[23] & 1.3--10&\footnotemarkB[3]\textsuperscript{,}\footnotemarkB[30]%sutinen1995,bicer2014  
\\
\quad Lysozyme & 10--22 &\footnotemarkB[20]\textsuperscript{,}\footnotemarkB[23]%rantonen2000, kumar2017a
&    3--250 &\footnotemarkB[3]\textsuperscript{,}\footnotemarkB[13]\textsuperscript{,}\footnotemarkB[31]%bicer2014,hatch1992,thompson1990
\\
\quad Mucins & 6--55 [22]&\footnotemarkB[6]%cheaib2013
& $\sim$1&\footnotemarkB[14]%henderson2014
\\
%\quad MUC5B (mucin MG1) & 16 %zhang2013
%&&& \\
%\quad MUC7 (mucin MG2) &&&& \\
\quad Transferrin            & 0.3--1.2%kang2018,suh2009
&\footnotemarkB[18]\textsuperscript{,}\footnotemarkB[29] &                         30--170 [105]     &\footnotemarkB[3]\textsuperscript{,}\footnotemarkB[25]%rennard1990,bicer2014
\\
%alpha-1-antitrypsin    &    & &                     &              \\
\quad SP-A                   & $<$0.1%schicht2015
&\footnotemarkB[27] & 1.9--41 [30]%hull1997,dargaville
&\footnotemarkB[7]\textsuperscript{,}\footnotemarkB[16]\\ %SP-A, 12% of the total protein  RTLF Bredberg2012
\quad SP-B                   & $<$0.1 %schicht2015
& \footnotemarkB[27] & 34--120 [77]&\footnotemarkB[16]  %hull1997
             \\[4pt]
%\quad Peroxidases            &    & &                     &              \\
%Glutathione Peroxidase &    & &                     &              \\
%\quad Catalase               &    & &                     &              \\
Urate                 & 0.5--21 [2]  &\footnotemarkB[20]  %kumar2017a 
&                      1.6--3.5
&\footnotemarkB[3]\textsuperscript{,}\footnotemarkB[13]\textsuperscript{,}\footnotemarkB[28]\textsuperscript{,}\footnotemarkB[32]    %bicer2014,vandervliet1999,hatch1992,slade1993
\\
Urea                   &12-70 [30]%kumar2017 (12-70) ,dwyer2004 (32.4), meurman (29.4-48)
&\footnotemarkB[12]\textsuperscript{,}\footnotemarkB[20]\textsuperscript{,}\footnotemarkB[22]  &     27                  &\footnotemarkB[12] %dwyer2004             
\\
%Fatty acids                   &10  &\footnotemarkB[1]%renke2016 &                       &              \\
%Ascorbate              &  & &                       7--50&%bicer2014,vandervliet1999,hatch1992,slade1993,cross1994,sun2001\\
\end{tabular}
\end{ruledtabular}
\raggedright{
\footnotetextB{\citet{ben-aryeh1986}}, %1
\footnotetextB{\citet{ben-aryeh1990}}, %2
\footnotetextB{\citet{bicer2014}}, %3
\footnotetextB{\citet{brandtzaeg1998}}, %4
\footnotetextB{\citet{bredberg2012}}, %5
\footnotetextB{\citet{cheaib2013}}, %6
\footnotetextB{\citet{dargaville1999}}, %7
\footnotetextB{\citet{dauletbaev2001}}, %8
\footnotetextB{\citet{dawes1969}}, %9
\footnotetextB{\citet{dawes1974}}, %10
\footnotetextB{\citet{dawes1995}}, %11
\footnotetextB{\citet{dwyer2004}}, %12
\footnotetextB{\citet{hatch1992}}, %13
\footnotetextB{\citet{henderson2014}}, %14
\footnotetextB{\citet{henskens1993}}, %15
\footnotetextB{\citet{hull1997}}, %16
\footnotetextB{\citet{iwasaki2006}}, %17
\footnotetextB{\citet{kang2018}}, %18
\footnotetextB{\citet{knowles1997}}, %19
\footnotetextB{\citet{kumar2017a}}, %20
\footnotetextB{\citet{larsson1996}}, %21
\footnotetextB{\citet{meurman2002}}, %22
\footnotetextB{\citet{rantonen2000}}, %23
\footnotetextB{\citet{renke2016}}, %24
\footnotetextB{\citet{rennard1990}}, %25
\footnotetextB{\citet{sarkar2019}}, %26
\footnotetextB{\citet{schicht2015}}, %27
\footnotetextB{\citet{slade1993}}, %28
\footnotetextB{\citet{suh2009}}, %29
\footnotetextB{\citet{sutinen1995}}, %30
\footnotetextB{\citet{thompson1990}}, %31
\footnotetextB{\citet{vandervliet1999}}, %32
\footnotetextB{\citet{veldhuizen1998}}%33
\newline
\footnotetextS[1]{Calculated sum of cholesterol and phospholipids.} \newline
DPPC: Dipalmitoylphosphatidylcholine,\newline Ig: Immunoglobulin, 
SP: Surfactant protein\newline
}
\setcounter{footnote}{0}
\end{table}

% \cite{Vejerano RN1610, Effros: RN1779, Davies: RN1922, Larsson: RN1776, Niazi: RN1817}.

% Ohar 2019: In the airways, the surface liquid comprises at least 2 layers: the mucus (gel) layer and the periciliary layer surrounding the cilia which consists of a gel mesh of cell-tethered mucins and polysaccharides. Mucus consists of a mixture of fluid and secretions from surface epithelium and submucosal glands and is predominantly comprised of water (95%) glycoproteins (2%–3%), proteoglycans (0.1% –0.5%), lipids (0.3% – 0.5%), proteins, and DNA.7 The glycoprotein component consists of secreted mucins, particularly the large polymeric structures MUC5AC and MUC5B, which account for the rheologic properties of mucus

% Bansil (RN1909): The physical state of the mucus, change in the concentration of secreted mucin, and the strong dependence of its physicochemical properties on environmental factors such as ionic strength and pH play an important role in many diseases.

\begin{figure}[ht]
  \includegraphics[width=7.6cm]{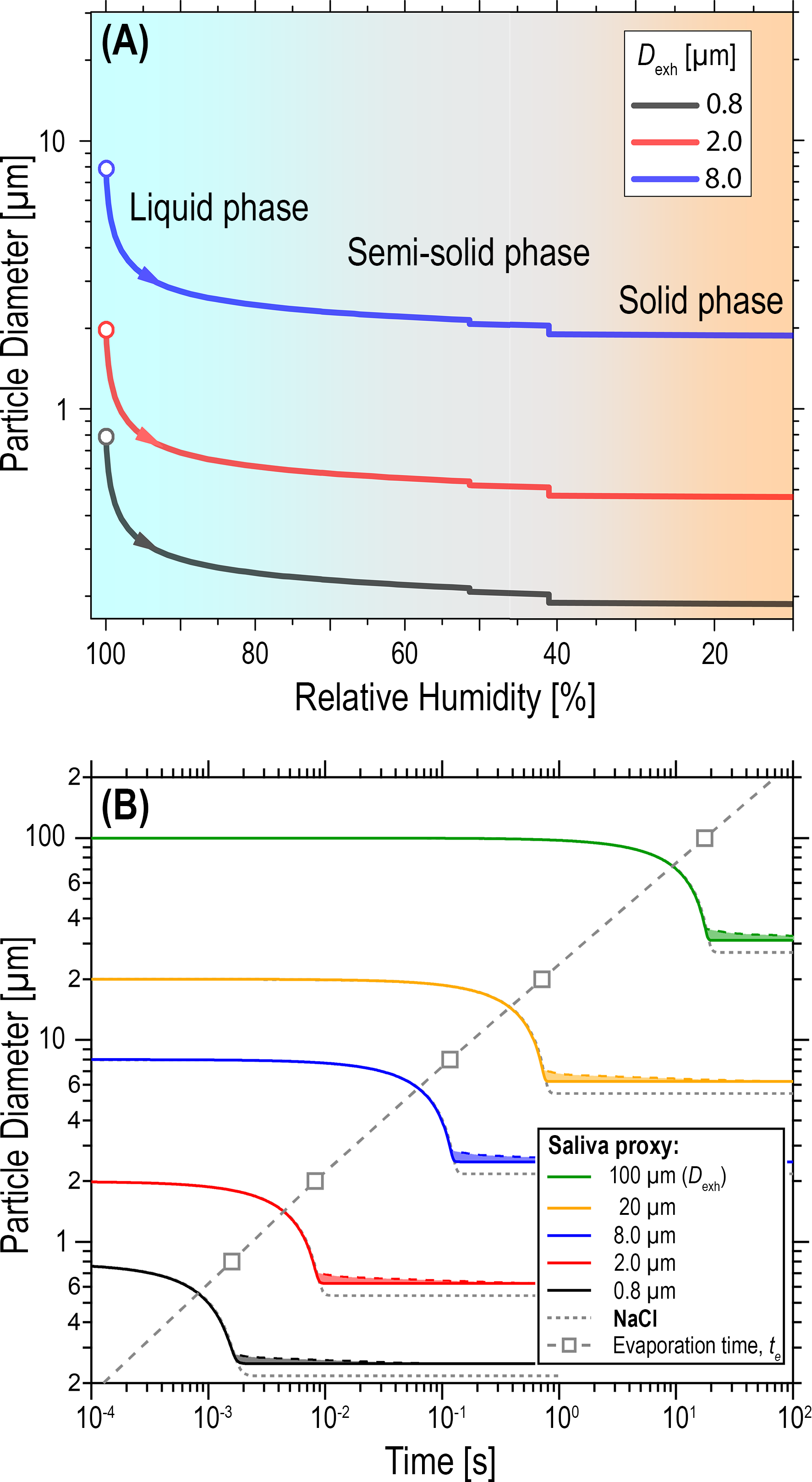}
  \caption{\label{fig:SalivaDrying} Drying of saliva proxy droplets consisting of aqueous NaCl, KCl, and protein (BSA) with dry volume fractions of $\epsilon_\text{NaCl}=0.1$, $\epsilon_\text{KCl}=0.1$, and $\epsilon_\text{BSA}=0.8$. \textbf{(A)} Thermodynamic modeling: equilibrium diameters plotted against relative humidity (RH) assuming different initial diameters, $D_\text{exh}$ (open circles), upon exhalation at 100\,\% RH. Steps near 50\,\% RH and 40\,\% RH correspond to efflorescence phase transitions of NaCl and KCl, respectively. \textbf{(B)} Kinetic modeling: droplet diameters plotted against time assuming different initial diameters $D_\text{exh}$ upon exhalation at 99.5\,\% RH into an environment at 55\,\% RH. Solid colored lines represent saliva proxy droplets with shadings indicating uncertainties related to different diffusivity parameterizations for semisolid phase state $\lesssim$ 80\,\% RH. Dotted grey lines represent aqueous NaCl droplets for comparison. Grey squares and dashed grey line represent the characteristic time of evaporation $t_e$ (e-folding time) plotted as a function of initial particle diameter $D_\text{exh}$. } 
\end{figure}\par
\clearpage

Figure\,\ref{fig:SalivaDrying} illustrates the drying process for saliva proxy droplets consisting of an aqueous mixture of salts and protein.\footnote{
Saliva proxy comprises salts NaCl and KCl and protein bovine serum albumin, BSA, with volume fractions $\epsilon_\text{NaCl}=0.1$, $\epsilon_\text{KCl}=0.1$, $\epsilon_\text{BSA}=0.8$. BSA was chosen as proxy protein for complex protein mixture in real saliva (compare Table\,\ref{tab:Composition}).
} It shows, that the droplets -- after emission from the respiratory tract with $\text{RH}\approx100\,\%$ into an environment with $\text{RH}<100\,\%$ -- shrink substantially in the initial phase of drying. The shrinkage curve in Fig.\,\ref{fig:SalivaDrying}A represents thermodynamic equilibrium states, however, particles might deviate from the equilibrium curve if changes in RH are rapid. A droplet's evaporation time is roughly proportional to $D_\text{exh}^2$ \cite{RN2087,RN2088,RN2086,RN433} according to
\begin{eqnarray}
D(t)^2=D_\text{exh}^2+\frac{8 D_\text{g} M_\text{w}}{\rho_\text{w}} (c_\text{g} - c_\text{gs}) \cdot t
\label{eq:D2rule}
\end{eqnarray}

\noindent in which $D_\text{g}$ is the diffusion coefficient of water in the gas phase, $M_\text{w}$ the molar mass of water, and $\rho_\text{w}$ the density of water. Note that this classical so-called $D^2$-law is based on the assumption of isolated, pure liquid droplets evaporating into a homogeneous environment of a given RH \citep{RN2026} and holds for multi-component mixtures only in the early stages of the evaporation process, when the difference between far-surface water concentration and near-surface water concentration $(c_\text{g} - c_\text{gs})$ is approximately constant \cite{RN2090,RN2101}. \par

%\begin{figure}[t]
 % \includegraphics[width=8cm]{figs/Fig_Saliva_Drying_Lifetime_v5.png}
%  \caption{\label{fig:SalivaDryingLifetime}Characteristic time of evaporation $t_e$ (e-folding time) for saliva proxy droplets plotted against initial droplet diameter $D_\text{exh}$ upon exhalation at 99.5\,\% RH into an environment with 55\,\% RH (KM-GAP model results, analogous to Fig.\,\ref{fig:SalivaDrying}B).}
%\end{figure}

The drying dynamics of saliva proxy particles are illustrated in Fig\,\ref{fig:SalivaDrying}B using calculations with the kinetic multi-layer model of gas-particle interactions in aerosols and clouds, KM-GAP \cite{RN2080}. Processes limiting particle evaporation are the gas-phase diffusion of water molecules away from the particle surface, heat transfer to the particle surface after evaporational cooling, and, potentially, bulk diffusion of water molecules through the partially dried-out salt-protein matrix. While small particles equilibrate on the millisecond time scale, large particles may take seconds to reach thermodynamic equilibrium, especially if the diffusion of water through the salt-protein matrix is slow. \par

The shadings in Fig\,\ref{fig:SalivaDrying}B illustrate the minor effect of three different diffusivity parameterizations ranging from fully liquid particles to particles undergoing a liquid-to-semisolid phase transition and significant drop in water diffusivity between 75 and 80\,\% RH.\footnote{
Diffusivity parameterizations employed for the calculations in this study include (i) a constant diffusivity of water that would be expected in pure liquid water droplets $D_\text{w} = 1\times10^{-5}$ cm$^2$ s$^{-1}$ (lower boundaries in Fig.\,\ref{fig:SalivaDrying}B); (ii) a Vignes-type mixing rule \cite{RN2083} $D_\text{w} = D_\text{w,w}^{x_\text{w}} \cdot D_\text{w,s}^{1-x_\text{w}}$ between pure liquid water diffusing at $D_\text{w,w} = 1\times10^{-5}$ cm$^2$ s$^{-1}$) and water in a semisolid salt-protein matrix diffusing at $D_\text{w,s} = 1\times10^{-10}$ cm$^2$ s$^{-1}$ (solid lines in Fig.\,\ref{fig:SalivaDrying}B); (iii) percolation theory \cite{RN409,RN2081,RN2082} assuming a coordination number of $Z=4$, a packing fraction $f=0.95$ and values for $D_\text{w,w}$ and $D_\text{w,s}$ as defined above (upper boundaries in Fig.\,\ref{fig:SalivaDrying}B).} A Vignes-type mixing rule between liquid water and a semi-solid salt-protein matrix yielded near identical results to the fully liquid scenario. \par

\begin{figure}[!ht]
  \includegraphics[width=7.4cm]{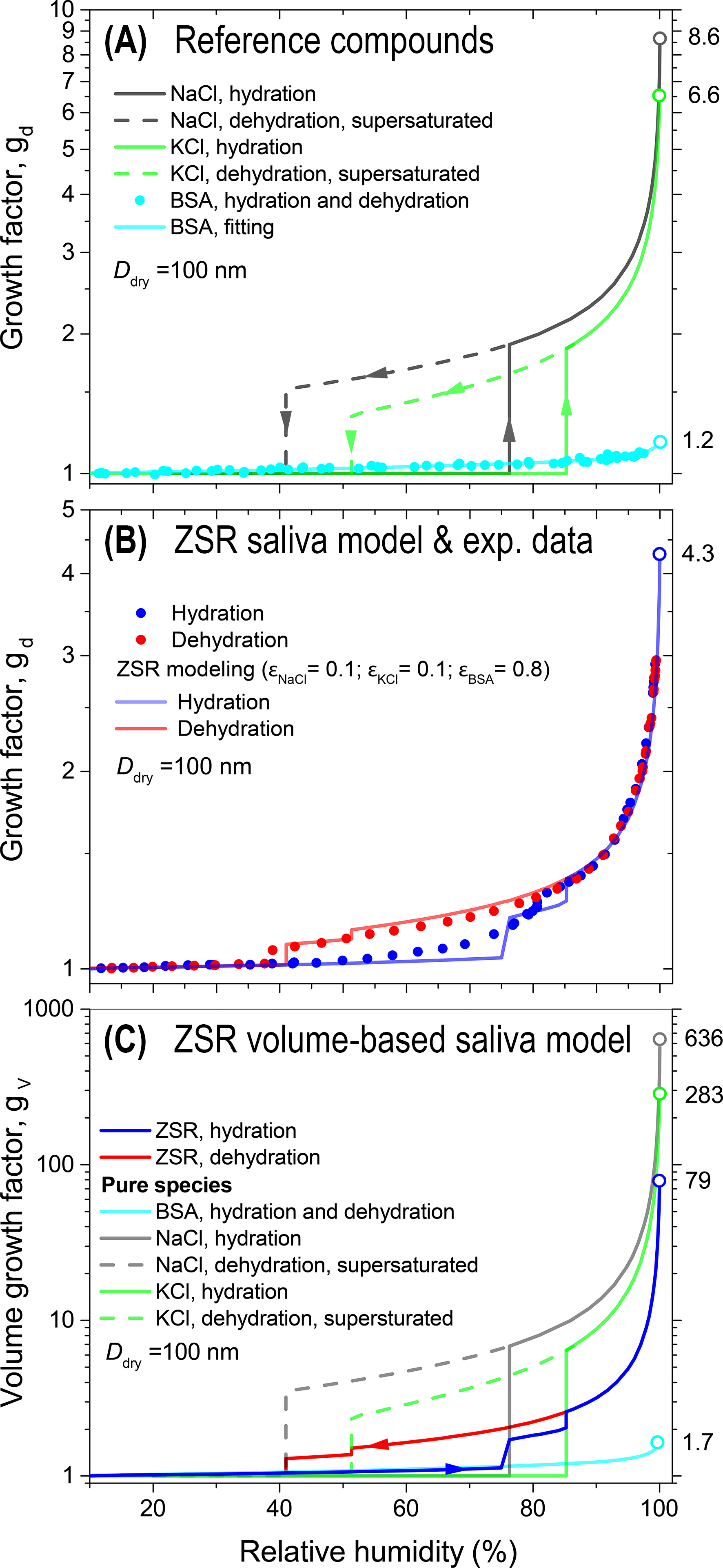}
  \caption{\label{fig:Hygros1} Hygroscopic growth curves as a function of RH (up to 99.5\,\%) for the reference compounds NaCl, KCl, and protein (BSA) \textbf{(A)} as well as for authentic saliva \textbf{(B,C)}, showing a hysteresis shape in growth factor $g_\text{d}$ with deliquescence and efflorescence phase transitions. Panel \textbf{(C)} shows the volume hygroscopic growth factor ($g_\text{V}$) corresponding to $g_\text{d}$ in \textbf{(B)}. The modelled data (solid and dashed lines) for pure compounds and saliva proxy system (with volume fractions $\epsilon_\text{NaCl}=0.1$, $\epsilon_\text{KCl}=0.1$, $\epsilon_\text{BSA}=0.8$) were obtained from the Aerosol Inorganic Model (AIM) \cite{RN1968} and the Aerosol Inorganic-Organic Mixtures Functional groups Activity Coefficients (AIOMFAC) models \cite{RN1971} \cite{RN1997,RN1809}. The experimental data (circular markers) in \textbf{(B)} were obtained from high humidity tandem differential mobility analyser (HHTDMA) analysis according to \citet{RN1809} for stimulated saliva from three healthy, non-smoking individuals.}
\end{figure}

The grey dotted lines in Fig.\,\ref{fig:SalivaDrying}B show the corresponding drying dynamics of pure NaCl solution droplets, as shown previously and in good agreement with other modelling approaches \cite{RN2063}. The KM-GAP model returns near-identical evaporation speeds for saliva proxy and NaCl particles. \footnote{Water activity parameterizations for NaCl particles were obtained using the Extended Aerosol Inorganics Model (E-AIM) \cite{RN1968}, water activity in saliva particles was parameterized according to the ZSR model presented in Fig.\,\ref{fig:Hygros1}B.} Due to the consumption of latent heat during evaporation, particles cool to a minimum temperature of $\sim291.7$\,K, irrespective of initial particle size. This temperature lies close to the wet bulb temperature, which is the lower limit at which net evaporation can still take place under the given ambient conditions. \par

The grey squares and dashed grey line in Fig.\,\ref{fig:SalivaDrying}B represent the characteristic e-folding time of evaporation $t_e$, which is the time needed for a droplet to shrink by a factor of 1/$e$ relative to its initial size $D_\text{exh}$: 
\begin{eqnarray}
    t_e\, =\, D_\text{exh}^2\cdot\frac{\left(e^{-2}-1\right)\rho_\text{w}}{8 D_\text{g}M_\text{w}(c_\text{g}-c_\text{gs})}
\label{eq:D3scaling}
\end{eqnarray}

\noindent It exhibits a $D^2$-dependence analogous to the droplet evaporation rate (Eq.\,\ref{eq:D2rule}), which applies as long as the evaporation kinetics are governed by gas diffusion (rather than bulk diffusion or phase transitions). \par

A comparison of these evaporation times with the sedimentation times of larger respiratory droplets -- as shown in Fig.\,\ref{fig:Re} -- reveals that the largest droplets (i.e., $\sim$100\,\micro m and larger) may settle too fast to reach an equilibrium state with the ambient RH \cite{RN1600,RN1822,RN1609,RN1817} and can stay in a water-rich state over their entire lifetime. \citet{RN2026} further showed that the influence of the local RH field around the droplets in the exhaled humid puffs has to be considered as it tends to delay the evaporation significantly (i.e., factor 30 or even larger). In this sense, the calculations in Fig.\,\ref{fig:SalivaDrying}B (assuming an instantaneous change in environment RH to 55\,\%) represent lower limits of the evaporation times (or upper limits for the equilibration rate/speed) of characteristic droplet sizes without the influence of the surrounding puff of humid exhalation air \cite{RN2084,RN2085}.\par  

Figure\,\ref{fig:SalivaDrying}A suggests a shrinkage factor of $D_\text{exh}/D_\text{dry}\approx4.5$ for complete drying of saliva as well as $D_\text{exh}/D_\text{wet}\approx4$ for drying to typical room RH levels (i.e., $\sim$40 to $\sim$80\,\% RH). This is consistent with the shrinkage expected based on typical salt concentrations in saliva (Table\,\ref{tab:Composition}): the drying of a saline solution with $\sim$1\,\% NaCl \cite[a typical value for saliva, from][]{RN1951} would result in $g_\text{d}\approx6$. In previous studies on respiratory particles, estimates or measurements of $g_\text{d}$ ranged from $\sim$2 and $\sim$6 \cite[e.g.,][]{RN1701,RN1600,RN1737,RN1822,RN1823,RN1658,RN1817,RN2019,RN2063,RN2079}.\par

The hygroscopic growth and shrinkage of the respiratory particles can be calculated by the K\"ohler theory, which describes thermodynamic water equilibrium states between the gas and aqueous phases \cite{RN1546}. Specifically, it describes the interplay of the enhancement in water saturation vapor pressure ($p_0$) over a curved relative to a flat surface and the reduction in $p_0$ over a solute surface relative to pure water. The K\"ohler theory has been broadly applied in atmospheric research as it allows to describe and model the water uptake and loss by ambient aerosol particles under variable RH conditions \cite{RN433}. It has been particularly important to describe the nucleation of cloud droplets by ambient aerosol particles \cite{RN330}. The corresponding K\"ohler equation 
\begin{eqnarray}
    s\, &&=\, a_\text{w}\,Ke\\
    \text{with}\quad Ke\, &&=\,\exp\left(\frac{4\,\sigma_\text{sol}\,M_\text{w}}{R\,T\,\rho_\text{w}\, D_\text{wet}}\right)\nonumber
\label{eq:Kohler_1}
\end{eqnarray}

\noindent expresses the necessary conditions for an aqueous solution droplet to be in equilibrium state with the water vapor of the surrounding gas. Specifically, it relates the water vapor saturation ratio ($s$) to the \textit{Raoult term}, which is the water activity in the aqueous solution ($a_\text{w}$) and describes the size and composition dependencies of the droplet's solute effect, as well as the \textit{Kelvin term} ($Ke$), which describes the increase in equilibrium water vapor pressure due to droplet's surface curvature \cite{RN1546,RN433}. In $Ke$, $\sigma_\text{sol}$ is the solution droplet's surface tension, $M_\text{w}$ is the molar weight of water, $\rho_\text{w}$ is the water density, $R$ is the universal gas constant, $T$ is the absolute temperature, and $D_\text{wet}$(RH) is the droplet diameter at a given $s$ or RH, with $s = RH/100\,\%$ \cite{RN433}. The effect of droplet curvature becomes important for $D_\text{wet}<0.1$\,\micro m (Eq.\,\ref{eq:Kohler_1}).\par 

Different approximations and parameterizations exist to describe $a_\text{w}$ and $\sigma_\text{sol}$ as a function of the droplet's chemical composition as outlined systematically in \citet{RN370}. Typically, $\sigma_\text{sol}$ is approximated by the surface tension of pure water. A commonly used parameterization of $a_\text{w}$ is based on the hygroscopicity parameter, $\kappa$, introduced by \citet{RN669} according to
\begin{equation}\label{eq:pk2}
    a_\text{w} = \left(1+\kappa\,\frac{V_\text{s}}{V_\text{w}}\right)^{-1}
\end{equation}

\noindent with $V$ as the volumes of the dry solute (s) and pure water (w). The parameter $\kappa$ reflects the chemical composition of the solutes in the droplets. It is widely used in atmospheric aerosol research \cite[e.g.,][]{RN709,RN895,RN1355} and (for ambient aerosol samples) typically ranges from $\sim$0.1 for organic solutes to $\sim$0.9 for salts \cite{RN330}. According to \citet{RN370} and \citet{RN669}, $\kappa$ can be related to fundamental properties of the water and solute as well as the solute's dissociation behavior through 
\begin{gather}
    \label{eq:rA27}
    \kappa = i_\text{s}\frac{n_\text{s} V_\text{w}}{n_\text{w} V_\text{s}} = i_\text{s}\frac{\rho_{\text{s}}M_\text{w}}{\rho_{\text{w}}M_\text{s}}
    \,\,\,\,\,\text{with}\,\,\,\,\,i_\text{s}\,\approx\,\nu_s \Phi_s
\end{gather}

\noindent with $n$ as the numbers of moles, $\rho$ as the densities, and $M$ as molar masses of the dry solutes (s) and pure water (w) as well as the van't Hoff factor of the solute ($i_\text{s}$) with $\nu_s$ as the stoichiometric dissociation number and $\Phi_s$ as the molar osmotic coefficient in aqueous solution. To calculate $\kappa$ based on $g_\text{d}$, it is convenient to transform Eq.\,\ref{eq:pk2} into
\begin{equation}
    \label{eq:gK}
    \kappa = (g_V-1)\frac{(1-a_\text{w})}{a_\text{w}}\,\,\,.
\end{equation}

Figure\,\ref{fig:Hygros1}A shows the modelled hygroscopic growth and shrinkage of pure 100\,nm NaCl and KCl particles, which are both main constituents of saliva (Table\,\ref{tab:Composition}). Both salts show a pronounced hysteresis in $g_\text{d}$ with sharp phase transitions at their deliquescence and efflorescence relative humidities (DRH and ERH). For increasing RH starting from dry particles (i.e., $<$ERH), a sudden deliquescence phase transition occurs at $\text{DRH}\approx75\,\%$ for NaCl and at $\text{DRH}\approx85\,\%$ for KCl. In the course of the drying from $\text{RH}>\text{DRH}$, the efflorescence phase transitions occur at $\text{ERH}\approx40\,\%$ for NaCl and at $\text{ERH}\approx50\,\%$ for KCl \cite{RN1977}. Further shown in Fig.\,\ref{fig:Hygros1}A is the hygroscopic growth curve of BSA as a proxy for the complex protein mixture in ELF and saliva. The protein (as well as organic compounds in general) shows a significantly lower $g_\text{d}$ than the salts and is further characterized by the absence of a hysteresis \cite{RN369,RN1313}. The experimental $g_\text{d}$ data points for BSA originate from \citet{RN1819} and were fitted by a polynomial three-parameter function according to \citet{RN1976} as follows
\begin{gather}
    g_d=\left[1+\left(k_1+k_2s+k_3s^2\right)\frac{s}{1-s}\right]^{1/3}%\\%\text{with}\,\,\,s=\frac{RH}{100\,\%}\nonumber
\end{gather}

\noindent using $k_1= 0.111$, $k_2= 0.0239$, and $k_3= -0.131$, which yields a good fit with $\text{R}^2 = 0.953$.\par

\begin{figure}[t]
  \includegraphics[width=8cm]{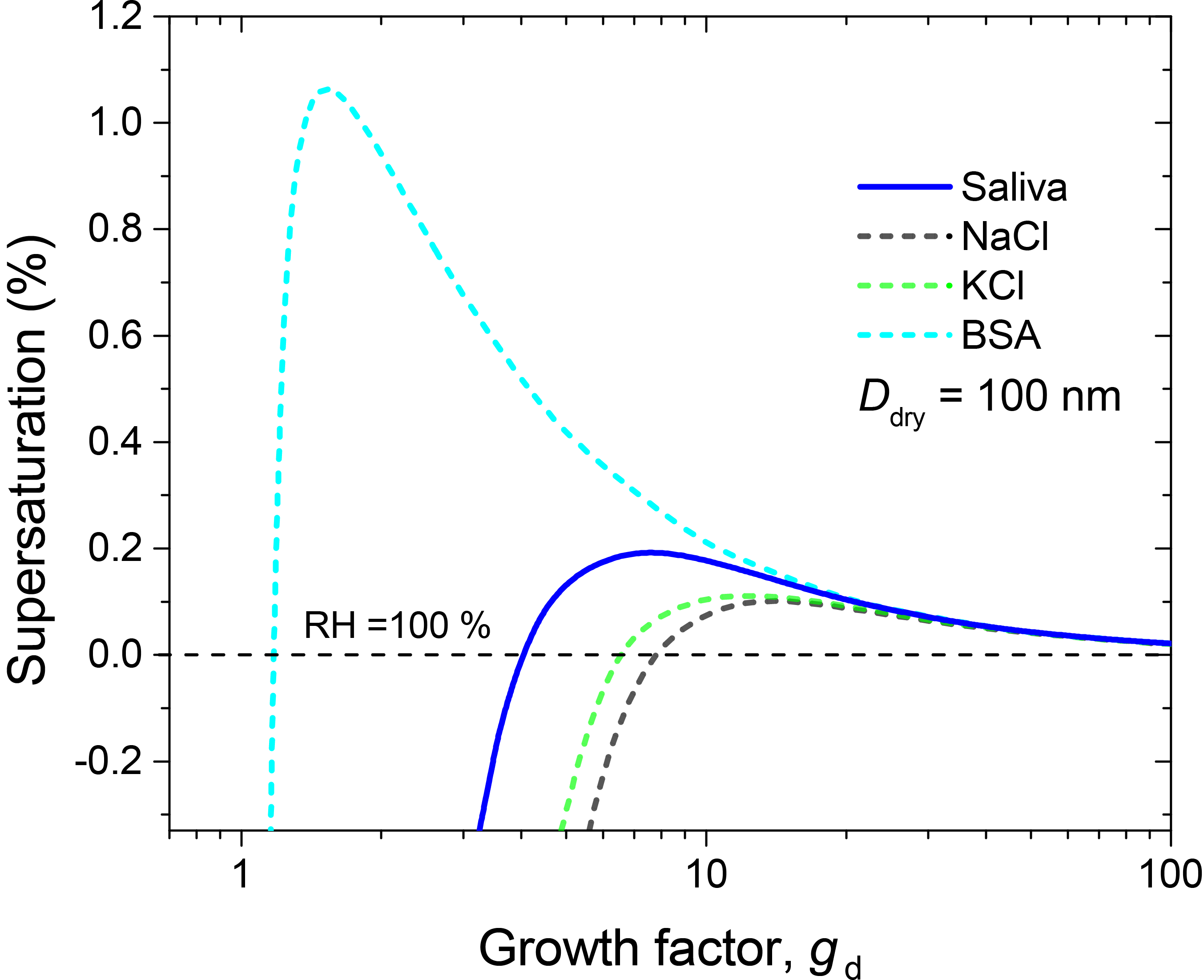}
      \caption{\label{fig:SuperSat} Hygroscopic growth behavior, represented by growth factor $g_\text{d}$, for the reference compounds NaCl, KCl, and the protein bovine serum albumin (BSA) under supersaturated water vapor conditions.}
\end{figure}

\citet{RN369} generally defined deliquescence as a transformation of a solid or semisolid substance into a liquid aqueous solution, with gas phase water being absorbed -- also called liquefaction/liquescence upon humidification/hydration. Efflorescence is a transformation of a substance from a liquid aqueous solution into a (semi-)solid phase upon water evaporation -- also called solidification upon drying/dehydration. Below DRH, the dissolved salts in the particles are supersaturated until the efflorescence phase transition occurs, which is a kinetically limited, homogeneous nucleation process and depends on particle size as well as impurities \cite{RN2008,RN2004}. Accordingly, for NaCl -- as a main constituent in saliva -- a range of ERH values from 37 to 50\,\% has been reported \cite[][and references therein]{RN704,RN1819,RN2004}. Figure\,\ref{fig:Hygros1}A and C show the NaCl ERH at $\sim$40\,\%, which is at the lower end of the aforementioned RH range. DRH values $>$50\,\% are commonly attributed to heterogeneous nucleation due to the presence of impurities \cite{tang1994,RN1819,RN2007}.\par

Figure\,\ref{fig:SuperSat} relates to Fig.\,\ref{fig:Hygros1}A and shows the modelled behavior of $g_\text{d}$ beyond water saturation into the supersaturated regime. Under certain conditions, such as breathing into cold and humid air, the respiratory particles can experience an episodic occurrence of supersaturated conditions \cite{RN2027}. Figure\,\ref{fig:SuperSat} shows that as long as the supersaturated conditions prevail, the exhaled particles might grow substantially before the onset of droplet drying in the course of dilution and dissipation of the exhaled puffs.\par

As a prediction of the expected hygroscopic behavior of saliva, the three-compound system with the main constituents NaCl, KCl, and BSA was modelled and is shown in Fig.\,\ref{fig:Hygros1}B. The overall $g_\text{d}$ of multi-compound systems can usually be approximated accurately as the additive influence of the $g_\text{d,i}$ from the individual compounds $i$ as well as their corresponding volume fractions $\epsilon_i$ in the mixture
\begin{equation}
    \label{eq:ZSR}
    g_d=\left(\sum_{i} \epsilon_{i}{g_{d,i}}^3\right)^\frac{1}{3}\,\,\,\text{and}\,\,\,\epsilon_i = \frac{{V}_{\text{s}i}}{{V}_\text{s}}
\end{equation}

\noindent with $V_{\text{s}i}$ as the volume of the individual components and $V_{\text{s}}$ as the total particle volume. This approach is based on the Zdanovskii–-Stokes–-Robinson (ZSR) model, which assumes independent water uptake of individual components in mixtures \cite{RN1969}.\par 

The resulting $g_\text{d}$ of the NaCl-KCl-BSA system in Fig.\,\ref{fig:Hygros1}B is characterized by a hysteresis with two subsequent deliquescence phase transitions, reflecting the influence of both salts. These model predictions were compared to the results of a 'proof-of-concept' hygroscopicity measurement of authentic saliva and the corresponding data points were added to Fig.\,\ref{fig:Hygros1}B.\footnote{
The experimental proof-of-concept data in Fig.\,\ref{fig:Hygros1}B were obtained from high humidity tandem differential mobility analyser (HHTDMA) analysis in the RH range from 2 to 99.5\,\% according to \citet{RN1809} for stimulated saliva from three healthy, non-smoking individuals. The collected saliva was diluted (1\,ml aliquot from each sample mixed with 75\,ml of pure water), filtered through a 5\,\micro m syringe filter (25 mm  GD/X, sterile, 6901-2504, GE Healthcare Life Science, Whatman), and nebulized for HHTDMA analysis. The HHTDMA procedure and data analysis is outlined in detail in \citet{RN1809} and \citet{RN1849}. Briefly, three operation modes are available for the HHTDMA instrument: a restructuring mode, a hydration mode, and a dehydration mode. The restructuring mode was used to specify the optimal RH range, in which initially irregular particles transform into compact spherical particles. In hygroscopic growth experiments, the restructuring mode was coupled in-situ with a conventional hydration or dehydration mode.
} 
The model and experimental results agree remarkably well, despite the fact that the NaCl-KCl-BSA system largely simplifies the chemical complexity of real saliva. The experimental results further underline that a hysteresis in the $g_\text{d}$ and $g_\text{V}$ curves can generally be expected in the hygroscopic behavior of saliva -- and presumably also of ELF as both fluids are characterized by similar salt concentrations (Table\,\ref{tab:Composition}). Hygroscopic growth curves for a simulated ELF system with $\text{DRH}\sim70\,\%$ and $\text{ERH}\sim50\,\%$ were recently reported by \citet{RN2063}, which are largely consistent with Fig.\,\ref{fig:Hygros1}B. Note in this context that the ERH of the saliva at $\sim$38\,\% is lower than the efflorescence RH of the constituents NaCl and KCl at $\sim$41\,\% and $\sim$52\,\% (Fig.\,\ref{fig:Hygros1}C). This can be explained by a suppression of the efflorescence phase transition by organic compounds and a corresponding shift of the ERH of the salts to lower values \cite[][and references therein]{RN1819}.
\par 

The occurrence of a hysteresis in $g_\text{d}$ has been discussed as a presumably important microphysical process in respiratory particles that can affect the viability of embedded pathogens \cite[e.g.,][]{RN1610,RN1817}. A hysteresis entails a 'bistability' in the humidity dependence within the intermediate humidity range $\text{ERH}<\text{RH}<\text{DRH}$, which corresponds to typical indoor RH levels (i.e, $\sim$40 to $\sim$80\,\%) \cite{RN2003}. If the particles are dried from $\text{RH}>\text{DRH}$, they shrink upon continuous evaporative water loss and retain a certain amount of water in a metastable salt supersaturation state until the ERH is reached. If the particles are humidified from $\text{RH}<\text{ERH}$, they remain as dried residues until the DRH is reached. This means that for $\text{ERH}<\text{RH}<\text{DRH}$, the pathogen-laden particles can be either dried residues or contain a certain amount of water, depending on the 'history' of RH change. It has been proposed that this effect might help to explain conflicting observations on pathogen survival in relation to RH \cite[e.g.,][]{RN1817,RN2003}.\par

\begin{figure}[t]
\includegraphics[width=8cm]{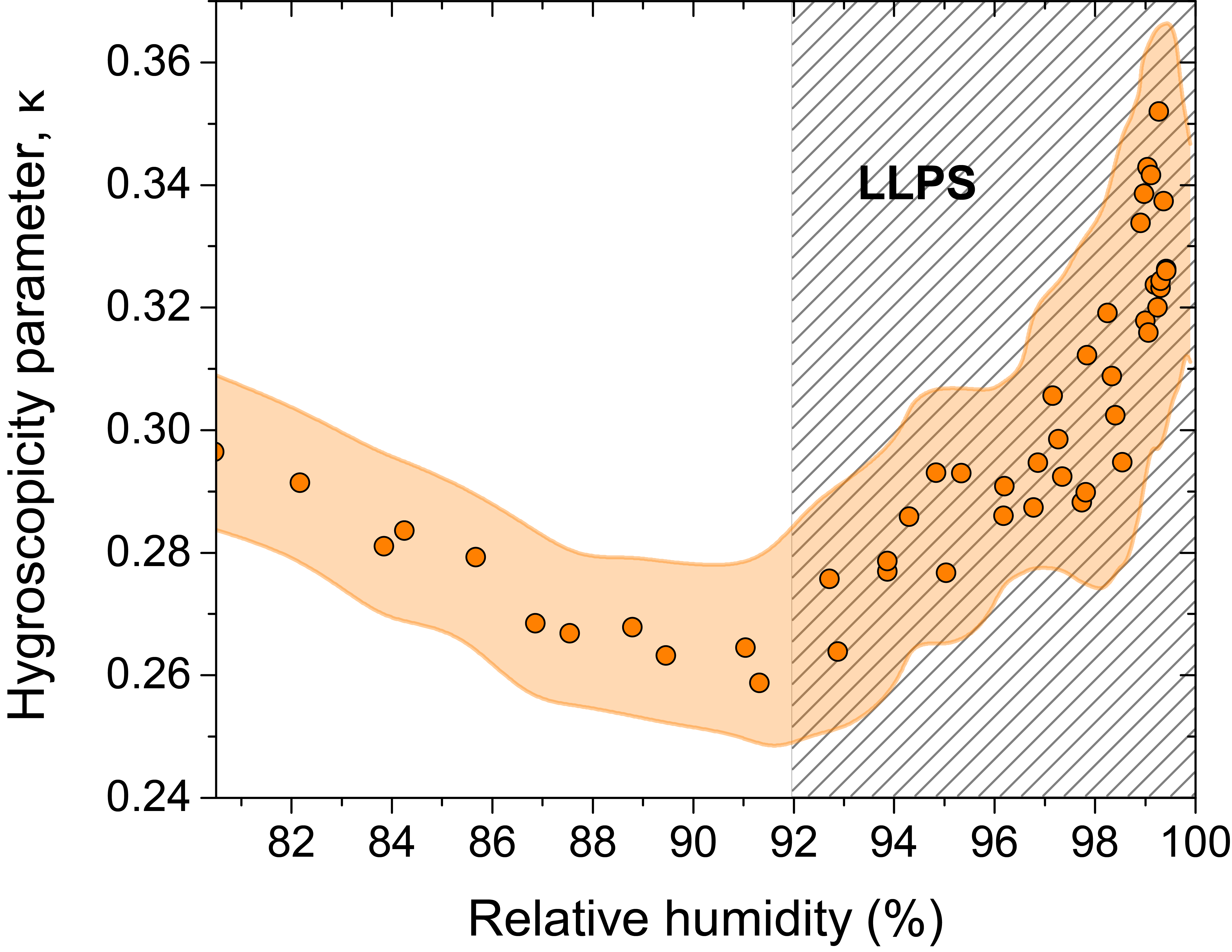}
\caption{\label{fig:Hygros3} Hygroscopicity parameter $\kappa$ in high RH regime obtained for dried and size-selected ($\sim$100\,nm) saliva particles based on HHTDMA measurements (see Fig.\,\ref{fig:Hygros1}B). Orange shading indicates uncertainty in $\kappa$ through error propagation as outline in \citet{RN1809}. Gray background shading $>$92\,\% RH with increasing $\kappa$ values suggests that particles underwent liquid-liquid phase separation (LLPS).}
\end{figure}

A further result of the saliva measurement is the RH-dependence of $\kappa$ in the high humidity regime calculated with Eq.\,(\ref{eq:gK}) as shown in Fig.\,\ref{fig:Hygros3}. A declining trend in $\kappa$ was observed $<$92\,\% RH, followed by a sudden increase $>$92\,\% RH. \citet{RN1551} and \citet{RN1849} reported very similar results for atmospherically relevant purely organic as well as mixed organic-inorganic aerosol particles. Both studies explained the increasing $\kappa$ for high RH with occurrence of liquid-liquid phase separation (LLPS) in the particles, which was found to be a common and important micro-physical process in atmospheric aerosols \cite[e.g.,][]{RN461,RN1479,RN1223,RN462}. This suggests that LLPS might also be a characteristic phenomenon in saliva and ELF, again with potentially important implications for pathogen viability \cite{RN1610,RN1817}.\par

\subsection{\label{sec:Emission}Formation mechanisms and sites of respiratory particles}

\begin{figure*}[ht!]
\includegraphics[width=\textwidth]{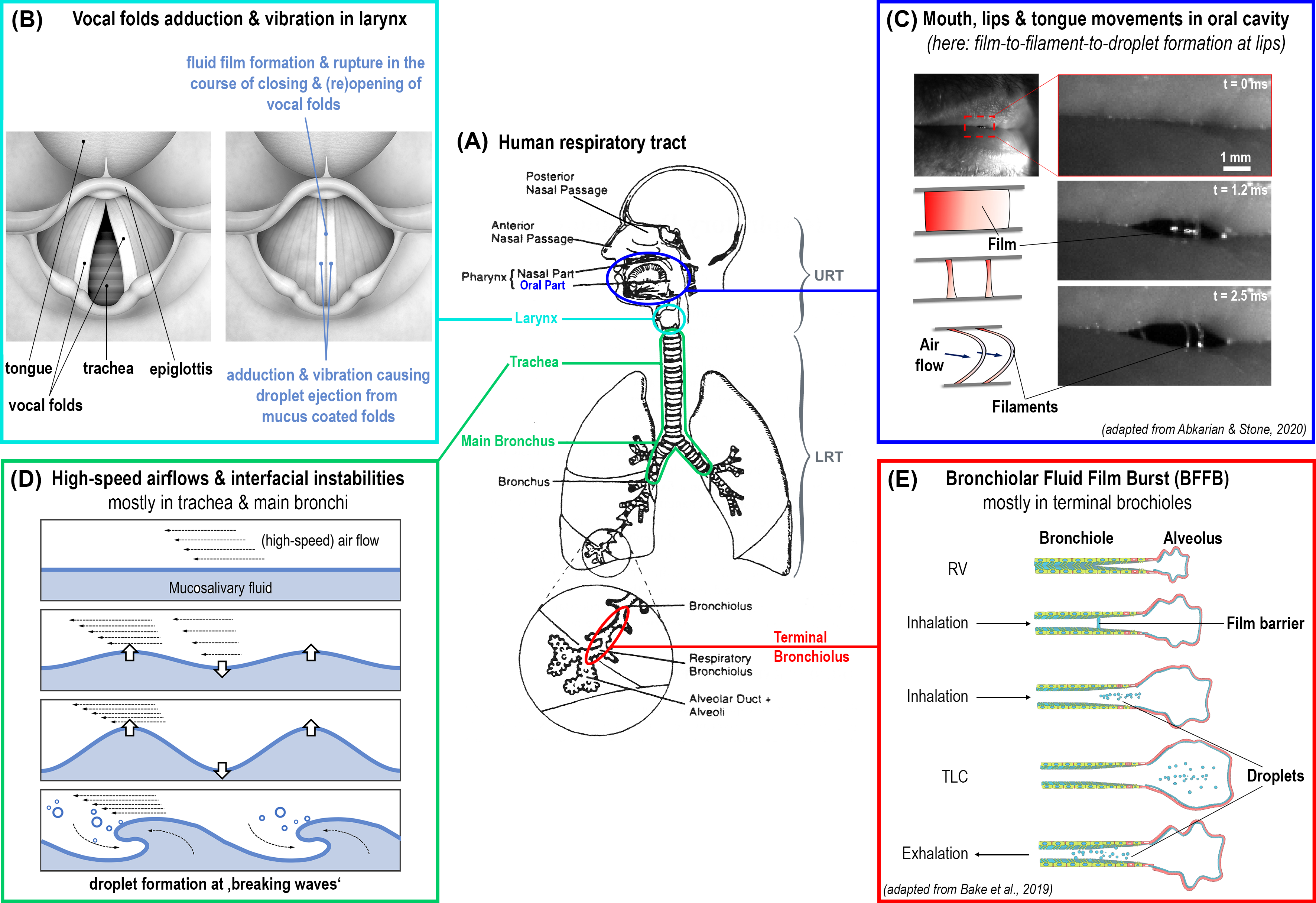} % [width=\textwidth]
\caption{Conceptual scheme summarizing particle formation mechanisms and sites \textbf{(B--E)} in human respiratory tract \textbf{(A)}. Scheme emphasizes the subdivision into the upper respiratory tract (URT) or extrathoracic airways, reaching from larynx to mouth and nose, vs the lower respiratory tract (LRT) or tracheobronchial airways, including trachea, bronchi, bronchioles, and alveoli. Note that the four particle formation categories \textbf{(B--E)} are distinguished in the literature and also here primarily by production sites, whereas the production mechanisms are governed by the same few fluid mechanical principles. Figure relates to earlier scheme in \citet{RN1817}. Scheme (A) adapted from \citet{RN1691} with permission from John Wiley \& Sons, Inc. Scheme (C) adapted from \citet{RN1913} with permission from the American Physical Society. Scheme (E) adapted from \citet{RN1684} with permission under CC Attribution License 4.0.}
\label{fig:RT}
\end{figure*}

\noindent Knowledge of the formation mechanisms of human-expired particles and the corresponding sites in the respiratory tract is essential for a mechanistic understanding of airborne pathogen transmission \cite[e.g.,][and references therein]{RN1681,RN1683,RN1663,RN2012,RN2017}. Generally, the formation is driven by a complex combination of shear and film rupture instabilities, ejecting parts of the mucosalivary films that cover the air-facing surfaces of the respiratory tract \cite{RN1907,RN1908,RN2093}. These processes are complex as they depend on a variety of factors, such as respiratory activities (e.g., breathe, speak, and cough), geometries and movements of the air-fluid interfaces in the respiratory tract, as well as composition and viscoelastic properties of the mucosalivary fluids \cite[e.g.,][]{RN1681,RN2030}. The formation site largely defines the composition of exhaled particles, comprising a variable mixture of ELF from the LRT and saliva from the URT (see Sect.\,\ref{sec:Hygros}). Moreover, pathogens preferentially colonize certain regions of the airways \cite[e.g.,][]{RN1730,RN1731}. Thus, expiration particles may carry particularly high pathogen loads if the site of infection is the same or very close to the site of particle formation \cite{RN1683,RN2017}. The four particle formation mechanisms and associated sites outlined below and summarized in Fig.\,\ref{fig:RT} are widely discussed and considered as most relevant.   

\subsubsection{\label{sec:BFFB}Bronchiole fluid film burst mechanism}

\begin{figure}[t]
  \includegraphics[width=\columnwidth]{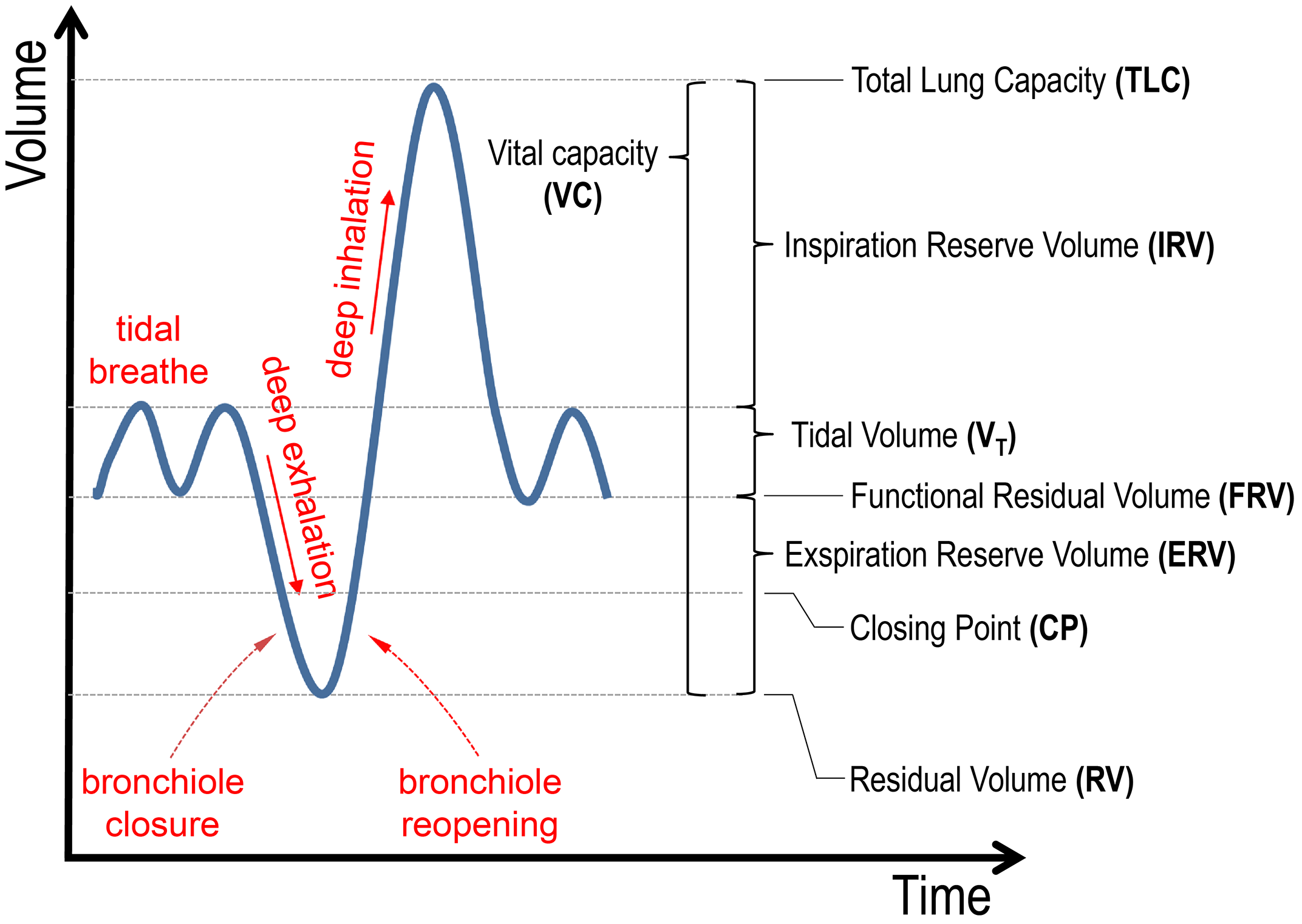}
  \caption{\label{fig:BreathPat}Conceptual figure showing breathing patterns, including 'normal' tidal breathing (corresponding to tidal volume) as well as deep inhalations and exhalations. Similar representations for a variety of breathing patterns can be found in previous studies \cite[e.g.,][]{RN1787,RN1614,RN2017}. 
  }
\end{figure}

The bronchiole fluid film burst (BFFB) mechanism -- or bronchiolar particle generation -- is illustrated in Fig.\,\ref{fig:RT}E. It produces comparatively small particles ($<1$\,\micro m) from collapsing liquid films (deep) in the lungs \cite{RN1617,RN1841,RN1846}. During exhalation, the airways partly close, which refers to a compression and an associated blockage of air passage when the airway walls get in contact \cite{RN1613}. Figure\,\ref{fig:BreathPat} shows conceptually the breathing patterns that involve the BFFB process. Closure begins in the lower lungs and progresses toward the upper lung regions with decreasing lung volume \cite{RN1729,RN1787}. During (e.g., normal tidal) breathing, the terminal bronchioles are considered as the primary site of airway closure. During the subsequent inhalation, the bronchioles reopen and films of the ELF are spanned across the passages, forming a blockage of liquid menisci \citep{RN1787}. When these films rupture, small particles form similar to soap film droplets \cite{RN1317,RN1724}. These particles are first drawn into the alveoli and subsequently exhaled, which explains why the concentrations of emitted particles increase towards the end of an individual expiration \cite{RN1787,RN1845}. Exhaled particle concentrations ($C_\text{N}$) decrease upon breath holding at high lung volume due to diffusion and sedimentation losses in the alveoli, whereas concentrations increase upon breath holding at low lung volume due to a closure of more and more bronchioles with breath holding time \cite{RN1617,RN1787,RN1818,RN1830}.  
    
The BFFB particle formation can be modulated by different breathing patterns, with the $C_\text{N}$ of exhaled particles scaling proportionally to the fraction of fully contracted bronchioles \cite{RN1663,RN1787,RN1614,RN1613,RN1615,RN1818}. Especially exhalation below functional residual capacity (FRC) is related to airway closure and enhanced particle emission \cite{RN1845,RN1847}. \citet{RN1845} showed that the $C_\text{N}$ is related to the ventilation ratio $V_\text{B}$/$V_\text{VC}$ through
\begin{equation} \label{eq:VentRatio}
    C_\text{N}\,=\,C_\text{T}\, \exp\left(b \,\frac{V_\text{B}}{V_\text{VC}} \right)
\end{equation}

\noindent with $V_\text{B}$ as the breathed air volume, $V_\text{VC}$ as the vital capacity (Fig.\,\ref{fig:BreathPat}, $C_\text{T}$ as the particle number concentration during low-volume tidal breathing, and $b$ as an empirical factor representing the shape of exponential function and typically ranging between 4 to 12 \cite{RN1845}. The ventilation ratio for 'normal' tidal breathing corresponds to $\sim$0.2 and approximates unity for airway closure breathing patterns. Further, \citet{RN1845} found a large intersubject variability in contrast to a high reproducibility in $C_\text{N}$ for the same individual and suggested that the properties of the BFFB particle emissions can serve as a fingerprint for the actual individual lung status. The BFFB particle production is active upon breathing and, therefore, also involved in all other respiratory activities (e.g., speak, cough) \cite{RN1663}. Thus, BFFB particle production is considered as potentially important for airborne pathogen transmission from symptomatic as well as presymptomatic, asymptomatic, and paucisymptomatic infected individuals \cite[e.g.,][]{RN1681,RN1841}.

\subsubsection{\label{sec:VocalFold}High-speed and turbulent airflows and airway compression and vibration}
    
Gas-fluid interactions and especially turbulence induced instabilities at the air-mucus interface (two fluids streaming at largely different speeds relative to each other) have been discussed widely as particle generation mechanisms, especially for short and vigorous exspirations such as coughing and sneezing \cite[e.g.,][]{RN1613,RN1681,RN1732,RN1897,RN2012,RN2017}. The central airways, such as the trachea and main bronchi, with their high-speed and often turbulent air flows are considered as the region, where this process is assumed to be most pronounced, whereas air flows in the small airways are mostly laminar due to the strong increase in cross-sectional area \cite{RN1613,RN1681,RN1617,RN1691,RN2017}. In such high-speed airflows, the associated interfacial shearing causes Kelvin-Helmholtz instabilities as interfacial waves of mucus with increasing amplitude (Fig.\,\ref{fig:RT}D). Eventually small droplets are torn off the crest of these waves or filaments of mucus are drawn thin and fragmented into droplets (Plateau-Rayleigh instability) \cite{RN1681,RN1908}. The critical air speed, which created instabilities at the ELF-air interface is defined by the ELF layer thickness (typically $5-10$\,\micro m \cite{RN1908}), along with its viscoelastic properties and surface tension \cite{RN1681,RN1732,RN2030}. Furthermore, dynamic compression and vibration of the airways "squeezes and loosens mucus and promotes expulsion of foreign material from the airways" \cite{RN1681}. For breath-related emissions, the turbulence-induced aerosolization has been excluded as a relevant mechanism since variations in flow rates had essentially no influence on droplet emissions \cite{RN1617,RN1845}.

\subsubsection{Larynx with vocal folds adduction and vibration}

Vocal folds adduction and vibration within the larynx -- or laryngeal particle generation -- is regarded as relevant mechanism during coughing and sneezing as well as during vocalizations such as speaking and singing \cite{RN1608} (Fig.\,\ref{fig:RT}B). It has been suggested that the narrowing of the mucus-bathed folds forms a flow restriction, in which a high-speed air-stream caused sufficient shear stress at the ELF-air interface to tear off droplets \cite{RN1681}. Additionally, vigorous vibration and energetic movement of the ELF-coated folds during vocalization may create instabilities in the ELF surface layer and result in particle formation \cite{RN1707,RN1732}. It has been further considered that -- similar to the BFFB mechanism -- ELF films may form and burst or ELF filaments may fragment in the open-close cycling of the glottic structure (i.e., the opening between the vocal folds) \cite{RN1608,RN1663,RN2017,RN2012}. Some studies suggest that the vocalization frequency (vocal pitch) modulates the particles' emission rate and diameters \cite{RN1608,RN1707}.
\vfill

\subsubsection{Oral cavity with mouth, lips, and tongue movements}

The upper respiratory tract, and particularly the oral cavity between lips and epiglottis as well as nasal passage, is regarded as the main site for large droplet formation \cite{RN1895}. Here, droplets through fragmentation of liquid sheets and filamentous structures that are formed from the permanently present saliva in the course of mouth, lips, and tongue movements (Fig.\,\ref{fig:RT}C) \cite{RN1663,RN1908}. Also pulsed and high velocity air flows, such as coughing and sneezing, cause significant shear forces in the throat, nasal and buccal passages, creating a droplet spray through Kelvin-Helmholtz instabilities \cite{RN1897}. The velocity and pressure fields involved strongly depend on anatomy of the airflow passages as well as the chemical and fluid properties of the saliva \cite{RN1897}.    

\subsubsection{Clinical aerosol-generating procedures}

So-termed "aerosol-generating procedures" (AGPs), e.g. intubation or high flow nasal oxygen treatments, are clinical procedures that are labeled as such, because they can emit aerosols from the patient \citet{RN2098}. Aerosols can be generated from the nose, mouth, throat, or lungs from a patient undergoing various kinds of invasive procedures. From a clinical perspective, these AGPs have often been considered the one of the only sources of infectious aerosols from patients with most respiratory diseases, including COVID-19. The distinction is vitally important, because protective guidelines are often very different for situations when healthcare workers expect to deal with AGPs versus every other situation. When aerosols are present that could contain pathogens, protective equipment (e.g. masks, respirators, eye coverings) need to be significantly improved to reduce inhalation of small airborne particles (i.e. FFP2 or N95 respirators). Thus, surgical-style procedure masks and face shields are not sufficient protection whenever aerosols are present, and so healthcare workers can be left at risk. A survey of AGPs and the physical properties of aerosols emitted by these procedures will not be discussed here, in large part because AGPs have been shown to generally produce fewer aerosols than standard respiratory activities like breathing, speaking, and coughing \cite[e.g.,][]{RN2095,RN2096,RN2097}. \par 
\clearpage

\section{\label{sec:MainSec}Multimodal size distributions of human respiratory particles and its parameterization}

\subsection{\label{sec:LitSyn}Literature data synthesis and development of a parameterization scheme}

Published, peer-reviewed studies with size distribution data on human respiratory particles along with information on sampling conditions and parameters have been collected and summarized in Table\,\ref{tab:LitSyn}. The PSDs from the individual studies have been either obtained from data tables, if available, or have been digitized from figures. Table\,\ref{tab:LitSyn} summarizes all studies with PSDs that were found. Since not all of them were equally appropriate for the development of a general parmetrization, the following flags specify in Table\,\ref{tab:LitSyn} to what extent the individual studies were used:
\begin{itemize}
    \item \textbf{A:} The study was used in the literature and data synthesis. The reported PSDs agree with the overall parameterization. The size resolution of the reported PSDs is sufficiently high and allows a multimodal fitting with comparatively low uncertainty. The reported PSDs were further used to calculate the average parameterizations in Fig.\,\ref{fig:overview} and averaged fit parameters in Table\,\ref{tab:Fit_Syn}. 
    
    \item \textbf{B:} The study was used in the literature and data synthesis of this work for comparison only. The reported PSDs agree with the overall parameterization. The size resolution of the reported PSDs is comparatively low and fitting entailed rather high uncertainties. Therefore, the reported PSDs were not used to calculate the average parameterizations in Fig.\,\ref{fig:overview} and averaged fit parameters in Table\,\ref{tab:Fit_Syn}.
    
    \item \textbf{C:} The corresponding study was not used in the literature and data synthesis of this work. The study was omitted for at least one of the following reasons: (i) data points from the original publication could not be unambiguously digitized \cite{RN1728,RN1710}; (ii) important information for the calculation of concentrations from the PSDs was missing \cite{RN1714,RN1835}, or (iii) fundamental open questions on the methodology or experimental limitations remained \cite{RN1607}.
\end{itemize}

\begin{table*}
\caption{\label{tab:LitSyn}Summary of studies in chronological order reporting aerosol number size distributions (NSDs) from human respiratory activities. Table summarizes analyzed respiratory activities, number of volunteers in study ($N$), sizing instruments or techniques, covered size range (as available in figures or tables for further use) as well as relative humidity (RH) and temperature ($\theta$) in the moment of droplet/particle measurement. Note that RH and $\theta$ are not clearly defined in all studies. For the definition of the flag in last column see text of Sect.\,\ref{sec:LitSyn}.}
\begin{ruledtabular}
\begin{tabular}{lrp{2.3cm}p{4cm}cccc}
% &\multicolumn{2}{c}{XYZ}&\multicolumn{2}{c}{XYZ}\\
Reference & $N$ & Respiratory & Instrument & Nominal size & RH & $\theta$ & Flag \\
 &&activity&\textit{(manufacturer, model)}& range [\micro m] & [\%]&[\degree\,C] &\\[2pt]
\hline & \\[-1.7ex]
  \citet{RN1701} & n.a. & speak, cough,\newline sneeze & impaction \& microscopy & 1\,--\,2000 & n.a. & n.a. & A\\ 
 
  \citet{RN1762,RN1763} & {3} & {speak, cough} & filter sampling or sedimentation \& microscopy & {2\,--\,2000} & n.a. & n.a. & {A}\\ 
  
  \citet{RN1728} & {5} & {breathe, speak,\newline cough} & {OPC  \textit{(Climet, CI-7300)} $+$\newline sedimentation \& microscopy} & {0.3\,--\,2.5} & {45} & {25} & {C}\\

  \citet{RN1607} & {54} & {cough} & {SMPS \textit{(TSI, 3934)} $+$\newline OPC \textit{(TSI, 3310A)}} & {0.6\,--\,30} & {$\sim$95 vs. $\sim$35} & n.a. & {C}\\

  \citet{RN1599} & {13} & {breathe} & {OPC \textit{(Airnet 310)}} & {0.3\,--\,5.0} & n.a. & n.a. & {B}\\
  
  \citet{RN1739,RN1707} & {15} & {breathe, speak,\newline cough} & {UV-APS  \textit{(TSI, 3312A)}} & {0.5\,--\,20} & {83\,-\,93} & {$\sim$27} & {A}\\

  \citet{RN1611} & {11} & {speak, cough} & {particle image velocimetry\,\&\newline interferometric Mie imaging} & {$2-2000$} & {$\sim$100\footnote{Referring to statement in study that "evaporation and condensation effects had negligible impact on the measured droplet size".}} & {$\sim$37} & {A}\\
 
  \citet{RN1682} & {1} & {breathe} & {OPC \textit{(Grimm, 1.108)}} & {$0.3-3.0$} & {$\sim$100\footnote{Estimated RH range. Study states that analysis was conducted "without altering particle size through evaporation or condensation".}} & {36} & {A}\\ % RH rather >95 than 100 ??

  \citet{RN1617} & {17} & {breathe} & {UV-APS  \textit{(TSI, 3312A)}} & {0.5\,--\,20} & {90\,$\pm$\,7} & {28\,$\pm$\,1} & {A}\\
  
  \citet{RN1750} & {7} & {speak, cough} & {sedimentation \& microscopy} & {10\,--\,1000} & {$\sim$70} & {$\sim$28} & {A}\\

  \citet{RN1614} & {10} & {breathe} & {OPC \textit{(Grimm, 1.108)}} & {0.3\,--\,3.0} & {$\sim$100\footnote{Estimated RH range. See \citet{RN1682}.}} & {36} & {A}\\
  
  \citet{RN1615} & {16} & {breathe} & {laser spectrometer\newline \textit{(PMS, Lasair II-110)}} & {0.1\,--\,5.0} & {$\sim$95} & {37} & {B}\\
  
  \citet{RN1613} & {16} & {breathe} & {SMPS \textit{(TSI, 3936)} $+$\newline OPC \textit{(Grimm, 1.108)}} & {0.01\,--\,4.0} & {$\sim$95} & {$\sim$35} & {A}\\
  
  \citet{RN1845,RN1830} & \makecell[tr]{21} & breathe & {laser spectrometer\newline\textit{(PMS, Lasair II-110)}} & {0.1\,--\,5.0} & {90\,--\,100} & {37} & {B}\\

  \citet{RN1786} & {8} & {shout} & {OPC \textit{(Lighthouse, 5016)}} & {0.5\,--\,10} & \makecell[tc]{n.a.\\($\sim$dried)} & \makecell[tc]{n.a. \\($\sim$RT)} & {B}\\

  \citet{RN1818} & {17} & {breathe} & {OPC \textit{(Climet, CI-550)}} & {0.3\,--\,10} & n.a. & n.a. & {B}\\
  
  \citet{RN1663} & {15} & {breathe, speak,\newline cough} & {UV-APS  \textit{(TSI, 3312A)} $+$\newline sedimentation \& microscopy} & {0.5\,--\,2000} & {90\,$\pm$\,7} & {28\,$\pm$\,1} & {A}\\
  
  \citet{RN1737} & {3} & {breathe} & {OPC \textit{(Grimm, 1.108)}} & {0.3\,--\,20} & {75 vs. 99.5} & {$\sim$35} & {A}\\

  \citet{RN1714} & {20} & {sneeze} & {laser diffraction droplet sizer \textit{(Spraytec, Malvern Inst.)}} & {0.1\,--\,1000} & {$>$32} & {23\,--\,24} & {C}\\
  
  \citet{RN1787} & {19} & {breathe} & {OPC  \textit{(Grimm, 1.108)}} & {0.3\,--\,3.0} & {$\sim$100} & {36} & {A}\\ % RH rather >95 than 100 ??

  \citet{RN1848} & {10} & {cough} & {SMPS \textit{(TSI, 3910)} $+$\newline OPS \textit{(TSI, 3330)}} & {0.01\,--\,10} & {30\,--\,50} & {21--25} & {B}\\  

  \citet{RN1608} & {48} & {speak} & {APS \textit{(TSI, 3321)}} & {0.5\,--\,20} & {45\,--\,80} & {$20-25$} & {A}\\
  
  \citet{RN1890} & {18} & {breathe, speak,\newline cough} & {laser particle counter\newline \textit{(LPC, Solair 3100)}} & {0.3\,--\,3.0} & {40\,$\pm$\,2\footnote{$\theta$ and RH are not specified in \citet{RN1890}. The values given here are those specified in the study by \citet{RN1710}, which used the same experimental setup.}} & {22\,$\pm$\,0.5} & {B}\\

  \citet{RN1835} & {7} & {speak, cough} & {laser diffraction droplet sizer \textit{(Spraytec, Malvern Inst.)}} & {1\,--\,2000} & {90\,--\,100} & n.a. & {C}\\

  \citet{RN1810} & {12} & {breathe, speak,\newline sing} & {APS \textit{(TSI, 3321)}} & {$0.5-10$} & {$\leq$40} & {$\sim$22} & {A}\\
  
  \citet{RN1710} & {8} & {breathe, speak,\newline sing} & laser particle counter \newline \textit{(LPC, Solair 3100)} & {0.3\,--\,10} & {40\,$\pm$\,2} & {22\,$\pm$\,0.5} & {C}\\

  \citet{RN1869} & {1} & cough\footnotemark[5] & {APS  \textit{(TSI, 3321)}} & {0.5\,--\,20} & {$54\pm5$} & {$25\pm1$} & {A}\\   
  
  \citet{RN1708} & 25 & breathe, speak,\newline sing & APS  \textit{(TSI, 3321)} & 0.5\,--\,20 & $\sim$45 & $\sim$20 & A\\
\end{tabular}
\end{ruledtabular}
\footnotetext{Sampling was conducted at three distances (i.e., 0.3\,m, 0.9\,m, and 1.8\,m) between inlet and volunteer. The sampling at distances of 0.3\,m and 0.9\,m were implemented here.}
\end{table*}

Different instruments and measurements strategies for respiratory aerosol characterization have been applied with specific strengths and limitations and were critically evaluated here. Moreover, the individual studies typically cover only part of the entire aerosol size range. The characterization of the comparatively low concentrations of respiratory aerosols requires online or offline techniques that provide full size distributions, ideally also at high time resolution (up to 1\,Hz). The most commonly applied online instruments are the scanning mobility particle sizer (SMPS, manufactured by TSI Inc., St. Paul, MN, USA) with a nominal size range typically from 0.01 to 0.4\,\micro m, the aerodynamic particle sizer (APS, TSI Inc.) with a nominal size range from 0.5 to 20\,\micro m, and the optical particle sizer/counter (OPS/OPC, from different manufacturers) with a nominal size range typically from 0.3 to 10\,\micro m. For larger particle fractions (i.e., $>$10\,\micro m), passive sampling through particle sedimentation or impaction of expelled droplets (often called droplet deposition analysis, DDA) or active sampling through droplet impaction on solid surfaces (e.g., glass slides or culture plates) followed by image analysis have been used. 
% \textcolor{red}{// mention instruments in Chao and Han //}      
% Chao: PIV = particle image velocimetry, IMI = interferometric Mie imaging

The APS has been a widely used instrument in respiratory aerosol characterization \cite[e.g.,][]{RN1708,RN1608,RN1617,RN1707,RN1663}. Different models of the APS have been used (see Table\,\ref{tab:LitSyn}). However, a variety of instrumental issues have been reported for different APS models and, thus caution is required when using APS-derived size distributions \cite{RN1726,RN1727,RN1660,RN1659}. While the sizing accuracy of the APS is generally acceptable, issues with the counting efficiency and instrument's unit-to-unit variability in certain size ranges have been reported \cite{RN1726,RN1727}. Generally, \citet{RN1727} showed that the counting efficiency can be strongly size-depend and varies between different APS models. Particularly relevant for the characterization of expiration aerosols upon release -- which means in humid state -- is a substantially decreased counting efficiency of liquid particles (i.e., declining from 75\,\% at 0.8\,\micro m to 25\,\% for 10\,\micro m) due to impaction losses in the instrument's flow system \cite{RN1660}. \citet{RN1726} recently showed a strongly increased unit-to-unit variability (up to 60\,\%) for particles smaller 0.9\,\micro m and larger 3\,\micro m, probably due to different detector sensitivities. Here we have chosen a conservative approach and limited the size range of reported APS data to the relatively narrow band from 0.9\,\micro m to 5\,\micro m, where the counting efficiency is relatively high and unit-to-unit variability relatively low. Note in this context that \citet{RN375} also reported a decline in APS counting efficiency for ambient aerosols (relative to an OPC) and limited the APS size range (i.e., to 0.8\,\micro m to 5\,\micro m). The instrumental limitations of the APS have been emphasized in previous studies on expiration aerosols \cite[e.g.,][]{RN1707,RN1617}. For all other instruments and sizing strategies the entire reported size ranges (see Table\,\ref{tab:LitSyn}) have been used.
     
For OPC instruments, polystyrene latex spheres (PSLs) are typically used for particle size calibration. The refractive index of PSLs differs from respiratory particles, however, which will effect calculated sizing. \citet{RN1613} provides a size correction of $\sim$1.6 based on diluted isotopic water solution of organic and inorganic solutes, which approximates the composition of exhaled particles. This correction was adopted to all OPS data, if not already implemented in the original studies.

The development of the parameterization followed the workflow below:

\begin{enumerate}
    \item Respiration PSDs were obtained from tables or digitized from figures of the original publications.\footnote{For digitization, the web application WebPlotDigitizer available under https://automeris.io/WebPlotDigitizer/ (last access 30 Jan 2021) has been used. Only those data points were collected that could unambiguously be picked.} The size range of the APS data was limited to 0.9\,--\,5\,\micro m for aforementioned reasons. For OPC data, the size bins were corrected according to difference in the refractive index of calibration and respiratory particles.
     
    \item The data was normalized by the size bin widths to $\text{d}C/{\text{dlog}D}$ and $\text{d}Q/{\text{dlog}D}$, if not already available as such in the original publications.
    
    \item The PSDs where fitted by a multimodal lognormal function with least square in number and volume (NSD and VSD) using Eq.\,\ref{eq:f_NSD}. The fitting was conducted with IGOR Pro (version 8.04, Wavemetrics, Inc., Portland, OR, USA). The fit functions were iteratively optimized to equally describe both, the NSD and VSD of a given data set. A fit for a given PSD was accepted if both NSD and VSD were described in a physically meaningful way with a maximal coefficient of determination ($\text{R}^2$). As a part of the fitting process, some fit parameters were constrained, described in detail in the corresponding parts of Sect.\,\ref{sec:MainSec}.

    \item After fitting, the NSDs and VSDs were normalized to allow a comparison between all PSDs of a certain category. Normalization to a specific area under the curve was conducted: (a) for small particles ($D\,<\, 5$\,\micro m), the overall trimodal PSD were normalized to the area under breath-related bimodal distribution (i.e., modes \textbf{B1} and \textbf{B2}, for details see Sect.\,\ref{sec:BreathResults}), (b) for large particles ($D\,>\, 5$\,\micro m), the overall bimodal PSD was normalized to the area under the second of both modes (i.e., mode \textbf{O2}, for details see Sect.\,\ref{sec:SpeakSingResults}).  
    
    \item Average PSDs were calculated by the arithmetic means of the corresponding fit parameters, based on all PSDs flagged with A in Table\,\ref{tab:LitSyn}. 

\end{enumerate}

\subsection{\label{sec:ParamSec}parameterizations of particle size distributions for specific respiratory activities}

\subsubsection{\label{sec:BreathResults}Breathing}

\begin{figure*}[htbp]
\includegraphics[width=11.5cm]{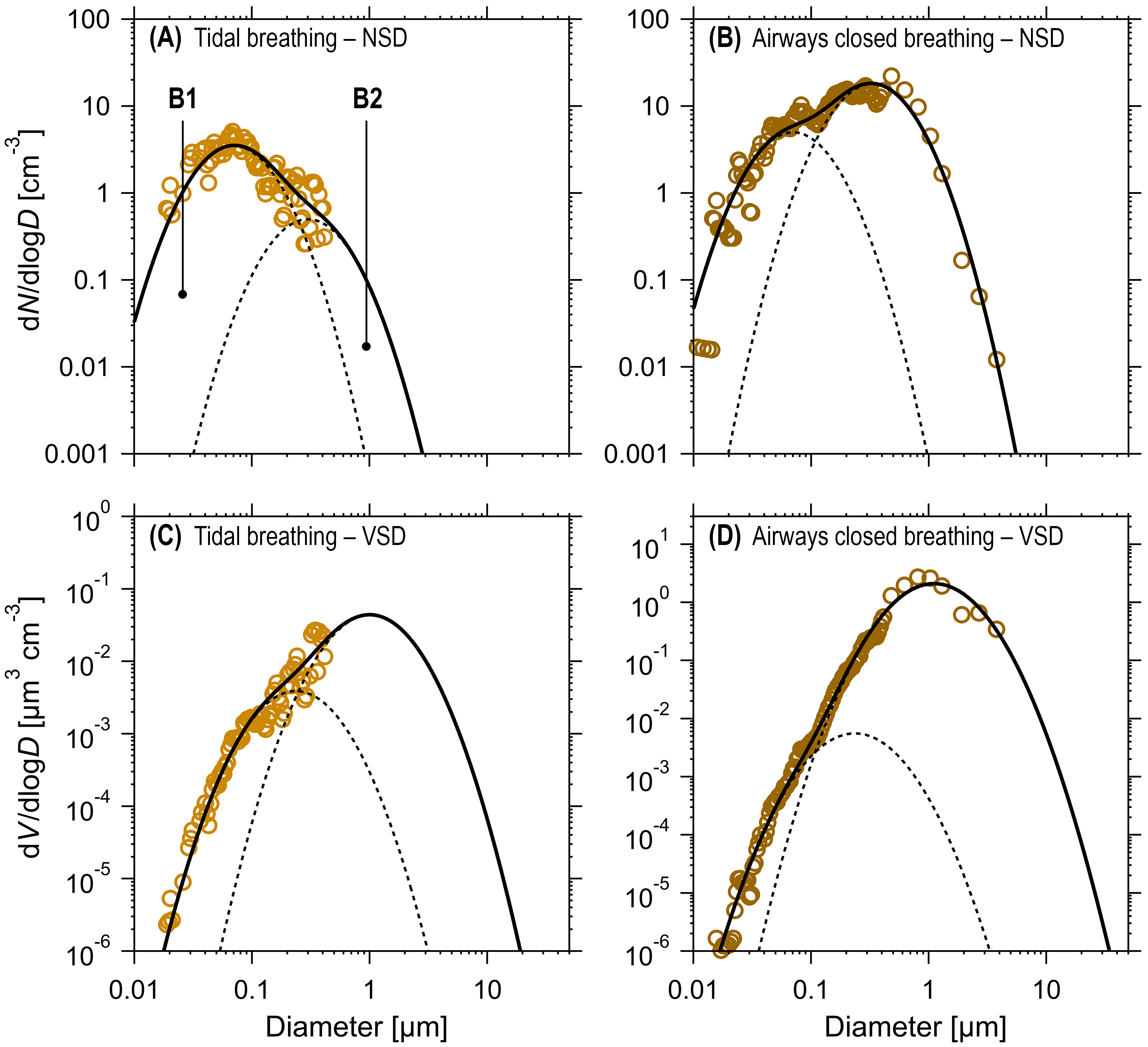} 
\caption{Number and volume particle size distributions (NSD and VSD) for tidal breathing (\textbf{A} and \textbf{C}) vs breathing with airways closure (\textbf{B} and \textbf{D}) adapted from \citet{RN1613} (colored markers) with bimodal lognormal fit functions for parameterization (solid line). The underlying lognormal modes \textbf{B1} and \textbf{B2} are shown as black dashed lines. In this representation, the individual PSDs were not normalized. Corresponding fit parameters are summarized in Table\,\ref{tab:breath_Holm}.}
\label{figs:breath_panel}
\end{figure*}

\begin{table*}[htb]
\caption{\label{tab:breath_Holm}Fit parameters for bimodal lognormal fits of NSDs and VSDs for (i) tidal breathing vs breathing with airways closure based study by \citet{RN1613} as shown in Fig.\,\ref{figs:breath_panel} and (ii) sea spray aerosol based on study by \citet{RN1317} as shown in Fig.\,\ref{fig:ssa}. Table specifies mode-specific particle number and volume concentrations ($C_\text{N}$, $C_\text{V}$), height ($A_\text{i}$), position ($D_\text{i}$), and width ($\sigma_\text{i}$). Coefficient of determination ($\text{R}_\text{N}^2$ vs $\text{R}_\text{V}^2$) shows quality of fits for number and volume representation of PSDs.}
\begin{ruledtabular}

\begin{tabular}{lcD{.}{.}{2.2}D{.}{.}{2.2}D{.}{.}{1.2}D{.}{.}{1.2}D{.}{.}{1.2}D{.}{.}{1.4}D{.}{.}{1.2}}
\multicolumn{1}{l}{\textbf{Breathe pattern}} & \textbf{Mode} & 
\multicolumn{1}{c}{\textbf{$C_\text{N}$}} & 
\multicolumn{1}{c}{\textbf{$A_\text{i}$}} &
\multicolumn{1}{c}{\textbf{$D_\text{i}$}} & 
\multicolumn{1}{c}{\textbf{$\sigma_\text{i}$}} & 
\multicolumn{1}{c}{\textbf{$\text{R}_\text{N}^2$}} &
\multicolumn{1}{c}{\textbf{$C_\text{V}$}} & 
\multicolumn{1}{c}{\textbf{$\text{R}_\text{V}^2$}} \\
&&
\multicolumn{1}{c}{[$\text{cm}^{-3}$]} &
\multicolumn{1}{c}{[$\text{cm}^{-3}$]} &  
\multicolumn{1}{c}{[\micro m]} & & &
 \multicolumn{1}{c}{[\micro$\text{m}^3\text{cm}^{-3}$]} & \\[2pt]
\hline & \\[-1.7ex]
  Tidal breathe         & B1 & 2.42 & 3.5 & 0.07 & 0.9 & 0.77& 0.003 &  0.69    \\
                        & B2 & 0.35 & 0.5 & 0.3  & 0.9 &     & 0.03  &          \\[1.2ex]
  Airways closure       & B1 & 3.46 & 5   & 0.07 & 0.9 & 0.84& 0.004 & 0.89     \\
                        & B2 & 12.47& 18  & 0.33 & 0.9 &     & 1.45  &          \\[2pt]    \hline & \\[-1.7ex]
SSA: Sintered glass     & 1 & 1.08  & 2   & 0.05 & 0.70& 0.93& 0.0002& 0.54     \\
                        & 2 & 0.05  & 0.07& 0.25 & 0.90& 0.93& 0.002 &          \\[1.2ex]
SSA: Plunging waterfall & 1 & 0.98  & 0.7 & 0.11 & 1.8 & 0.90& 0.99  & 0.69     \\
\end{tabular}
\end{ruledtabular}
\end{table*}

\begin{figure*}[htbp]
\includegraphics[width=17.5cm]{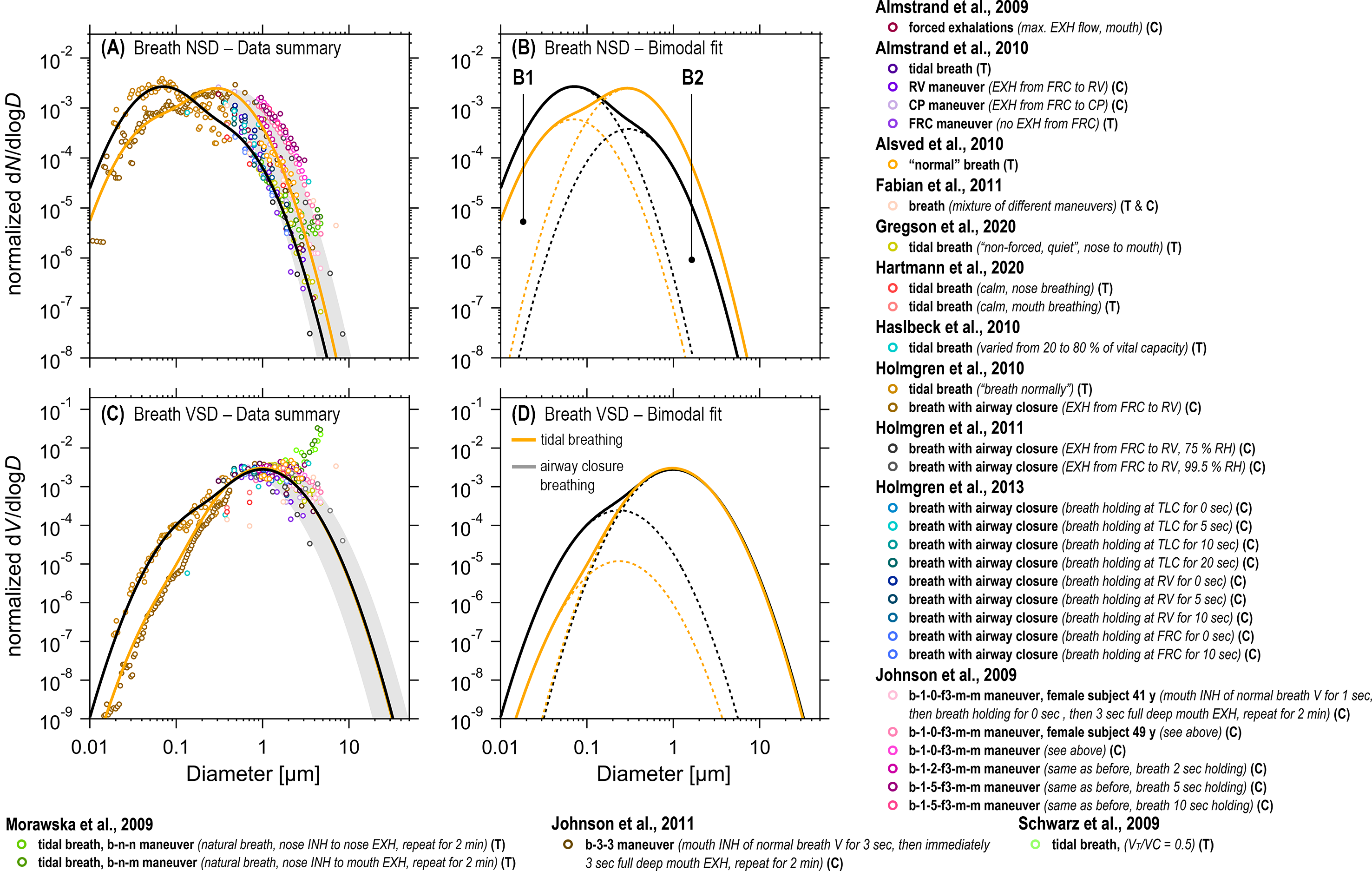}
\caption{Breath aerosol number and volume size distributions obtained from multiple studies \textbf{(A,C)} and bimodal fit functions with underlying lognormal modes \textbf{(B,D)} for parameterization of NSDs and VSDs. Tidal breathing (black line) vs airway closure breathing (orange line) maneuvers were discriminated according to \citet{RN1613}. In the legend, T specifies tidal-like breathing patterns and C specifies breathing patterns with airway closure. After fitting, all NSDs and VSDs were normalized to the area under the curve to account for the widely variable breathe aerosol concentrations. The underlying lognormal modes \textbf{B1} and \textbf{B2} are shown as dashed lines. Grey shading in \textbf{(A)} and \textbf{(C)} shows influence of aerosol humidification state based on breathe aerosol characterization at 75\,\% vs 99.5\,\% RH in \citet{RN1737}. Legend summarizes information on specific breathing maneuvers conducted in the individual studies in relation to Fig\,\ref{fig:BreathPat}. For further details, refer to the original articles directly.}
\label{figs:breath}
\end{figure*}

\begin{figure}[htb]
  \includegraphics[width=6.5cm]{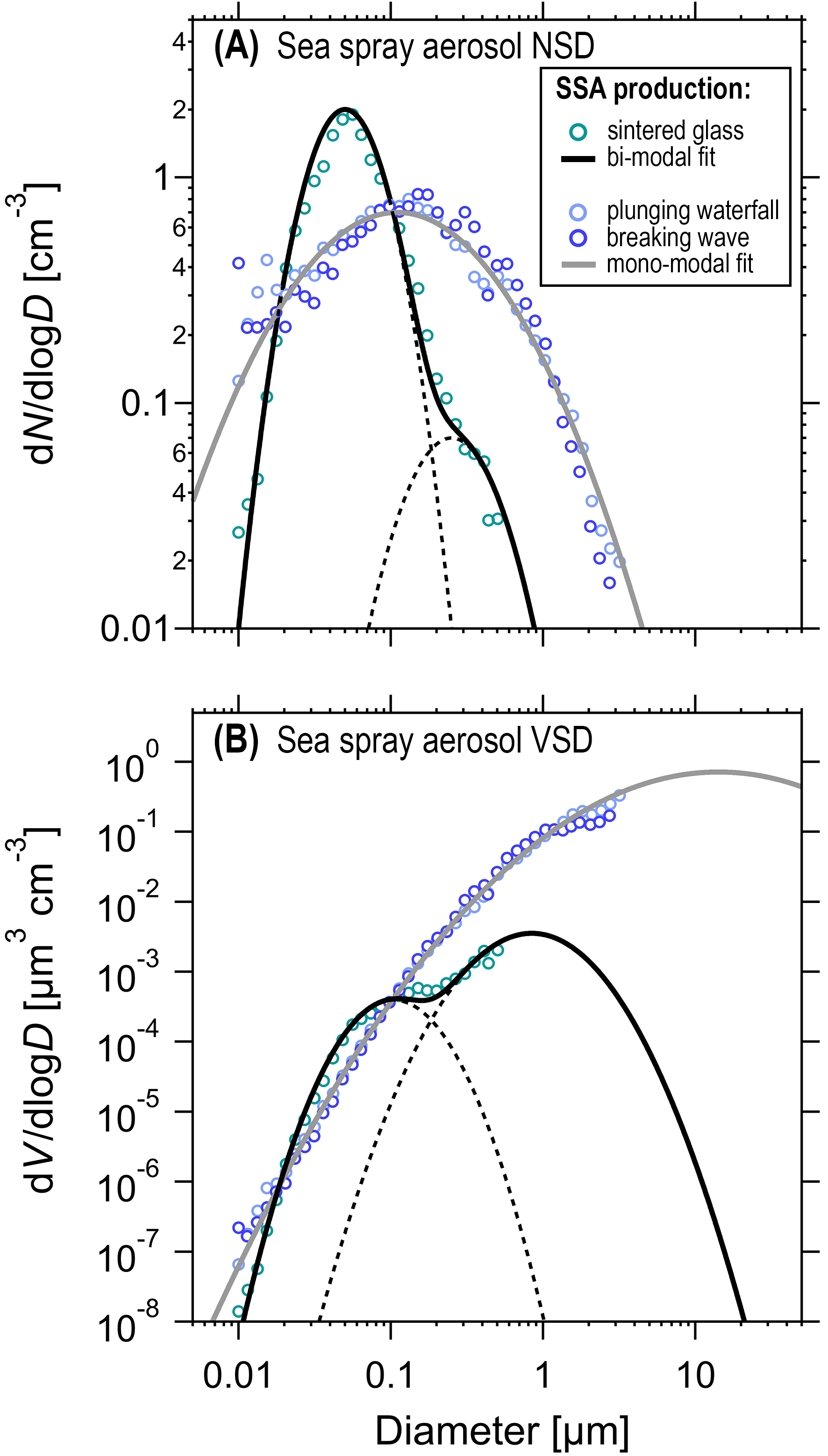}
  \caption{\label{fig:ssa}Number and volume particle size distributions (NSD and VSD) for laboratory generated sea spray aerosols (SSA) based on study by \citet{RN1317} with lognormal fit functions (solid lines). The size distributions were not normalized. Sea spray aerosols in \citet{RN1317} were generated with three different methods: (i) sintered glass, (ii) plunging waterfall, and (iii) breaking waves (see differently colored markers). Note that the PSDs of SSA react sensitively to the bubble size distributions. Data from plunging waterfall and breaking wave were combined and fitted jointly with monomodal lognormal functions as size distributions were similar. Sintered glass data was fitted with separate bimodal lognormal fit and shows remarkable similarity to breath aerosol (Fig.\,\ref{figs:breath}. The corresponding fit parameters are summarized in Table\,\ref{tab:breath_Holm}.
  }
\end{figure}

Breathing is the most fundamental, frequent, and continuous respiratory activity. Speaking, laughing, singing, coughing, and sneezing involve breathing as an underlying mode. Hence, the development of a parameterization of respiration PSDs should start with a robust representation of breath-related emissions. Several studies have investigated aerosol formation in relation to different breathing patterns (see Fig.\,\ref{fig:BreathPat}), and it is widely supposed that the BFFB mechanism is primarily -- if not exclusively -- responsible for the breath-related particle formation (Sect.\,\ref{sec:Emission}) \cite{RN1684,RN1617,RN1788,RN1613,RN1734}. Here, we integrate the published experimental data to parameterize the shape of the breath PSD. The largely variable particle concentrations ($C_\text{N}$, $C_\text{V}$) and emission rates ($Q_\text{N}$, $Q_\text{V}$) of breath-related emissions, in relation to other respiratory activities, are addressed in Sect.\,\ref{sec:ConcRates}. This analysis shows that -- after fitting and normalization -- all published breath PSDs agree remarkably well, despite differences in methodology and variable RH conditions during the measurements. 

The experimental basis for the parameterization comprises 34 breath PSDs from 14 studies (see Table\,\ref{tab:LitSyn}). Good data coverage exists for the particle size range between 0.3 and 5\,\micro m due to the wide use of the instruments APS \cite{RN1810,RN1708,RN1617,RN1663,RN1707} (i.e., 0.9 to 5\,\micro m, see Sect.\,\ref{sec:LitSyn}) as well as different OPC models \cite{RN1614,RN1682,RN1818,RN1599,RN1890,RN1615,RN1787,RN1613,RN1845,RN1830,RN1710} (i.e., 0.3 to 3\,\micro m) (see\,Table\,\ref{tab:LitSyn}). Studies with data coverage $<$0.3\,\micro m are sparse. \citet{RN1613} published the only study so far with highly size-resolved (SMPS) measurements with $D<0.3$\,\micro m, which covers the peak and overall shape of the breath PSD. Therefore, we used this data to parameterize the lower end of the breath PSD. Figure\,\ref{figs:breath_panel} shows the PSDs from \citet{RN1613} and the corresponding bimodal lognormal fits for tidal breathing vs breathing with airways closure. The fits represent the PSDs accurately in number and volume representation. Relating to \citet{RN1663}, we refer to both \textit{bronchiolar} breathing modes as \textbf{B1} (peaking $D\,<\,$0.1\,\micro m in NSD) and \textbf{B2} (peaking $D\,>\,$0.1\,\micro m in NSD). In the fits, the position ($D_\text{i}$) and the height ($A_\text{i}$) of both lognormal modes were free parameters, whereas the width ($\sigma_\text{i}$) of the modes was fixed to 0.9, This $\sigma_\text{i}$ value was iteratively optimized in this work. In fact, an important result of the fitting approach overall is the observation that $\sigma_\text{i}\,=\,0.9$ describes the width of all lognormal modes of respiratory PSDs across multiple studies well. For comparison, the width of the characteristic modes in the ambient aerosol is typically smaller with $\sigma$ ranging from $\sim$0.4 to $\sim$0.6 \cite{RN944}. The data in the studies by \citet{RN1615} and \citet{RN1845} -- both with sizing data down to 0.1\,\micro m, though with much lower size resolution -- generally support the bimodal character of the breath PSD.\par

Figure\,\ref{figs:breath_panel} and Table\,\ref{tab:breath_Holm} show that the resulting $D_\text{i}$ values for both modes are rather consistent for both breathing patterns and that the difference in PSD shape is mainly determined by a variable height of mode \textbf{B2}. \citet{RN1613} attributed mode \textbf{B2} to the BFFB mechanism in the terminal bronchioles as it increased strongly for airway closure relative to tidal breathing (i.e., $A_\text{B2}$ increased by factor 36). They further speculate that mode \textbf{B1} may originate from a similar film bursting e.g., at the alveoli openings in the course of alveolar dynamics during respiration, which (to the best of our knowledge) so far has remained an unverified hypothesis \cite{RN1972,RN1973,RN1974}. The difference in breath PSD shape can be well described by the ratio of both mode heights in NSD representation with $A_\text{B1}\,/\,A_\text{B2}\,=\,7.0$ for tidal breathing vs $A_\text{B1}\,/\,A_\text{B2}\,=\,0.3$ for breathing with airway closure. These fixed mode height ratios have been used in the course of our further analyses to constrain multimodal fits for PSD data that cover (part of) the size range of mode \textbf{B2} but not of mode \textbf{B1}.

Figure\,\ref{figs:breath} summarizes all breath PSDs found in the literature after fitting and normalization. It shows a consistent picture with essentially all data points falling within a relatively narrow 'corridor'. Only the VSDs from \citet{RN1707} deviate, for unknown reasons. At first glance, the good agreement of all breath PSDs is remarkable given that multiple studies with different instruments and data from a large number of volunteers in the experiments are combined here. In fact, Fig.\,\ref{figs:breath} suggests that breath aerosols are associated with a characteristic bimodal shape of the PSD without a clear inter- or intra-subject variability. This is in stark contrast to the breath-related $C_\text{N}$ and $Q_\text{N}$, which are widely variable across individuals (Sect.\,\ref{sec:ConcRates}) \cite{RN1710,RN1845}. At second glance though, the consistent PSD shape appears less surprising as the size distribution -- after BFFB emission deep in the lung -- is 'shaped' by the transmissibility of the respiratory tract upon exhalation \cite{RN1845}. This means that on their way to the mouth or nose certain small particle fractions are removed by diffusional losses and certain large particle fractions are removed by sedimentation \cite{RN1323}. In fact, established lung deposition models, such as the International Commission on Radiological Protection (ICPR) model \cite{RN1691}, show a penetration maximum of the respiratory tract between 0.3 and 0.4\,\micro m, which corresponds with the size distribution maximum in Fig.\,\ref{figs:breath}. Accordingly, the characteristic shape of the breath PSD is presumably determined by both, the BFFB emission mechanism as well as particle losses upon exhalation.

The fit functions in Fig.\,\ref{figs:breath} were calculated as the averages of all fit parameters from the individually fitted NSDs and represent a robust parameterization of breath PSDs. The average fit parameters are summarized in Table\,\ref{tab:Fit_Syn}. Across all individual NSDs, mode \textbf{B2} ranges between $D_\text{B2}$\,=\,0.15\,\micro m and 0.53\,\micro m (average $D_\text{B2}$\,=\,0.31\,\micro m). The variability in $D_\text{B2}$ can presumably be explained by two effects: First, the measured droplet diameters equilibrate quickly as a function of RH (Sect.\,\ref{sec:Hygros}). Therefore, the difference in $D_\text{B2}$ of the individual NSDs reflect to some extent the (large) differences in RH under the corresponding measurement conditions (see Table\,\ref{tab:LitSyn}). Note that temperature and RH are poorly defined or undefined in some studies, which complicates the comparison of different NSDs. For many studies in Table\,\ref{tab:LitSyn}, RH levels ranged between 80\,\% and water saturation -- a humidity regime, in which a small $\Delta$RH causes a large $\Delta D$ (Fig.\,\ref{fig:SalivaDrying} and Fig.\,\ref{fig:Hygros1}). A reference point in this context has been provided by \citet{RN1737}, who characterized the RH-related $\Delta D$ by measuring comparable breath NSDs (involving airway closure) at 75\,\% vs 99.5\,\% RH. Here, the NSD at 75\,\% corresponds to $D_\text{B2}$\,=\,0.18\,\micro m and the NSD at 99.5\,\% corresponds to $D_\text{B2}$\,=\,0.45\,\micro m, yielding an increase in $D$ by a factor of 2.5 in good agreement with Fig.\,\ref{fig:Hygros1}. The RH-related difference in mode \textbf{B2} for both PSDs is shown in Fig.\,\ref{figs:breath}A and C as background shading. It shows that the RH-related variability in $D$ corresponds well with the overall scattering of the data points. Second, the different breathing pattern summarized in Fig.\,\ref{figs:breath} and the associated differences in air residence time in the respiratory tract might also cause certain modulations in the overall PSD shape \cite[e.g.,][]{RN1845,RN1787,RN1617}. This effect seems to be smaller than the RH influence, though further studies are needed.

\begin{figure*}
\includegraphics[width=16.3cm]{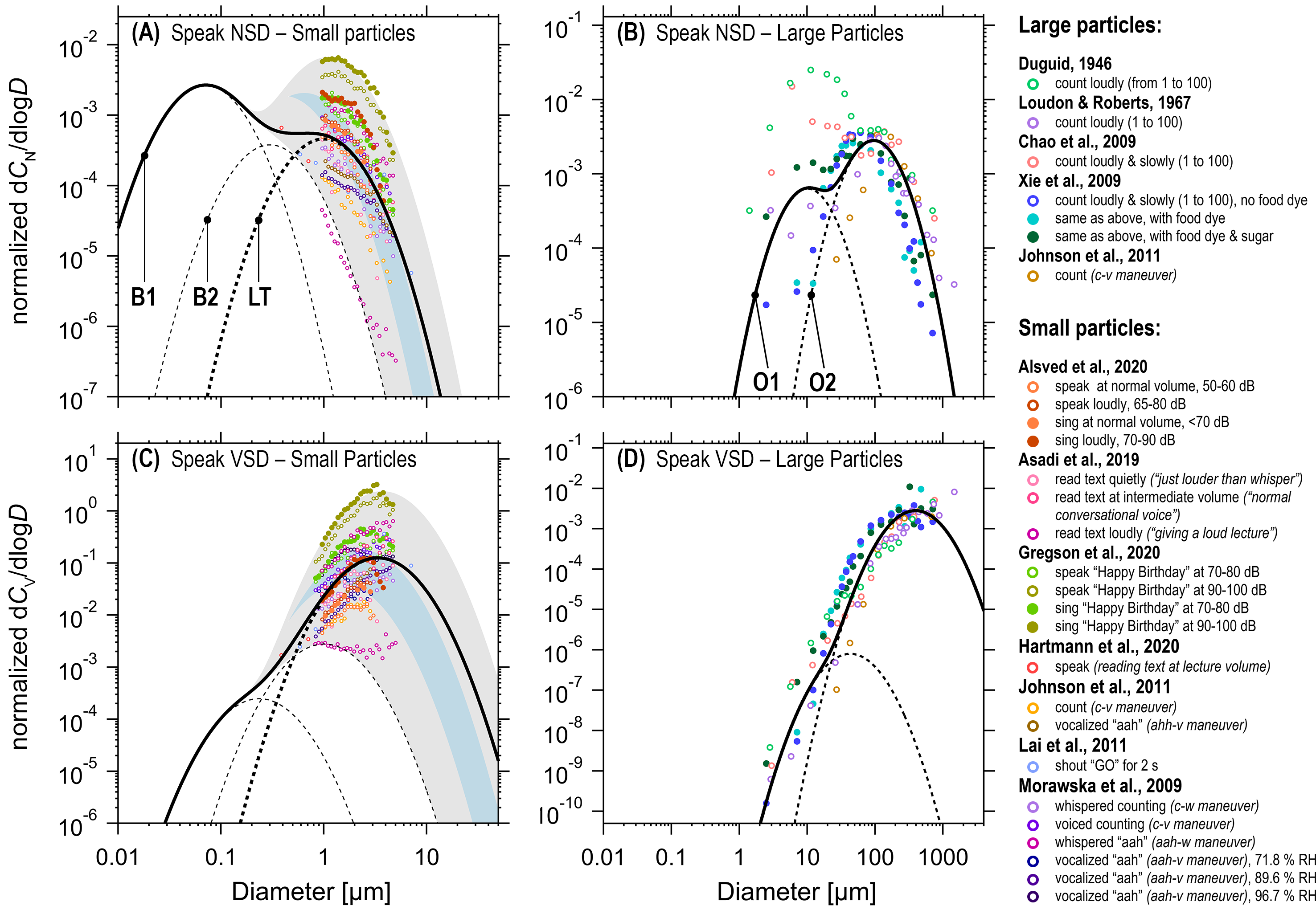}
\caption{Combined literature data on speaking- and singing-related particle NSDs \textbf{(A,B)} and VSDs \textbf{(C,D)} with parameterization based on five lognormal modes. The modes \textbf{B1}, \textbf{B2}, and \textbf{LT}, centered $<$5\,\micro m, are described by a trimodal lognormal fit \textbf{(A,C)}. The trimodal distributions in \textbf{(A)} and \textbf{(C)} were normalized to the area under the breath-related bimodal curve (combined area of modes \textbf{B1} and \textbf{B2}) to emphasize the relative contribution of mode \textbf{LT}. Grey shading shows variability of mode \textbf{LT}. The influence of varying RH (i.e., 71.8\,\% vs 96.7\,\%) based on experimental data from \citet{RN1707} is illustrated as light blue shading in \textbf{(A)} and \textbf{(C)}. Note that this study is an outlier in terms of \textbf{LT} mode position (i.e., $D_\text{LT}$\,=\,0.45 lowest value observed, which, however, does not diminish its value in specifying the RH-related variability. The modes \textbf{O1} and \textbf{O2}, centered $>$5\,\micro m, are described by mono- or bimodal lognormal fits \textbf{(B,D)}. All NSDs and VSDs in \textbf{(B)} and \textbf{(D)} were normalized to the area under mode \textbf{O2} to emphasize the relative variability of mode \textbf{O1}. Legend summarizes information on specific speak and sing activities conducted in the individual studies. For further details, refer to the original articles directly.}
\label{fig:speak_syn}
\end{figure*}

\begin{table*}[!htp] %
\caption{\label{tab:Fit_Syn}Average fit parameters for breathing (see Fig.\,\ref{figs:breath}), speaking and singing (see Fig.\,\ref{fig:speak_syn}) as well as coughing (see Fig.\,\ref{fig:cough_syn}) size distributions in $\text{d}N/\text{dlog}D$ representation. Table specifies the mode-specific parameters: height ($A_\text{i}$), position ($D_\text{i}$), width ($\sigma_\text{i}$), integral particle number and volume concentrations ($C_\text{N}$, $C_\text{V}$) as well as emission rates ($Q_\text{N}$, $Q_\text{V}$). $Q_\text{N}$ and $Q_\text{V}$ were obtained from $C_\text{N}$ and $C_\text{V}$ by applying Eq.\,\ref{eq:Q_NV_2} with $q$ from Table\,\ref{tab:RespParam}.}

\begin{ruledtabular}
\begin{tabular}{lcD{.}{.}{6}D{.}{.}{6}D{.}{.}{1.2}D{.}{.}{6}D{.}{.}{7}D{x}{\cdot}{-1}D{x}{\cdot}{-1}}
\multicolumn{1}{l}{\textbf{Respiratory activity}} & \textbf{Mode} & \multicolumn{1}{c}{\textbf{$A_\text{i}$}} & \multicolumn{1}{c}{\textbf{$D_\text{i}$}} &
\multicolumn{1}{c}{\textbf{$\sigma_\text{i}$}} & \multicolumn{1}{c}{\textbf{$C_\text{N}$}} & \multicolumn{1}{c}{\textbf{$C_\text{V}$}} & \multicolumn{1}{c}{\textbf{$Q_\text{N}$}} &
\multicolumn{1}{c}{\textbf{$Q_\text{V}$}} \\
&&\multicolumn{1}{c}{[$\text{cm}^{-3}$]} &\multicolumn{1}{c}{[\micro m]} &  &\multicolumn{1}{c}{[$\text{cm}^{-3}$]} &\multicolumn{1}{c}{[\micro$\text{m}^3\text{cm}^{-3}$]} &
\multicolumn{1}{c}{[$\text{h}^{-1}$]} &\multicolumn{1}{c}{[\micro$\text{m}^3\text{h}^{-1}$]} \\[2pt]
\hline & \\[-1.7ex]
Tidal breathe                 & B1 & 7.7 & 0.07 & 0.90 & 5.33   & 5.93\cdot10^{-3} & 1.92\,x\,10^{6} & 2.13\,x\,10^{3}  \\ [1.2ex]
                              & B2 & 1.1 & 0.30 & 0.90 &  0.76 & 6.67\cdot10^{-2} & 2.74\,x\,10^{5} & 2.40\,x\,10^{4}  \\[1.2ex]
Breathe with airway closure   & B1 & 2.00\,\cdot\,10^{1}  & 0.07 & 0.90 & 1.39\,\cdot\,10^{1}  & 1.54 \,\cdot\,10^{-2} & 4.99\,x\,10^{6} & 5.54\,x\,10^{3} \\
                              & B2 & 2.60\,\cdot\,10^{1} & 0.30 & 0.90 & 1.80  \,\cdot\,10^{1}  & 1.57   & 6.48\,x\,10^{3} & 5.67\,x\,10^{5} \\[1.2ex]
Speak                         & B1 & 9.8 & 0.07 & 0.90 & 6.79   & 7.54\,\cdot\,10^{-5} & 4.75\,x\,10^{6} & 5.28\,x\,10^{3}  \\
                              & B2 & 1.4 & 0.3  & 0.90 & 0.97   & 8.48\,\cdot\,10^{-2} & 6.79\,x\,10^{5}  & 5.94\,x\,10^{4}  \\
                              & LT & 1.7 & 1    & 0.90 & 1.18 & 3.81 & 8.24\,x\,10^{5}  & 2.67\,x\,10^{6} \\
                              & O1 & 0.03 & 1.00 \,\cdot\,10^{1}  & 0.98 & 2.26\,\cdot\,10^{-2} & 1.03\,\cdot\,10^{-2} & 1.58\,x\,10^{4}  & 7.20\,x\,10^{7}  \\
                              & O2 & 0.17 & 9.60 \,\cdot\,10^{1}  & 0.97 & 0.127 & 4.46\cdot10^{5} & 8.88\,x\,10^{4} & 3.12\,x\,10^{11}\\[1.2ex]
Cough                         & B1 & 2.62~\cdot\,10^{2} & 0.07 & 0.90 & 1.81 \,\cdot\,10^{2} & 0.20 & 2.72\,x\,10^{6}  & 3.03\,x\,10^{5}\\
                              & B2 & 3.70~\cdot\,10^{-1}  & 0.3  & 0.90 & 2.60 \,\cdot\,10^{-2}  & 2.24 & 3.84\,x\,10^{5}  & 3.36\,x\,10^{4} \\
                              & LT & 4.3 & 1    & 0.98 & 3.24   & 1.47 \,\cdot\,10^{1}   & 4.87\,x\,10^{4}   & 2.21\,x\,10^{5}  \\
                              & O1 & 1.4 & 1.10~\cdot\,10^{1}   & 0.95 & 1.02 & 5.44\,\cdot\,10^{3}   & 1.54\,x\,10^{4}    & 8.15\,x\,10^{7}  \\
                              & O2 & 0.5 & 1.28~\cdot\,10^{2}  & 1    & 0.38 & 4.0\,\cdot\,10^{6}  & 5.77\,x\,10^{3}    & 6.0\,x\,10^{10} \\
  
\end{tabular}
\end{ruledtabular}
\end{table*}

In Sect.\,\ref{sec:Emission}, the mechanistic analogy between respiratory aerosol formation through the BFFB process and other natural bubble bursting processes, such as sea spray aerosol (SSA) formation through bubble bursting at the air-ocean interface, has been pointed out \citet{RN1613}. For further illustration, Fig.\,\ref{fig:ssa} shows SSA size distributions from \citet{RN1317}, generated in the laboratory with natural sea water through three different processes. Worth noting here is that the breath aerosol vs SSA PSDs span across a similar size range and further show some analogies in terms of mode position and shape (i.e., similar bimodal shape of sintered glass PSD vs tidal breath PSD). This might reflect the mechanistic relationship of both aerosol formation processes, however, should probably not be over-interpreted. In this context, the similarity underlines the general plausibility of the breath PSD parameterizations in Fig.\,\ref{figs:breath} in the relation to the relatively well characterized SSA size distributions \cite[e.g.,][]{RN1305,RN1119,RN1579}.

\subsubsection{\label{sec:SpeakSingResults}Speaking and singing}

\begin{figure*}
\includegraphics[width=14.2cm]{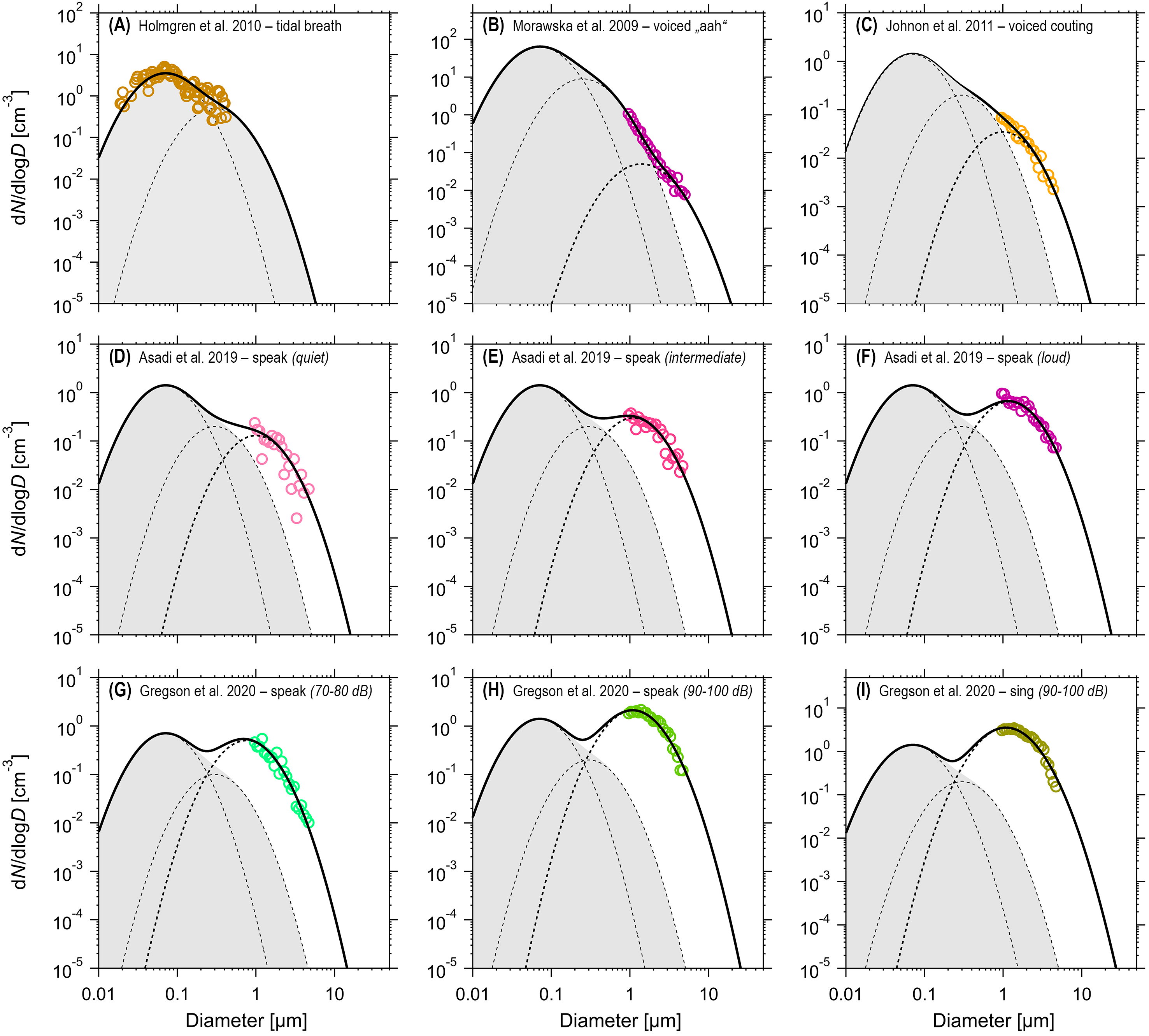}
\caption{Number aerosol size distributions (NSD) for different speaking and singing maneuvers from multiple studies with trimodal lognormal fit functions (solid black line). Panel \textbf{(A)} shows tidal breathing with underlying modes \textbf{B1} and \textbf{B2} in relation to Fig.\,\ref{figs:breath_panel}A to emphasize the emerging third speaking- and singing-related mode \textbf{LT}. Strength of mode \textbf{LT} appears to increase with increasingly 'vigorous' speaking and/or singing. As tidal breathe underlies all speaking and singing maneuvers, tidal breath-related bimodal fit from Fig.\,\ref{figs:breath_panel}A (with fixed ratio $A_\text{B1}\,/\,A_\text{B2}\,=\,7.0$) has been used in all trimodal fits here. Underlying modes \textbf{B1} and \textbf{B2} are marked by grey shading in all panels. The individual size distributions here were not normalized. All three modes \textbf{B1}, \textbf{B2}, and \textbf{LT} are shown as black dashed lines.}
\label{figs:speak_panel_N}
\end{figure*}

Speaking and singing involve opening or closing of the glottis as well as tensing and vibrating of the vocal folds in the larynx, producing -- together with mouth, lips, and tongue movements -- a wide spectrum of sounds \cite[e.g.,][]{RN1895}. Both, speaking and singing, are widely variable with, e.g., speaking spanning from whispering to shouting. Speaking is a relatively frequent and semi-continuous respiratory activity. In relation to speaking, singing is typically characterized by continuous vocalization, higher sound pressure, higher frequencies, deeper breaths, higher peak air flows and more articulated consonants \cite{RN1810}. In airborne pathogen transmission, both speaking and singing have been considered as significant driving forces as they are major sources of respiratory particles \cite[e.g.,][]{RN1710,RN1608,RN1708}. Mechanistically, particle formation through speaking and singing is still somewhat uncertain \cite{RN1617,RN1663}. Each likely involves two or even three formation mechanisms and sites: (i) laryngeal particle generation, (ii) particle formation through mouth, lips, and tongue movements, and (iii) under vigorous conditions probably also high-speed shear forces at the ELF-air interface in the trachea \cite[e.g.,][]{RN1617,RN1663}. The breath-related bronchiolar particle formation is involved as an underlying process.\par

Figure\,\ref{fig:speak_syn} summarizes the existing experimental data on speaking- and singing-related particle emissions, spanning across a wide size range from $>$10\,nm to $\sim$1000\,\micro m. We found that a parameterization based on five lognormal modes represents the experimental data well. Two of these modes are the bronchiolar modes \textbf{B1} and \textbf{B2} since breathing is inherently involved in speaking and singing. Beyond the modes \textbf{B1} and \textbf{B2} (located at $D_\text{B1}$\,=\,0.07\,\micro m and $D_\text{B2}$\,=\,0.3\,\micro m), a third mode emerges, which presumably originates from particle formation in larynx and trachea (Fig.\,\ref{fig:speak_syn}A,C) according to \citet{RN1663}. We termed the third mode \textbf{LT} (L and T for larynx and trachea), accordingly. Based on the individual NSDs in Fig.\,\ref{fig:speak_syn}, mode \textbf{LT} ranges between $D_\text{LT}$\,=\,0.7\,\micro m and 1.5\,\micro m (average $D_\text{LT}$\,=\,1.0\,\micro m). Note that we omitted one outlier at $D_\text{LT}$\,=\,0.45 based on the study by \citet{RN1707} in this general statement. In the size range of large particles, the individual PSDs can be described well by either one or two lognormal modes (Fig.\,\ref{fig:speak_syn}B,D). As these large droplets presumably originate from mouth, lips, and tongue movements, we refer to them as oral modes \textbf{O1} and \textbf{O2} \cite{RN1663}. Mode \textbf{O1} ranges between $D_\text{O1}$\,=\,8\,\micro m and 13\,\micro m (average $D_\text{O1}$\,=\,10\,\micro m) and mode \textbf{O2} is located between $D_\text{O2}$\,=\,60\,\micro m and 130\,\micro m (average $D_\text{O2}$\,=\,96\,\micro m). A summary of all average fit parameters can be found in Table\,\ref{tab:Fit_Syn}. Overall, the five modes have significant overlap and constitute a rather continuous PSD across more than five orders of magnitude in $D$.\par

The parameterization of speaking- and singing-related PSDs has been developed separately for two size regimes: a small particle range $<$5\,\micro m with modes \textbf{B1}, \textbf{B2} and \textbf{LT} (Fig.\,\ref{fig:speak_syn}A,C) vs a large particle range $>$5\,\micro m with modes \textbf{O1} and \textbf{O2} (Fig.\,\ref{fig:speak_syn}B,D). This two-step approach was chosen because the existing experimental data originates from two 'groups' of instruments that focus primarily either on the small or the large particle regime. For the small particle range, the data coverage is comparatively good between 0.9 and 5\,\micro m, due to predominant use of the APS \cite{RN1810,RN1608,RN1708,RN1663,RN1707}. Data $<$0.9\,\micro m is relatively sparse and so far based on three studies with OPC measurements only \cite{RN1890,RN1786,RN1710}. Overall, we collected 17 speaking-related PSDs from 7 studies and 4 singing-related PSDs from 2 study for the range $<$5\,\micro m (see Table\,\ref{tab:Fit_Syn}). For the analysis of the large particle range between $\sim$5 to $>$1000\,\micro m, instruments such as an open-path laser diffraction droplet sizer \cite{RN1714,RN1835}, sedimentation sampling in combination with microscopy \cite{RN1762,RN1763,RN1750,RN1663,RN1701}, and interferometric Mie imaging \cite{RN1611} have been used. We found 7 speaking-related PSDs from 5 studies and no singing-related PSD for the size range $>$5\,\micro m (see Table\,\ref{tab:Fit_Syn}).\par 

In the trimodal lognormal fits of the modes \textbf{B1}, \textbf{B2}, and \textbf{LT}, some of the 9 fit parameters had to be fixed because (i) the available experimental data cover only a comparatively narrow band of the relevant size range and (ii) the modes are not separately resolved due to significant overlap. Specifically, the parameters $D_\text{B1}$, $D_\text{B2}$, $\sigma_\text{B1}$, and $\sigma_\text{B2}$ were fixed as they were adapted from the average breath PSD in Fig.\,\ref{figs:breath}. The only exception are two studies \cite{RN1810,RN1707}, in which the data points resolved part of the mode \textbf{B2} sufficiently well so that $D_\text{B2}$ could also be implemented as a free fitting parameter. Moreover the ratio $A_\text{B1}\,/\,A_\text{B2}\,=\,7.0$ for tidal breathing was fixed, whereas the overall height of this bimodal fit for \textbf{B1} and \textbf{B2} was a free parameter. For the mode \textbf{LT}, the parameters $D_\text{LT}$ and $A_\text{LT}$ were free, whereas $\sigma_\text{LT}$ was fixed to 0.9 for reasons outlined in Sect.\,\ref{sec:BreathResults}. In the bimodal lognormal fits of the modes \textbf{O1} and \textbf{O2}, all fit parameters (i.e., $D_\text{O1}$, $D_\text{O2}$, $A_\text{O1}$, $A_\text{O2}$, $\sigma_\text{O1}$, and $\sigma_\text{O2}$) were free because the experimental data cover the relevant size range well and the modes \textbf{O1} and \textbf{O2} were sufficiently separated. Note that also in these fits, where $\sigma_\text{i}$ was implemented as a free parameters, $\sigma_\text{O1}$ and $\sigma_\text{O2}$ values close to 0.9 were obtained, which underlines that this lognormal peak width is a good representation for respiration PSD modes overall.\par    

Figure\,\ref{fig:speak_syn} shows the variability of mode \textbf{LT} on top of the modes \textbf{B1} and \textbf{B2}. We found no systematic changes in $D_\text{LT}$ as a function of speak or sing activities as illustrated in Fig.\,\ref{figs:speak_panel_N}. This suggests that the shape of the trimodal distribution is generally consistent for different speak and sing activities, which is in line with previous studies reporting no major changes in PSD shape in relation to vocalization, loudness, and language spoken \cite{RN1608,RN1708,RN1710}. In contrast, the ratio of the modes \textbf{B2} and \textbf{LT} varies widely from $A_\text{B2}\,/\,A_\text{LT}$\,=\,180 (dominated by the breath-related mode \textbf{B2}) to 0.06 (dominated by the speaking- or singing-related mode \textbf{LT}). Here, the calculation of $A_\text{B2}\,/\,A_\text{LT}$ is associated with some uncertainty since most available speaking- and singing-related PSDs do not reach below 0.9\,\micro m and, therefore, cover mode \textbf{B2} only partly. Figure\,\ref{figs:speak_panel_N} suggests that the ratio $A_\text{B2}\,/\,A_\text{LT}$ depends on the intensity of the speaking or singing activities, which is also in line with previous evidence \cite{RN1608,RN1708}. Here, the tidal breathe example in Fig.\,\ref{figs:speak_panel_N}A serve as a reference case for the range of speak and sing activities from whispering in Fig.\,\ref{figs:speak_panel_N}B to load singing in Fig.\,\ref{figs:speak_panel_N}I. In relation to the general variability of the mode \textbf{LT}, Fig.\,\ref{fig:speak_syn}A and C emphasize that the RH influence on mode position and shape is comparatively small.\pagebreak

Mode \textbf{O2} was found for all individual PSDs in Fig.\,\ref{fig:speak_syn}B and D. The presence of mode \textbf{O1} is less conclusive though. It is clearly resolved in three studies \cite{RN1701,RN1762,RN1611} and appears to be absent in one study \cite{RN1663}. Specifically, in the three PSDs in \citet{RN1750}, mode \textbf{O1} was clearly resolved in one of them and absent in the other two. The ratio of the mode heights $A_\text{O1}/A_\text{O2}$ varies largely from $\sim$10 in \citet{RN1701} to $\sim$0.14 in \citet{RN1750}. The ambiguous finding on mode \textbf{O1} can potentially be explained by either experimental issues such as particle drying dynamics and, therefore, observations in different non-equilibrium states or different sensitivities of the instruments used in this particular size range. Further, the rather variable appearance of mode \textbf{O1} could potentially also be caused by mechanistic difference in the course of droplet formation in the mouth. The remarkable differences among the PSDs in \citet{RN1750} (which were taken under comparable conditions and with the same technique) suggest that the presence of mode \textbf{O1} indeed depends on differences in e.g., saliva composition. This particular study focused on the PSDs in relation to the use of food dye in the mouth for better microscopic detection of settled particles. Here, mode \textbf{O1} was present when sugar and dye were used, whereas mode \textbf{O1} was absent when no dye was used. Clearly, further studies are needed to extent our knowledge on speaking- and singing-related emission in this large particle range.\par 

Regarding the drying dynamics and measurements under non-equilibrium states, Fig.\,\ref{fig:SalivaDrying}B shows that the evaporation times of particles in the smaller modes \textbf{B1}, \textbf{B2}, and \textbf{LT} is comparatively short, which implies that these particle were likely dried to a significant extent under the experimental conditions of most studies in Table\,\ref{tab:LitSyn}. For the particles in the larger modes \textbf{O1} and \textbf{O2}, however, Fig.\,\ref{fig:SalivaDrying}B shows comparatively long evaporation times -- especially if evaporation delays as shown in \citet{RN2026} are taken into account -- which implies that these particles were likely measured under non-equilibrium conditions. Accordingly, these kinetic effects probably entail that the extent of drying decreases from mode \textbf{B1} towards \textbf{O2}. For Fig.\,\ref{fig:speak_syn}, this presumably entails that the modes \textbf{O1} and \textbf{O2} are shifted to larger diameters and would move closer towards the mode \textbf{LT}, if they were dried to the equilibrium state under the given RH conditions. \par

\begin{figure*}[t]
\includegraphics[width=12.5cm]{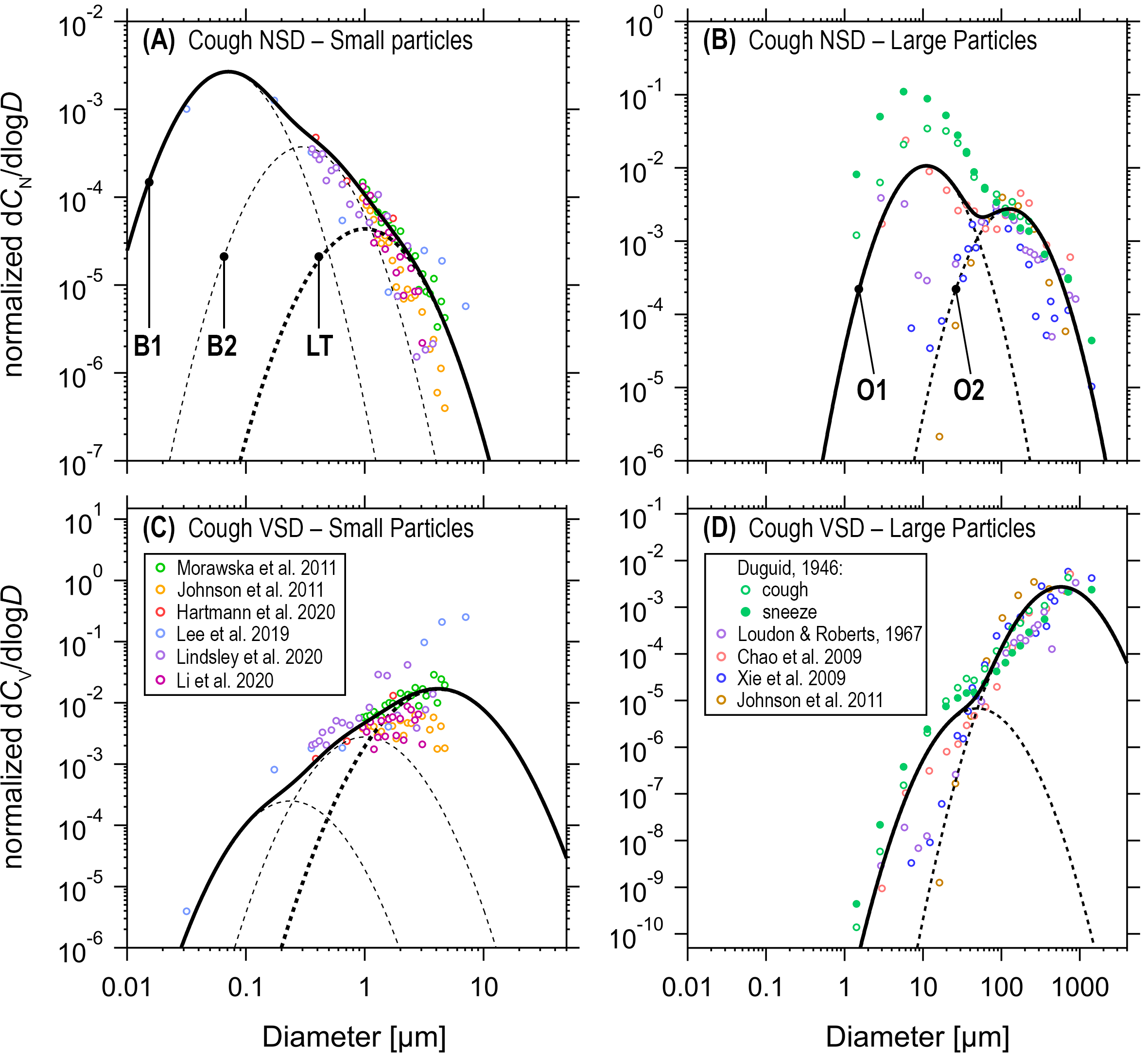}
\caption{Combined literature data on cough-related particle number size distributions, NSDs, \textbf{(A,B)} with calculated volume size distributions, VSDs, \textbf{(C,D)} and parameterization based on five lognormal modes. Figure is closely related to Fig. \ref{fig:speak_syn}. The modes \textbf{B1}, \textbf{B2}, and \textbf{LT}, centered $<$5\,\micro m, are described by a trimodal lognormal fit \textbf{(A,C)}. The trimodal distributions in \textbf{(A)} and \textbf{(C)} were normalized to the area under the breath-related bimodal curve (combined area of modes \textbf{B1} and \textbf{B2}) to emphasize the relative contribution of mode \textbf{LT}. The modes \textbf{O1} and \textbf{O2}, centered $>$5\,\micro m, are described by mono- or bimodal lognormal fits \textbf{(B,D)}. All NSDs and VSDs in \textbf{(B)} and \textbf{(D)} were normalized to the area under mode \textbf{O2} to emphasize the relative variability of mode \textbf{O1}.}
\label{fig:cough_syn}
\end{figure*}

\subsubsection{\label{sec:CoughSneezeResults}Coughing and sneezing}

Coughing is caused by an abrupt parting of the vocal folds and an associated sudden expulsion of air \cite{RN1707,RN2093}. Sneezing is a sudden violent spasmodic expiration with a blast of air being driven through the nasal and mouth chambers \cite{RN1896,RN1897,RN1714}. Both are short, non-continuous, and vigorous respiratory activities associated with an ejection of ELF and saliva. Traditionally, they have been regarded as main drivers for pathogen transmission \cite[e.g.,][]{RN1681,RN1720}. Similar to speaking and singing, (i) laryngeal aerosol generation, (ii) droplet formation through mouth, lips, and tongue movements, and (iii) high-speed shear forces at the ELF-air interface in the trachea and (main) bronchi are probably the relevant formation mechanisms and sites \cite{RN1617,RN1663,RN2017}. The breath-related bronchiolar particle formation is involved as an underlying process. Compared to breathing and speaking, the available data on cough- and sneezing-related particle formation is relatively sparse: for the size range $<$5\,\micro m, we found 5 PSDs from 5 studies for coughing and no PSD for sneezing. For the size range $>$5\,\micro m, we found 5 PSDs from 5 studies for coughing and 1 PSD for sneezing.\par

Figure\,\ref{fig:cough_syn} summarizes the experimental data on cough-related particle emissions in the size range from $>$10\,nm to $\sim$1000\,\micro m. The PSD shape is similar to the speaking- and singing-related PSDs and can also be described by five lognormal modes (Fig.\,\ref{fig:speak_syn}). In addition to the bronchiolar modes \textbf{B1} and \textbf{B2}, the \textbf{LT} mode emerges, similar to speaking and singing. The mode \textbf{LT} is located between $D_\text{LT}$\,=\,0.8\,\micro m and 1.2\,\micro m (on average $D_\text{LT}$\,=\,1.0\,\micro m). The mode height ratio $A_{B2}/A_{\text{LT}}$ ranges from 20 to 7.5 and is therefore higher than the ratio $A_{B2}/A_{\text{LT}}$ for speaking- and singing-related PSDs, which means that the cough-related mode \textbf{LT} is weaker than the speaking- and singing-related mode \textbf{LT}, relative to mode \textbf{B2} (at least within the scope of the data reviewed here).\par 

The large particle size range can be represented by two oral modes \textbf{O1} and \textbf{O2} as for speaking and singing \cite{RN1663}. Mode \textbf{O1} is located between $D_\text{O1}$\,=\,8\,\micro m and 13\,\micro m (on average $D_\text{O1}$\,=\,11\,\micro m). Mode \textbf{O2} is located between $D_\text{O2}$\,=\,90\,\micro m and 200\,\micro m (on average $D_\text{O2}$\,=\,128\,\micro m). The mode height ratio $A_\text{O1}/A_\text{O2}$ is rather variable, ranging from 0 in \citet{RN1701} to $\sim$16 in \citet{RN1762,RN1763}. Similar to speaking and singing, mode \textbf{O2} is present in all five studies, whereas mode \textbf{O1} is present in three studies \cite{RN1701,RN1611,RN1762} and absent in two studies \cite{RN1750,RN1663}. The PSD for sneezing for the large particle size range based on \citet{RN1701} shows mode positions $D_\text{O1}$\,=\,8 and $D_\text{O2}$\,=\,130 as well as a mode height ratio $A_\text{O1}/A_\text{O2} \,=\,48$. A summary of all average fit parameters can be found in Table\,\ref{tab:Fit_Syn}.\par 

A remarkable observation is the similarity between speaking-/singing-related vs coughing-related PSDs in terms of properties of all five modes (compare Figs.\,\ref{fig:speak_syn} and \ref{fig:cough_syn}). This suggests that for both respiratory activities the modes \textbf{LT}, \textbf{O1}, and \textbf{O2} are formed by the same or at least similar mechanisms in the respiratory tract. Another remarkable observation is that the sneezing-related PSD in Fig.\,\ref{fig:cough_syn}B and D resembles the corresponding coughing-related PSDs. At least in the range of modes \textbf{O1} and \textbf{O2}), both respiratory activities can be described by a similar PSD, at least within the scope of data in Fig.\,\ref{fig:cough_syn}. Note that PSD data on sneezing is generally sparse. Besides \citet{RN1701}, also \citet{RN1714} reported sneeze PSDs, which, however, could not be implemented in Fig.\,\ref{fig:cough_syn} due to difficulties with normalization.
\footnote{Since the area under mode \textbf{O2} was used for normalization in Fig.\,\ref{fig:cough_syn}B and D and not all PSDs in \citet{RN1714} showed mode \textbf{O2}, the data could not consistently be normalized. Furthermore, the PSDs were only provided in relative and not in absolute terms so that no mode-specific particle number and volume concentrations could be retrieved for the further steps of our analysis.  
} 
Nevertheless, qualitatively the data from \citet{RN1714} are mostly consistent with Fig.\,\ref{fig:cough_syn} as about half of the their PSDs show a bimodal PSDs with one mode at $\sim$70\,\micro m, corresponding to mode \textbf{O2}. In their data, another mode is reported at $\sim$400\,\micro m, indicating the presence of an additional mode \textbf{O3}, although comprising very low particle concentrations. The other half of PSDs in \citet{RN1714} is monomodal showing mode \textbf{O3} only.

\subsection{\label{sec:ConcRates}Number concentrations and emission rates}

\begin{figure*}
\includegraphics[width=16.5cm]{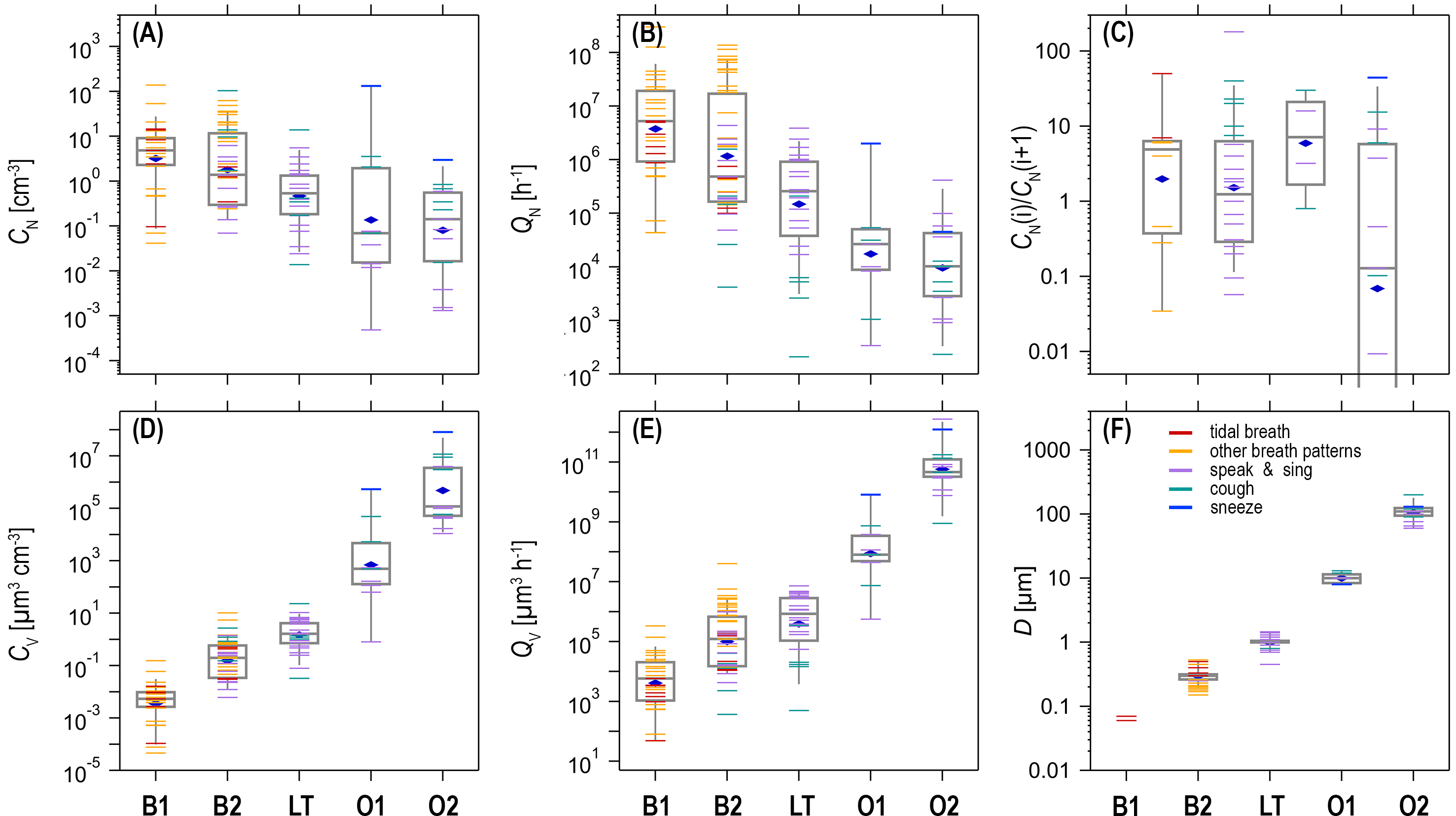}
\caption{Statistical summary of mode-specific particle concentrations ($C$), emission rates ($Q$), and peak diameters ($D$) for individual modes \textbf{B1}, \textbf{B2}, \textbf{LT}, \textbf{O1}, and \textbf{O2}. Figure illustrates the wide variability of $C$ and $Q$ across several orders of magnitude. Particle concentrations and emission rates are provided for number and volume ($C_\text{N}$ vs $C_\text{V}$ and $Q_\text{N}$ vs $Q_\text{V}$). Further, the ratios of $C_\text{N}$ of adjacent modes ($C_\text{N}(i)/C_\text{N}(i+1)$), derived from individual studies with PSD data covering two or more modes (i.e., $C_\text{N}\text{(B1)}/C_\text{N}\text{(B2)}$, $C_\text{N}\text{(B2)}/C_\text{N}\text{(LT)}$, $C_\text{N}\text{(LT)}/C_\text{N}\text{(O1)}$, and $C_\text{N}\text{(O1)}/C_\text{N}\text{(O2)}$) are shown. Colored markers (short vertical lines) distinguish between respiratory activities (see legend). In the box-whisker plots, the boxes represent the 25 and 75 percentiles, the grey lines the medians, the blue diamond markers the means and the whiskers the 9 and 91 percentiles.}
\label{fig:conc_syn}
\end{figure*}

\begin{figure*}[htb]
\includegraphics[width=17cm]{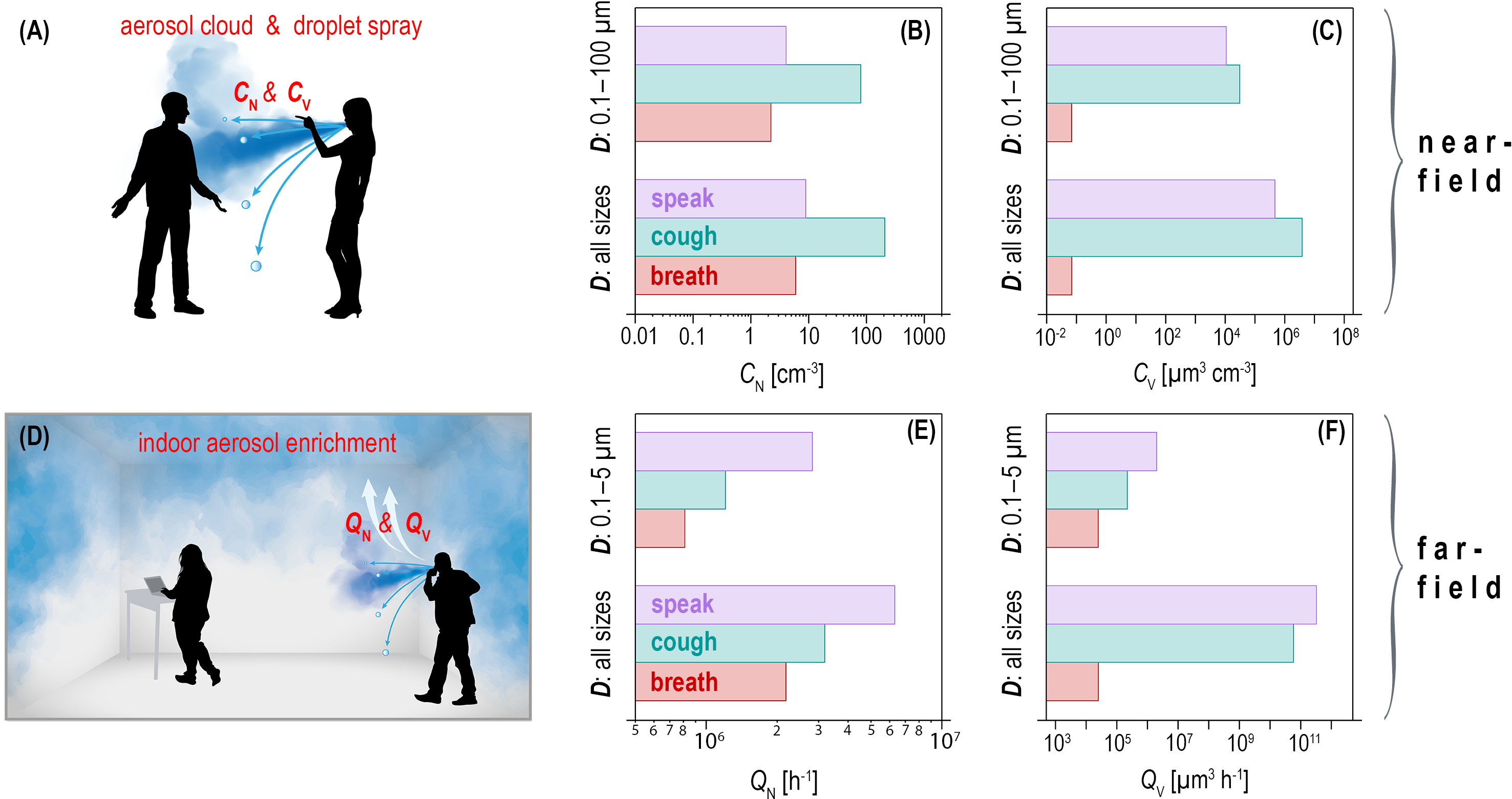}
\caption{Total number and volume concentrations ($C_\text{N}$, $C_\text{V}$) as well as emission rates ($Q_\text{N}$, $Q_\text{V}$) of exhaled particles for different respiratory activities. Two different particle size ranges were chosen to compare conditions under near- vs far-field disease transmission scenarios: (i) In the near-field \textbf{(A--C)}, $C_\text{N}$ and $C_\text{V}$ in the exhaled puff(s) prior to dilution and dissipation are primarily relevant. Here, a lower threshold of 0.1\,\micro m (typical size of virus pathogens, Fig.\,\ref{fig:overview_syn}) and an upper threshold of 100\,\micro m (upper limit of common aerosol definitions, see Sect.\,\ref{sec:Def}) were chosen. (ii) In the far-field \textbf{(D--F)}, $Q_\text{N}$ and $Q_\text{V}$ are primarily relevant, especially if emission extends over longer periods. Here, a lower threshold of 0.1\,\micro m and an upper threshold of 5\,\micro m (particles below this threshold have particularly long residence time in air, Fig.\,\ref{fig:Re}) were chosen. In addition, $C$ and $Q$ across the entire size range of the parameterization (i.e., 0.002 to 4000\,\micro m) are shown in all cases for reference. The total $C_\text{N}$ and $C_\text{V}$ as well as $Q_\text{N}$ and $Q_\text{V}$ values were derived as the integral under the average multimodal PSDs (e.g., Fig.\,\ref{fig:overview_syn}B and C).}
\label{fig:barplot}
\end{figure*}

The previous sections focused on the shape of the PSDs and developed a consistent parameterization of their multimodal character. Here, we used the multimodal parameterization to calculate and summarize the statistics of mode-specific properties in Fig.\,\ref{fig:conc_syn}, such as number and volume concentrations ($C_\text{N}$, $C_\text{V}$) as well as number and volume emission rates ($Q_\text{N}$, $Q_\text{V}$) based on the available studies with quantitative PSDs (flag A in Table\,\ref{tab:LitSyn}). Consistent with previous studies \cite[e.g.,][]{RN1613,RN1608,RN1708,RN1750,RN1734,RN1830,RN1710}, the obtained $C$ and $Q$ levels are highly variable and span across several orders of magnitude. The large variability in $Q$ among infected individuals has been considered as a potential explanation for the existence of superspreaders in infectious disease transmission as they emit much more potentially pathogen-laden particles than others \cite[e.g.,][]{RN1608,RN2037,RN1771}.\par 

On average, the highest $C_\text{N}$ and $Q_\text{N}$ levels were found for the modes \textbf{B1} and \textbf{B2}, which then decrease via mode \textbf{LT} to modes \textbf{O1} and \textbf{O2}. The opposite trend -- a strong increase from mode \textbf{B1} to \textbf{O2} -- was found for $C_\text{V}$ and $Q_\text{V}$ since the volume scales with $D^3$ and, therefore overcompensates the decreasing trend in number representation. Figure\,\ref{fig:conc_syn}F shows the comparatively low variability of the peak position of individual modes and further emphasizes that some modes (i.e., \textbf{B2} and \textbf{LT}) overlap significantly, whereas others (i.e., \textbf{O1} and \textbf{O2}) appear rather well separated. The ratios $C_\text{N}(i)/C_\text{N}(i+1)$ in Fig.\,\ref{fig:conc_syn}C shows that the variability of the relative strength of individual modes is rather high (compare with Sect.\,\ref{sec:SpeakSingResults} and Sect.\,\ref{sec:CoughSneezeResults}).\par

Both the particle concentration and emission flux for a given respiratory activity are relevant quantities for a better understanding of airborne disease transmission, which can in principle occur in the near-field or far-field (Fig.\,\ref{fig:scheme}). Figure\,\ref{fig:barplot} provides total $C$ and $Q$ levels for selected particle size ranges to emphasize the relative importance of breathing, speaking, and coughing in near- and far-field scenarios of disease transmission. \par  

In a potential near-field aerosol transmission scenario, the recipient is located closely (few meters) to the emitter and inhales her/his concentrated particle puff(s) relatively soon after release (Fig.\,\ref{fig:barplot}A). Thus, the time span between emission and inhalation is presumably too short for particle cloud dilution as well as large particle sedimentation to occur to a significant degree. Thus, Fig.\,\ref{fig:barplot}B and C compare the total $C_\text{N}$ and $C_\text{V}$ levels for the size range from 0.1 to 100\,\micro m. Here, 0.1\,\micro m was chosen as a lower limit as it is the physical size of, e.g., SARS-CoV-2 and infuenza virions (see Sect.\,\ref{sec:Modality}). Coughing as well as presumably also sneezing (not shown here due to sparse data availability) causes the highest concentrations in the near-field aerosol cloud. For $C_\text{N}$ in the range from 0.1 to 100\,\micro m, coughing clearly exceeds speaking, which is closely followed by breathing. For $C_\text{V}$ in the range from 0.1 to 100\,\micro m, coughing is highest, comparatively closely followed by speaking, and breathing being significantly smaller. These trends in number and volume representation result from the differences in the multimodal shape of the corresponding PSDs and have to be considered in the choice of either $C_\text{N}$ or $C_\text{V}$ for risk assessments.\par         

In a potential far-field aerosol transmission scenario, an infected individual emits pathogen-laden particles and those can accumulate over time in confined spaces, such as a restaurant room or public transport (Fig.\,\ref{fig:barplot}D) \cite[e.g.,][]{RN1986,RN1717,RN2066,RN2065,RN1807}. If critical pathogen concentrations can build up in room air, depends on multiple factors, such as particle size, room size, ventilation rates etc \cite[e.g.,][and references therein]{nazaroff2016indoor,riley2002indoor,lai2002particle,RN2105}. Infection risks depend on the airborne pathogen concentrations, exposure times, and the recipient's susceptibility. Here, the source strength, which is represented by the particle emission rate as a function of respiratory activities, is of primary relevance. To emphasize the different source strengths of breathing, speaking, and coughing, we focus on the particle size range from 0.1 to 5\,\micro m, comprising those particles with particularly long airborne residence times. Figure\,\ref{fig:barplot}E and F shows that in a far-field scenario speaking clearly dominates in $Q_\text{N}$ and $Q_\text{V}$ in the range from 0.1 to 5\,\micro m, followed by coughing and then breathing. This suggests that speaking might be a particularly important driver for airborne transmission as it represents a comparatively strong particle source and further occurs more frequent/continuous that, e.g., coughing. Note further that the transmission of SARS-CoV-2 occurs mostly pre-symptomatically, which further diminishes the supposed relevance of coughing and emphasizes the supposed importance of speaking.

% mention that 5 micron is often the upper limit in measurements; text: maybe this is the "reason" for medical definition of "aerosol vs. droplet")

% Note that all PSDs implemented here were measured under laboratory conditions for partly artificial respiratory maneuvers and, thus do not necessarily reflect real life conditions (see Table\,\ref{tab:LitSyn}).

\subsection{\label{sec:Modality}Multi-modality of size distributions in airborne disease transmission}

\begin{figure*}
\includegraphics[width=13cm]{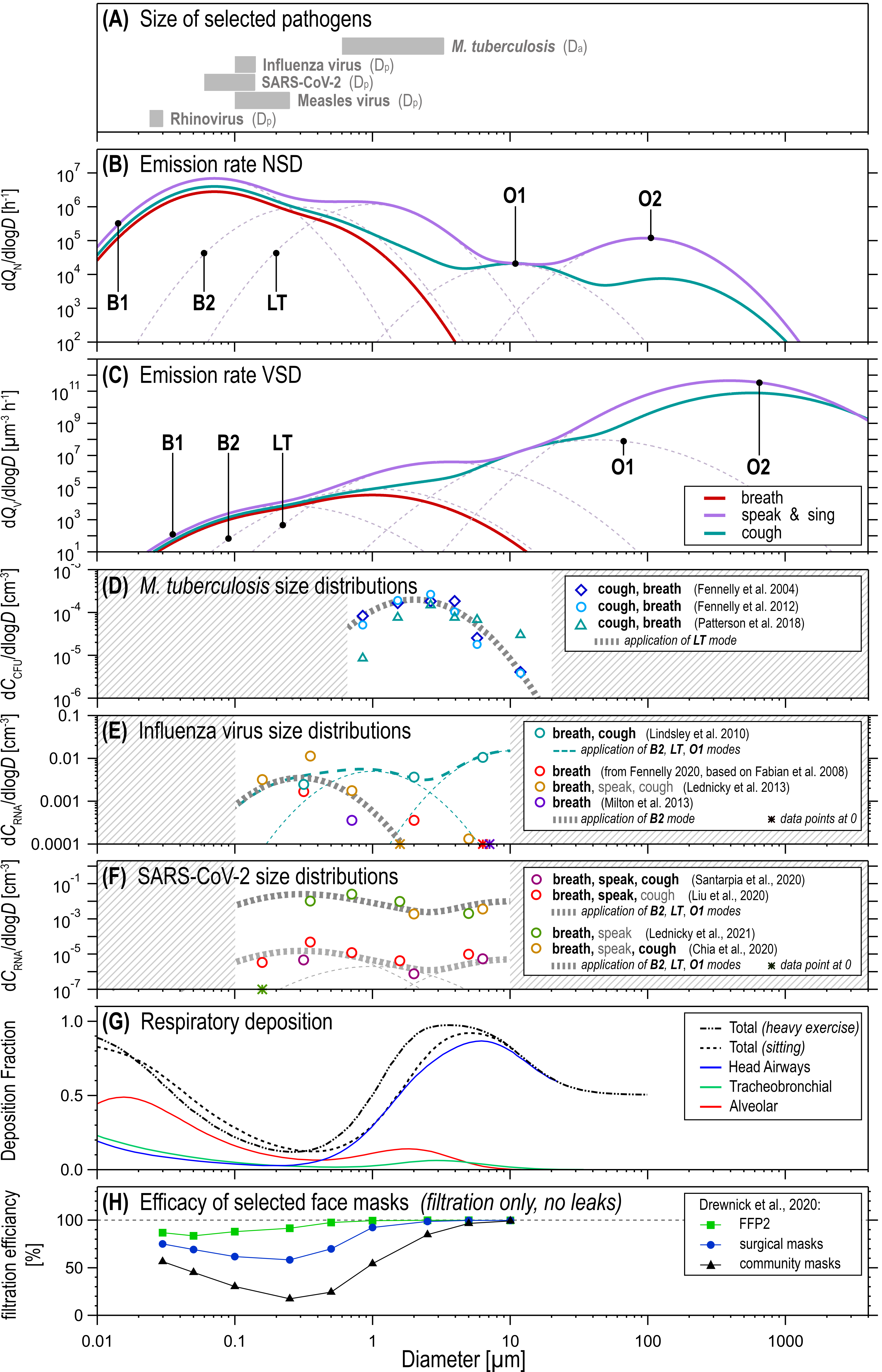}
\caption{Synthesis figure emphasizing the relevance of the multimodal character of the exhaled particle size distribution (PSD) in airborne disease transmission. Figure combines the PSD parameterization (\textbf{B}\,\&\,\textbf{C}) with the size of selected pathogens (physical diameter, $D_\text{p}$, for all viruses and aerodynamic diameter, $D_\text{a}$, for \textit{M. tuberculosis}) as a lower limit for pathogen carriage (\textbf{A}), measured size distributions of selected pathogens embedded in respiratory particles (\textbf{D}\,--\,\textbf{F}), deposition properties of the respiratory tract (ICPR model) (\textbf{G}), and filtration efficacy of face masks (\textbf{H}). \textbf{D}\,--\,\textbf{F}: The pathogen size distributions were obtained from the original studies (see legend), bin-normalized, and converted in pathogen entities per air volume. The legend further specifies involved respiratory activities (bold, black font) and activities that might be involved as well (gray font). The parameterization was applied to the pathogen size distributions by adapting the position(s) and width(s) of modes \textbf{B2}, \textbf{LT}, and \textbf{O1} (with only slight changes) and freely adjusting the mode height(s). \textbf{H}: Face mask transmission curves were measured at high air velocity, with surgical mask transmissibility representing the average of three types of surgical masks and community mask transmissibility representing the average of three materials: velvet polyester, 2 layers of cotton jersey, and thin silk.}
\label{fig:overview}
\end{figure*}

Figure\,\ref{fig:overview} combines main findings of this review and data synthesis and emphasizes that a detailed understanding of the multimodal shape of the respiration PSDs provides novel mechanistic insight into airborne disease transmission and the effectiveness of preventative measures \cite{RN1663}. An important parameter for the transport of viruses, bacteria, as well as bacterial or fungal spores by respiratory particles is the pathogen's size. Figure\,\ref{fig:overview}A, summarizes the size of selected aerosol transmissible pathogens, such as rhinoviruses (one of the agent groups causing the 'common cold') \cite[e.g.,][]{RN1818,RN1828}, measles morbilliviruses \cite[e.g.,][]{RN1667}, SARS-CoV-2 \cite[e.g.,][]{RN1832,RN1833,RN1834}, influenza viruses \cite[e.g.,][]{RN1836,RN1837,RN1842} and \textit{M. tuberculosis} \cite[e.g.,][]{RN1838}. The pathogen size defines the lower limit for the respiratory particle size range that is relevant for pathogen transport. Figure\,\ref{fig:overview}B and C combine the parameterizations for different respiratory activities, in relation to the average mode-specific emission rates from Fig.\,\ref{fig:conc_syn} and Fig.\,\ref{fig:barplot}. The breath-related bimodal PSD is shown in relation to the speaking- and singing-related as well as cough-related pentamodal PSDs, yielding a coherent overall picture. The comparison shows that all viruses in Fig.\,\ref{fig:overview}A are in principle small enough to be transmitted with all five modes. For the larger \textit{M. tuberculosis}, the modes \textbf{B1} and \textbf{B2} can presumably be excluded as potential carriers, which leaves the modes \textbf{LT}, \textbf{O1}, and \textbf{O2} for transmission.

Particularly valuable -- but also sparse since difficult to obtain -- is experimental data on the actual pathogen size distributions in the carrier aerosol as introduced in Sect.\,\ref{sec:Mixing}. It provides evidence on the modes, in which the pathogens are actually being transmitted. The available data from size-resolved sampling of airborne \textit{M. tuberculosis}, influenza virions, and SARS-CoV-2 virions are shown in Fig.\,\ref{fig:overview}D, E and F. Generally, the available data suggest that these pathogens are enriched in a certain size range of the overall PSD, which presumably reflects the sites of infection and particle formation in the respiratory tract. Note however, that the number of existing studies is small and certain parts of the overall size range (i.e., $>10$\,\micro m) have not been investigated with size-resolved sampling approaches, yet.\footnote{
Different aerosol samplers were used in the studies summarized in Fig.\,\ref{fig:overview}D, E and F: (i) the Sioutas five-stage personal cascade impactor (SKC, Inc.) \cite[described in][]{RN2041,RN2042} and \cite[used in][]{RN2099,RN1937,RN1651}; (ii) the cascade cyclone of the National Institute for Occupational Safety and Health (NIOSH) sampler \cite[described in][]{RN2043} and \cite[used in][]{RN1855,RN1652,RN1705}; (iii) the Andersen six-stage cascade impactor \cite[described in][]{RN2044} and \cite[used in][]{RN1695,RN1856,RN2018}, and (iv) a two-stage slit impactor with an intermediate condensational growth step as part of the G-II sampling setup \cite[described in][]{RN2100} and \cite[used in][]{RN1605}. Strengths and limitations of the individual sampling approaches have to be kept in mind when results are compared \cite[further information can be found in][]{RN2047}.
}  \par 

Under the assumption that airborne pathogen transmission is driven by specific respiratory activities and their associated PSD modes, we applied the previously defined parameterization to the pathogen size distributions. Specifically, we described the size-resolved concentrations of the pathogens -- i.e., ribonucleic acid (RNA) copies for virions and culture forming unit (CFU) for \textit{M. tuberculosis} -- by the previously defined mode properties (see Table\,\ref{tab:Fit_Syn}). Here, the position ($D_\text{i}$) and width ($\sigma_\text{i}$) of the mode(s) were mostly constrained and only slightly adjusted, whereas the number of modes (one or two) as well as their heights ($A_\text{i}$) were scaled to represent the pathogen size distributions best. In all cases, the pathogen size distributions could be described consistently with one to three of the modes \textbf{B2}, \textbf{LT}, and \textbf{O1}. This approach holds uncertainties, especially due to the small number of data sets as well as the comparatively coarse size resolution (e.g., three to six size bins only). Nevertheless, Fig.\,\ref{fig:overview}D, E and F show conceptually that the pathogen size distributions can be described consistently with an underlying modal structure in agreement with the overall PSD parameterization.\par  

\textit{M. tuberculosis} in Fig.\,\ref{fig:overview}D was sampled by \citet{RN1695,RN1856} and \citet{RN2018} from infected and coughing individuals and the resulting pathogen size distributions can be fitted well by the cough-related mode \textbf{LT}. The agreement between the \textit{M. tuberculosis} size distributions and mode \textbf{LT} is remarkably consistent. The \textit{M. tuberculosis} size distribution decreases towards the modes \textbf{B1} and \textbf{B2}, which is consistent with the fact that those exhaled particles are too small for pathogen transport. Moreover, the \textit{M. tuberculosis} size distribution decreases (rather steeply) towards the modes \textbf{O1} and \textbf{O2}, which suggests that also these modes may not play a primary role in TB transmission. It has to be kept in mind, however, that large particles are particular prone to impaction and sedimentation losses in the sampling setup and that the potential influence of such sampling artifacts on the shape of the pathogen size distribution has to be critically evaluated \cite{RN1817,RN1323,RN2047}.\par            
% Knibbs 14 with Adersen samples of P. aeruginosa: more CFU below 3.3 µm
% Jones-Lopez 13: used cough aerosol to predict TB transmission; minority of TB patients produced cough aerosol with pathogen; high aerosol loads: >=10 CFU (up to 400 !) vs low aerosol loads 1--9 CFU;  

The size distributions of airborne influenza virions, obtained from breathing \cite{RN1599,RN1783,RN1605} and coughing \cite{RN1855} individuals as well as from mixed respiratory activities \cite{RN2099} are shown in Fig.\,\ref{fig:overview}E.\footnote{
The upper limit of the largest size bin is not specified on all studies or pre-defined by the design of the samplers. It is needed, however, for the normalization of concentrations to the bin widths. \citet{RN1937} and \citet{RN1651} define 10\,\micro m as upper limit of the largest size bin. For all other influenza- and SARS-CoV-2-related studies \cite{RN1652,RN1705,RN2099,RN1605} 10\,\micro m were chosen here as well for consistency. Note also that 20\,\micro m was tested instead of 10\,\micro m as an upper size limit and did not affect the bin width-normalized concentration of the corresponding size bin significantly. The lower size threshold of the smallest size bin was set to 0.1\,\micro m, which is given by the average size of the SARS-CoV-2 virions.} 
The breath-related influenza size distributions suggest a predominance of mode \textbf{B2} without strong contributions by mode \textbf{LT}. This underlines a statement by \citet{RN1841} that "breathing is enough" for the transmission of influenza. Furthermore consistent with our results (Sect.\,\ref{sec:CoughSneezeResults}) is that the cough-related influenza size distribution involves the mode \textbf{B2} as well as the modes \textbf{LT} and \textbf{O1}, which both occur in relation to coughing. The low size resolution bears rather large uncertainties regarding the relative contributions from modes \textbf{LT} and \textbf{O1}. Nevertheless, a co-occurrence of both these modes would be expected for coughing according to Sect.\,\ref{sec:CoughSneezeResults}. The upper size limit of the sampling at $\sim$10\,\micro m does not allow solid conclusions on the influenza abundance in modes \textbf{O1} and particularly \textbf{O2}. Overall, these results underline previous observations of influenza virions especially in the small aerosol size range and emphasize that essentially all respiratory activities might be driving forces in the spread of this virus \cite[e.g.,][]{RN1605,RN1781,RN1766,RN1864,RN1771,RN1602}.\par

The four available SARS-CoV-2 size distributions with at least two size bins based on studies by \citet{RN1652}, \citet{RN1651}, \citet{RN1937}, and \citet{RN1705} are summarized in Fig.\,\ref{fig:overview}F. In the studies by \citet{RN1652}, \citet{RN1651}, and \citet{RN1705}, the sampling did not target specific respiratory activities, but was rather conducted inside hospital rooms with COVID-19 patients. Presumably, the sampling probed a mixture of breathing, coughing (a common symptom of COVID-19), and probably also speaking. \citet{RN1705} specified that the sampled patients were coughing. In the study by \citet{RN1937}, aerosol samples were collected inside a car driven by a COVID-19 patient with a mild clinical course of the disease and, thus an unknown mixture of respiratory activities was probed. The size distributions show a similar bimodal shape and the mode \textbf{B2} could be allocated to them rather clearly, which suggests that COVID-19 is also transmissible by breathing \cite{RN1841}. Further a second mode centered between the modes \textbf{LT} and \textbf{O1} is resolved. This suggests that also the modes \textbf{LT} and/or \textbf{O1} from speaking/singing or coughing can be drivers in COVID-19 transmission. Generally, the cough-related influenza and both SARS-CoV-2 size distributions show a rather similar bimodal shape, which might indicate mechanistic analogies in transmission mechanisms of both diseases.\par        
% Matricardi 20: different COVID-19 clinical courses, 'common' mild course --> mostly URT emission expected; more rare severe course, also LRT emission expected; 

\begin{figure}[t]
  \includegraphics[width=\columnwidth]{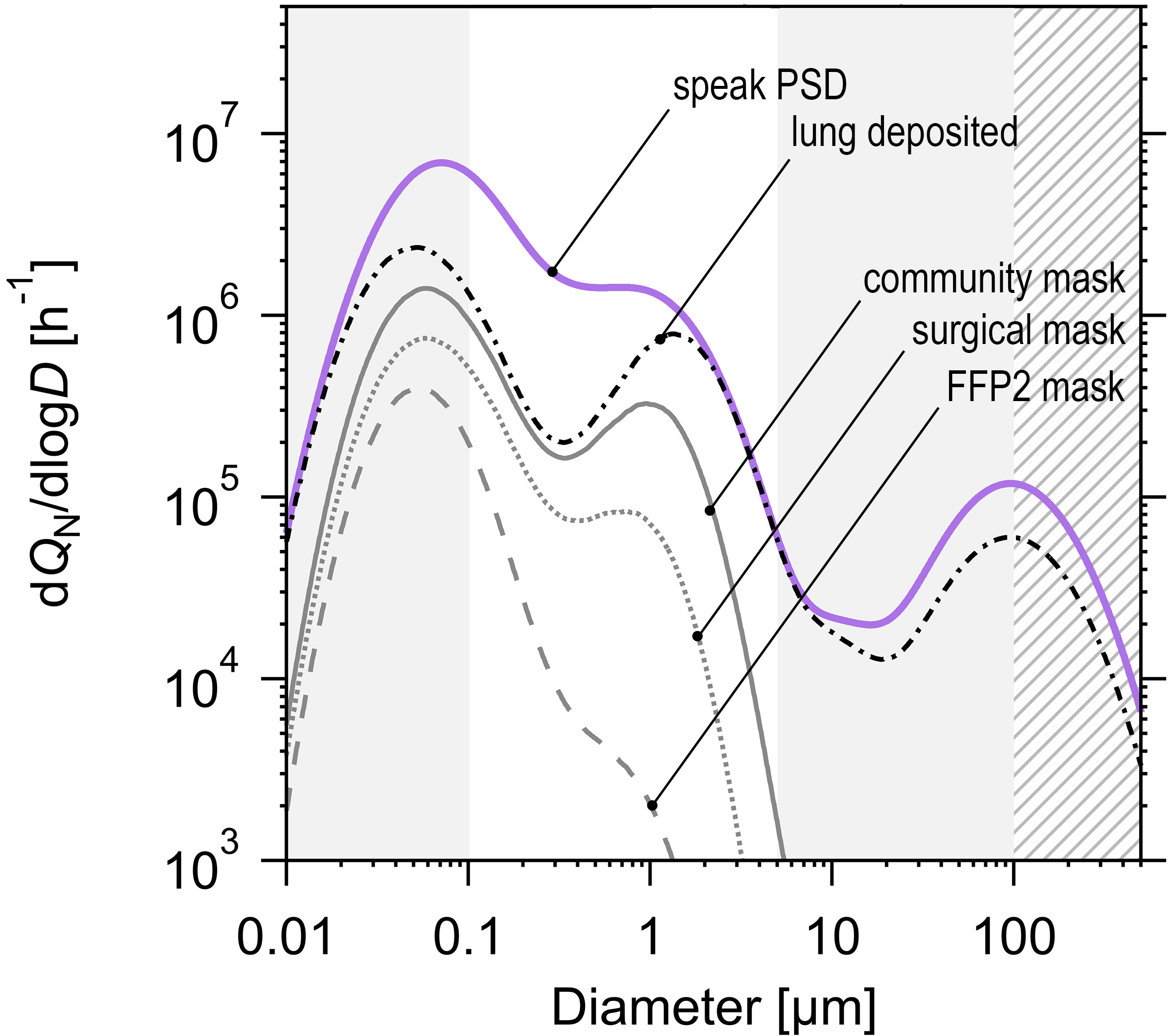}
  \caption{\label{fig:overview_syn} Effects of particle deposition in respiratory tract and face mask filtration on speaking-related particle size distribution (PSD, here number emission rate, $Q_\text{N}$, adapted from Fig.\,\ref{fig:overview}B). Figure compares (i) PSD after emission, (ii) fraction of PSD deposited in respiratory tract (compare Fig.\,\ref{fig:overview}G), and (iii) fraction of PSD deposited in respiratory tract after filtration through community masks, surgical masks, and FFP2 masks (compare Fig.\,\ref{fig:overview}H). Only the mask filtration and not the mask fitting properties (loose vs tight) are taken into account. Background shading divides overall size range according to common definitions (up to 100\,\micro m) and a size spanning from the physical size of influenza and SARS-CoV-2 virions ($\sim$0.1\,\micro m) to $\sim$5\,\micro m, which the fraction being particularly relevant for far-field transmission.}
\end{figure}

Beyond aerosol emission and transport from host to host, the size-dependent deposition of pathogen-laden particles throughout the human respiratory tract is of critical importance. Figure \ref{fig:overview}G shows the established ICPR deposition model, distinguishing deposition in the alveolar, tracheobronchiolar, and head airways, which is driven by the size-dependent influences of diffusion, impaction, and sedimentation particle losses. Here, a detailed understanding of the modality of the inhaled PSD (Fig.\,\ref{fig:overview}B and C) in combination with knowledge on those modes that presumably carry most of a given pathogen (Fig.\,\ref{fig:overview}D, E and F) allows to assess to what extend the pathogens reach their corresponding target sites in the respiratory tract. For example, alveolar macrophages are the target site for airborne infection with \textit{M. tuberculosis} \cite{RN1878,RN1910}. The combination of Fig.\,\ref{fig:overview}D and G shows that the carrier mode \textbf{LT} is co-located with a secondary maximum of alveolar deposition, underlining that the bacteria could reach the target directly in airborne state. Note further, that the modes \textbf{O1} and \textbf{O2}, for which Fig.\,\ref{fig:overview}D suggests that the bacteria are only sparsely present, corresponds with a size range in which the alveolar deposition approaches to zero. In this sense, mode \textbf{LT} appears to be the 'evolutionary optimized' vehicle for the airborne spread of TB directly to its target sites. \par  

Finally, Fig.\,\ref{fig:overview}H completes the picture by illustrating the characteristic size-dependent filtration efficiencies of selected face mask materials. Wearing face masks is a main measure to decelerate the spread of diseases, given that (semi)ballistic droplet spray and aerosol transmission are major infection routes. Accordingly, the literature on the influence of face masks - especially in the context of the current COVID-19 pandemic -- has grown considerably \cite[e.g.,][]{RN1867,RN1870,RN1869,RN1702,RN1868,RN1703,RN2031,RN2032,RN2033,RN2035,RN2059}. The similarity between the size-dependent deposition in the respiratory tract and the face mask filtration efficiency is obvious, which can be explained by the fact that the relevant particle loss mechanisms (i.e., diffusional, impaction, and sedimentation losses) are comparatively ineffective in the range roughly between 100 to 500\,nm. In the field of atmospheric aerosol physics, this size band is known as the location of the so called accumulation mode because particles tend to accumulate in here due to the minimum in particle loss/removal efficiencies \cite[e.g.,][]{RN433,lai2002particle,RN2102,RN2103}. \citet{RN1841} emphasized that this size band is particularly important as it is the co-location of the minimum in face mask efficiency and the maximum of the atmospheric lifetime of pathogen-laden particles. The peak of mode \textbf{B2} falls within this range, which implies that it has on average the longest lifetime in the air as well as the largest transmissibility through common face mask materials (especially so called community masks). Face masks with high filtration efficiencies (e.g., N95/FFP2) show a much lower transmissibility in this particular size range and are thus particularly useful for preventing the airborne transmission of diseases such as COVID-19 \citep[e.g.,][]{RN1703,RN1894,RN1986,RN2058}. \par

Figure\,\ref{fig:overview_syn} visualizes the filtration effect of face masks on the size distribution of speaking-related particles emission rates as shown in Fig.\,\ref{fig:overview}B. All PSDs -- the initial speaking-related PSD and the PSDs after face mask filtration -- have been multiplied by the total deposition curve of the respiratory tract as shown in Fig.\,\ref{fig:overview}H. Thus, the (black) PSDs in Fig.\,\ref{fig:overview_syn}B compare the deposited particle fractions with and without face mask filtration. Note that this assumes mask filtration without leaks (i.e., a tight fit of the mask). Beyond the filtration efficiency of the mask materials, the tightness of the fit and potential leakage between mask and face is of crucial importance for the overall mask performance \cite[e.g.,][]{RN2104}. Clearly, all masks have a high efficacy for large particles (i.e., larger $\sim$5\,\micro m). For the small particle range (i.e., smaller $\sim$5\,\micro m), however, the transmission curves of community, surgical, and FFP2 masks show significant differences. Note that these differences overlap with the size range of the modes \textbf{B2} and \textbf{LT}, which are likely involved in the transmission of the pathogen shown in Fig.\,\ref{fig:overview}D, E, and F.\vfill

\section{Summary and conclusions}

\subsection{Aerosol and droplet transmission of infectious human diseases}

\begin{figure*}[t]
\includegraphics[width=16cm]{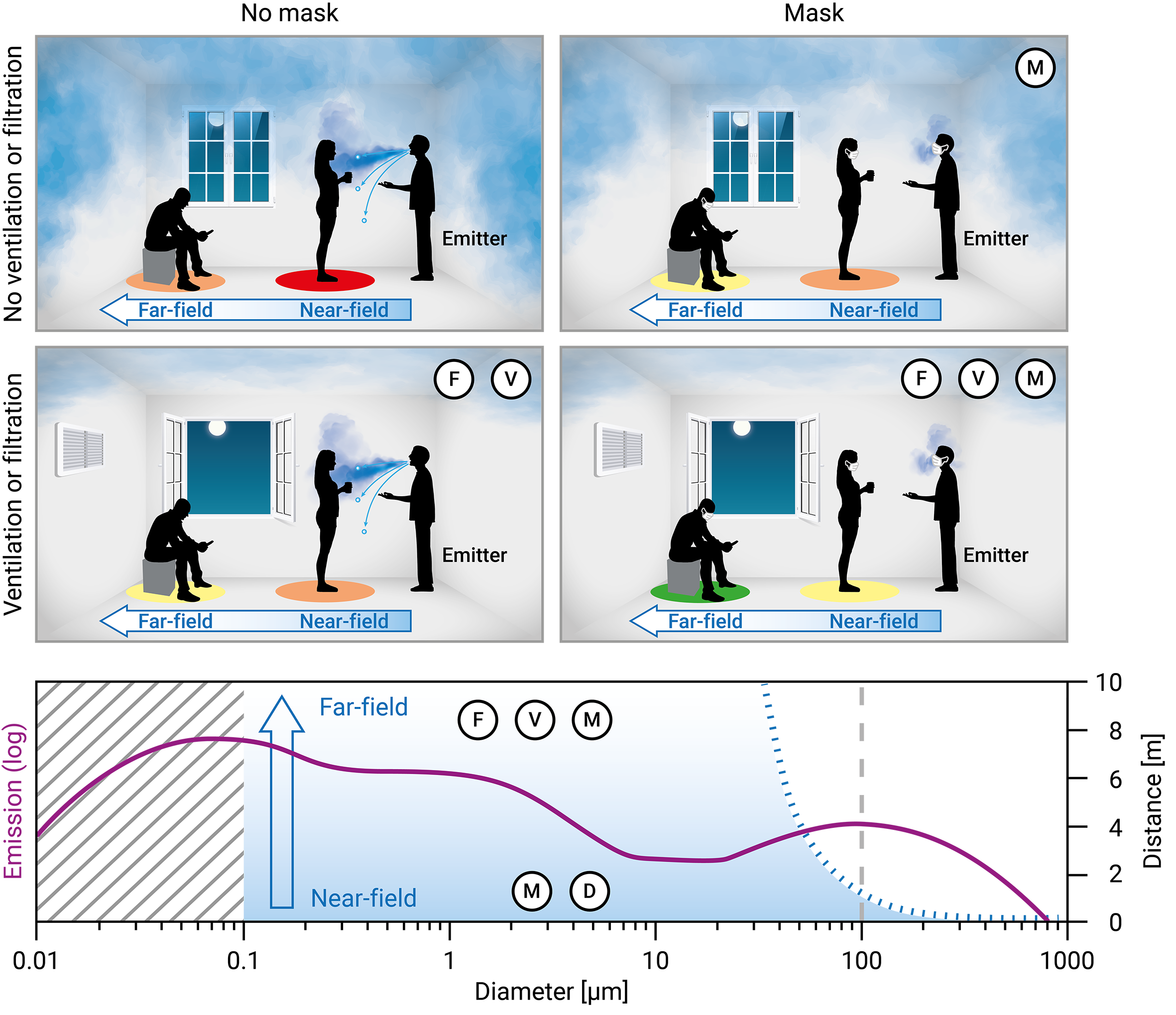}
\caption{Illustration of near- and far-field aerosol transmission in relation to the respiratory particle size distribution (PSD) and preventive measures: \textbf{M}\,$=$\,mask, \textbf{D}\,$=$\,distance, \textbf{F}\,$=$\,filtration, and \textbf{V}\,$=$\,ventilation. Only the aerosol emission from the infected emitter is visualized. The approximated infection risks for the recipients are illustrated by colored circles (red\,$=$\,highest risk\,$>$\,orange\,$>$\,yellow\,$>$\,green\,$=$\,lowest risk). Bottom panel shows speaking-related PSD (adapted from Fig.\,\ref{fig:overview_syn}B), which represents particle concentration (most relevant in near-field) and emission rate (most relevant in far-field) that can be converted into each other by a constant factor (see Table\,\ref{tab:RespParam}). Blue dashed line in PSD panel shows that the travel range of exhaled particles (for an exemplary emission velocity of $20\,\text{m\,s}^{-1}$ and emission height of 1.5\,m) in still air is strongly size-dependent ($D^2$ relationship): large particles (i.e., $>100$\,\micro m) are exclusively relevant in near-field, whereas smaller particles (i.e., $<100$\,\micro m) are relevant in both, near- and far-field. Blue background shading in PSD panel illustrates qualitatively that particle concentrations decrease with distance from emitter due to dilution and mixing. Threshold at 100\,nm marks the physical size of SARS-CoV-2 and influenza virions.}
\label{fig:overview_scem}
\end{figure*}

\textbf{(1)} A critical review of definitions, nomenclature, and concepts regarding airborne or contact-free disease transmission was presented. Through this process we clarified critical aspects of the discussion for efficient transdisciplinary exchange and collaboration between the medical and physical sciences (e.g., use of the term airborne, the continuity and overlap of respiratory aerosol and droplet size ranges, as well as droplet desiccation and hygroscopicity). Moreover, we highlighted important knowledge gaps that currently hamper the quantitative assessment and prediction of the spread of pathogens. Improved connection between infectious disease medicine and a physical or mechanistic understanding of respiratory emissions thus provides basis for improved control of disease transmission, including non-pharmaceutical intervention strategies (e.g. mask type, ventilation, filtration), especially for diseases for which no effective vaccination exists. While we endeavor to provide tools to bring groups of scientists closer together, we acknowledge disciplinary bias is implicit and have written primarily from the perspective of physical scientists.\par \vspace{3mm}

\textbf{(2)} Based on a comprehensive review and synthesis of the scientific literature, we developed a parameterization of particle emissions from the full respiratory tract. The parameterization requires only five lognormal modes for the size range from $>$10\,nm to $\sim$1000\,\micro m to efficiently describe all PSDs available in published literature. The parameterization was optimized in the course of a 'multi-dimensional fitting' to represent both number and volume size distributions. As a further development of \citet{RN1663}, the five lognormal modes can be subdivided in three categories that elucidate unique and different particle formation mechanisms and sites in the respiratory tract. The bronchiolar modes \textbf{B1} and \textbf{B2} are formed in the terminal brochioles through film bursting processes. The mode \textbf{LT} is presumably formed through vibrations and shear forces in the larynx and/or high turbulence and shear forces in the trachea. Modes \textbf{O1} and \textbf{O2} are formed by mouth, tongue, and lips movements. \par \vspace{3mm}  
% \textcolor{violet}{Note that particles smaller and larger $\sim$ 5 \micro m was measured with different methods, while larger particles was measured direct after exhalation where supersaturated conditions can be expected. These supersaturated conditions can lead to a growth of particles in the \textbf{LT} mode into the \textbf{O1} size range. For these reasoned it is important to future investigate the intercept of the two measurement methods.}

\textbf{(3)} Breathing is associated with a bimodal PSD (\textbf{B1} \& \textbf{B2}), whereas speaking and singing as well as coughing are associated with pentamodal PSDs (\textbf{B1}, \textbf{B2}, \textbf{LT}, \textbf{O1}, \& \textbf{O2}). Overall, the PSDs of all respiratory activities yield a consistent pattern. Respiratory activities show a characteristic, multimodal PSD shape, while the associated particle concentrations and emission rates vary across orders of magnitude. \par \vspace{3mm}  

\textbf{(4)} The variability among PSDs within a given respiratory activity can be explained by (i) the fact that the particles were measured under different RH conditions and thus different particle drying states after exhalation as well as by (ii) differences in mode height relative to each other in relation to breathing patterns and vocalization volumes. \par \vspace{3mm}

\textbf{(5)} The emission of pathogen-laden particles in the course of all respiratory activities can transmit diseases in the near- and far-field. For near-field aerosol transmission, the particle concentration in the exhaled puffs is of primary relevance. Here, our parameterization shows the following order for number concentrations ($C_{\text{N}}$): $C_{\text{N}}\text{(cough)}>C_{\text{N}}\text{(speak)}\approx C_{\text{N}}\text{(breathe)}$ and for volume concentrations ($C_{\text{V}}$): $C_{\text{V}}\text{(cough)}\approx C_{\text{V}}\text{(speak)}>C_{\text{V}}\text{(breathe)}$. For far-field aerosol transmission, the particle emission rate -- which strongly depend on the frequency of a given respiratory activity -- is of primary relevance. Here, our parameterization shows the following order for number emission rates ($Q_{\text{N}}$): $Q_{\text{N}}\text{(speak)}>Q_{\text{N}}\text{(cough)}> Q_{\text{N}}\text{(breathe)}$ and for volume emission rates ($Q_{\text{V}}$): $Q_{\text{V}}\text{(speak)}>Q_{\text{V}}\text{(cough)}> Q_{\text{V}}\text{(breathe)}$.  \par \vspace{3mm} 

\textbf{(6)} An understanding of the multimodal shape of the respiration PSDs provides novel mechanistic insight into airborne disease transmission and the efficacy of preventative measures. For common aerosol-transmissible diseases, modes were identified through which the pathogens are most likely transmitted: e.g., cough-related mode \textbf{LT} for tuberculosis; breath-, speaking-/singing-, and cough-related modes \textbf{B2}, \textbf{LT}, and \textbf{O1} for influenza and SARS-CoV-2. \par \vspace{3mm} 

\textbf{(7)} Public health recommendations can benefit from an improved understanding of the respiration PSDs parameterized here and through an updated view of the mechanisms of particle mixing and transmission as illustrated in Fig.\,\ref{fig:overview_scem}. Larger droplets ($>$\,100\,\micro m) are not generally inhalable, have rapid settling velocity, and are thus relevant only in the near-field where ballistic trajectories can launch them onto nearby surfaces or directly onto mucosal membranes of the recipient. For respiratory particles in this size regime, even relatively loose fitting masks (i.e. surgical-style) or face shields are sufficient for protection \cite[e.g.,][]{RN2074}. Smaller aerosols ($<$\,100\,\micro m) can be inhaled, have lower settling velocity and so can remain suspended for many minutes to hours, and are thus relevant in both the near- and far-fields. In the near-field, aerosols are concentrated in the breathing zone of the emitting person, but rapidly mix into the full room volume. In both near- and far-field, improved masks (i.e. FFP2 or N95 style respirators) are most appropriate to guard against aerosol inhalation \cite[e.g.,][]{RN1703,RN2091,RN1771}, and physical barriers are not very effective. In the far-field, aerosols that build up in poorly ventilated rooms \cite[e.g.,][and references therein]{lai2002particle,RN2103,RN2105} can also be removed via ventilation and filtration, but these added controls provide little benefit in the near-field. Appropriate control measures and non-pharmaceutical intervention tools, such as use of high-quality face masks, physical distancing, ventilation, and room filtration should thus be matched to the type of space involved and the sizes and properties of the respiratory particles relevant for a given respiratory disease. Other engineering controls, such as upper-room germicidal UV light, can also be deployed as one of a layered strategy for virus inactivation, with primary application against far-field aerosols \cite[e.g.,][]{RN2057}. It is also important to note that far-field aerosol transmission is almost exclusively an indoor challenge, because dilution as a function of increasing distance from emitter becomes so large outdoors that infection risk is dramatically lower (although not to zero, depending on wind patterns and other variables) \cite[e.g.,][]{RN2072,RN2073}. The near-field aerosol transmission, however, is relevant under both, indoor and outdoor settings.

\subsection{Aerosol and droplet transmission of COVID-19 via SARS-CoV-2 virus}

\textbf{(1)} No definitive proof exists that SARS-CoV-2 (or almost any other individual viral pathogens) is transmitted by a specific transmission mechanism \cite{RN2092}. That said, considerable evidence suggests that the transmission of SARS-CoV-2 proceeds primarily through aerosol particles in the size range of $\sim$0.1 to $\sim$10\,\micro m, possibly dominated by near-field exposure, but with important contribution from far-field mixing in rooms \cite[e.g.,][]{RN2019,RN1740,RN2025,RN1937,RN1741,RN1814,RN2022,RN1657,RN1744,RN2065,RN2066,RN2067,RN2068,RN1871,RN1899}. The parameterization here suggests that the source of these particles is PSD modes \textbf{B2}, \textbf{LT} and/or \textbf{O1}, originating from multiple respiratory activities, including breathing, speaking/singing, and coughing. Fecal material aerosolized in broadly similar size ranges ($\lesssim$10\,\micro m) have also been suggested to be important for disease spread in some cases \cite[e.g.,][]{RN2039,RN2061}, but was not discussed in detail here. \par \vspace{3mm}
    
\textbf{(2)} The relevance of SARS-CoV-2-carrying aerosol particles in the size range of $\sim$0.1 to $\sim$10\,\micro m, combined with the variable filtration efficiency and poor average fit quality of so-called community face masks in this size range imply that tight-fitting, high efficiency masks such as N95/FFP2 are particularly important for mitigating the airborne transmission of COVID-19 \cite[e.g.,][]{RN1703,RN1702}. The small size of these particles also supports the suggestion that increased ventilation and room-filtration will provide community health benefit against COVID-19 \cite[e.g.,][]{RN1935,RN2069,RN2070,RN2071,RN2105}. \par \vspace{3mm}
    
\textbf{(3)} With regard to studies suggesting that the upper respiratory tract (i.e., the nose) is the initial target site for SARS-CoV-2 \cite[e.g.,][]{RN1850,RN1853}, the particle modes \textbf{LT} and \textbf{O1} may be particularly important because they overlap with the size range of highest deposition probability in the upper airways. \par \vspace{3mm} 
    
\textbf{(4)} Speaking is a frequent activity of everyday life, often with other persons in close proximity, and is also associated with high particle emission rates. Therefore, speaking can play a particularly important role in both near- and far-field transmission of COVID-19. A summary of the aerosol emission rates and particle sizes, as shown supports the ideas that speaking is likely an stronger driver of the pandemic than was generally considered early in the development of the COVID-19 pandemic. \par \vspace{3mm}  
%(Note: This is consistent with/confirm other observations that it is the driver loud speaking/shouting is also the worst in Lelieveld et al. 2020). 
     
\textbf{(5)} Remarkably similar bimodal distributions observed for influenza and SARS-CoV-2 in the size range of the modes \textbf{B2}, \textbf{LT} and/or \textbf{O1} indicate analogies in the airborne transmission routes of these two pathogens. This is a critical piece in context of available preventative against each, because it shows that observations about mask-wearing and ventilation can be leveraged for tremendous public health benefit against seasonal influenza as well as other emerging respiratory diseases. \par \vspace{3mm}  

\textbf{(6)} Together, these findings (e.g., Fig.\,\ref{fig:overview_scem}) should be implemented in updated and sophisticated hygiene concepts allowing common work and life in times of pandemics.\vspace{-2ex}
\subsection{Open questions and research perspectives}
\textbf{(1)} Experimental data on respiratory PSDs, including relevant variability, should be extended to reduce uncertainties across individuals and with respect to different respiratory diseases. Data coverage of the smallest mode \textbf{B1} (centered at $\sim$70\,nm) and largest modes \textbf{O1} and \textbf{O2} (centered at $\sim$10\,\micro m \& $\sim$100\,\micro m) is especially sparse. Further, most available measurements have focused on a limited portion of the full range of respiratory particle sizes ($<$10\,nm to $>$1000\,\micro m). We propose that systematic experiments across the full size should be conducted for different respiratory activities and a large cohort of volunteers (including healthy and individual disease groups) to consolidate our knowledge on the overall modality, mode ratios and their variability as a function of respiratory activities. A number of instruments will be required to span the full range of particle sizes \cite[e.g.,][]{RN1663}. Within the design of these experiments, care should be taken so that instrument PSDs overlap for cross-validation. Instruments with high size resolution are preferred to resolve details of the PSDs, which ideally can be acquired at high time-resolution to also observe differences in PSD properties as a function of rapidly changing emission mechanisms. \par \vspace{2mm}
    
\textbf{(2)} Existing measurements have been conducted under widely different experimental conditions, which hampers the comparison across studies and individuals. Standardized operational procedures for the analysis of respiratory aerosols would therefore be desirable \cite[e.g., as outlined for atmospheric aerosols in][]{RN1904}. One effort in this area was recently conducted as a part of a journal special issues focused on standardizing bioaerosol measurements, however, this was only a first step, with significant follow-up required \cite[e.g.,][]{RN2045,RN2047,RN2049,RN2046,RN1581,RN2048,RN2050}. Important aspects of such a standardization might be: (i) defined drying conditions e.g., $<$40\,\% RH, which is a rather reproducible state, (ii) isokinetic and isoaxial aerosol sampling, especially for large particles, (iii) corrections for particle losses in the experimental setup \cite{RN1323}, (iv) documentation of the precise sampling conditions (i.e., RH, temperature), and (v) a documentation of basic physiological, spirometric, and demographic parameters. \par \vspace{2mm}    

\textbf{(3)} Recorded measurements of respiratory activities have thus far been mostly standardized maneuvers and, therefore reflect 'normal live' emissions only imperfectly. Therefore, continuous measurements of respiratory emission from volunteers during different everyday activities (e.g., conversation, office work, at school, gym workout) would be very useful to better assess infection risks under real-world conditions. \par \vspace{2mm} 
    
\textbf{(4)} \citet{RN2037} showed significant differences in SARS-CoV-2 aerosol emission rate as a function of age and several physiological factors, but otherwise data is still relatively sparse regarding the physical basis to explain why some individuals act as viral aerosol superspreaders. For example, it is still relatively uncertain to what extent the superspreading of many diseases by certain individuals correlates with particularly high emission rates of exhaled particles overall and why this might be the case. Dedicated studies on the high inter-subject variability of emission rates would be important to close gaps in this area of knowledge. \par \vspace{2mm}

\textbf{(5)} Data on the chemical composition of saliva and ELF and its variability is sparse. Further experiments are needed as the chemical microenvironment plays a primary role for the decay of viability of the embedded pathogens during airborne transport. \par \vspace{2mm}  

\textbf{(6)} Data on the hygroscopic properties of mucosalivary particles and their shrinkage or growth under changing RH conditions, including the influence of potential hysteresis effects upon efflorescence and deliquescence, is sparse. Instruments reaching high RH, such as a high-humidity tandem differential mobility analyzer (HHTDMA) \cite{RN1809,RN1849} or similar, may allow investigation of these parameters. \par \vspace{2mm} 

\textbf{(7)} The precise pathogen emission mechanisms and sites are still largely uncertain. In this sense, our assignment of certain modes and likely emission mechanisms and sites (i.e., \textbf{B1}, \textbf{B2}, \textbf{LT}, \textbf{O1}, and \textbf{O2}) can be considered as preliminary estimates. While the processes behind mode \textbf{B2} are rather well documented, the origin of mode \textbf{B1} is widely unknown. Additionally, the interplay of emission from the larynx and trachea, which we combined here in mode \textbf{LT}, present large uncertainties. Finally, the precise mechanisms and sites of droplet formation in the mouth that generate modes \textbf{O1} and \textbf{O2} are poorly understood. \par \vspace{2mm}

\textbf{(8)} Further experimental studies on the pathogen size distributions and their mixing in the carrier PSD are highly needed as they (i) provide suggestions on droplet formation mechanisms and sites in the respiratory tract, (ii) allow the assessment of which PSD modes are most relevant as drivers of pathogen spread, and (iii) help the identification of the most likely deposition sites in the respiratory tract in relation to data on pathogen tropism \cite[e.g.,][]{RN1851,RN1850}. \par \vspace{1.5mm} 

\textbf{(9)} A better understanding of respiratory particle transport and dilution in air as a function of particle size will be required to help model the relative risk of infection in both indoor and outdoor settings. Without the ability to detect the small fraction of respiratory particles, it is impossible to use particle sizing instruments alone to observe respiratory emissions in a room. Thus, improved instrumental techniques will be required to selectively analyze respiratory particles amidst the overwhelming pool of existing aerosols in any room or outdoor air volume. Further, an improved understanding will be required on the quantitative range of how many pathogen entities (e.g., virions) for each disease are necessary for infection via the inhalation of the aerosol phase or deposition of the droplet phase \cite{RN1797,RN2055,RN2056}. \par \vspace{1.5mm}

\textbf{(10)} Improved understanding of environmental viability of pathogens, as a function of aerosol size and composition, RH, temperature, and UV flux, will aid the estimation of infectivity as a function of real-world parameters \cite[e.g.,][]{RN1595,RN2060,RN1936}.

\begin{table}%[hb!]
\caption{\label{tab:abbreviations}List of abbreviations.}
\begin{ruledtabular}
\begin{tabular}{ll}
Abbreviation&
Description\\[2pt]
\hline & \\[-1.7ex]
  APS & aerodynamic particle sizer\\
  ASL & airways surface layer\\
  BFFB & bronchiole fluid film burst\\
  BSA & bovine serum albumin\\
  B1 & bronchiol mode 1\\
  B2 & bronchiol mode 2\\
  C & breathing with airway closure\\
  CFU & culture forming unit\\
  COVID-19 & coronavirus disease 2019\\
  CP & airway closing point\\
  DDA & droplet deposition analysis\\
  DRH & deliquescence relative humidity\\
  ELF & epithelial lining fluid\\
  ERH & efflorescence relative humidity\\
  ERV & expiration reserve volume\\ 
  EXH & exhalation\\
  FFP2 & \makecell[tl]{filtering facepiece respirator, filtering $\leq$94\,\% of\\airborne particles}\\
  FRC & functional residual capacity\\
  HHTDMA & \makecell[tl]{high-humidity tandem differential\\mobility analyzer}\\
  ICPR & \makecell[tl]{International Commission on\\Radiological Protection}\\
  IRV & inspiration reserve volume\\
  INH & inhalation\\
  LLPS & liquid-liquid phase separation\\
  LRT & lower respiratory tract\\
  LT & larynx \& trachea mode\\
  MERS-CoV & \makecell[tl]{middle east respiratory syndrome\\coronavirus}\\
  MeV & Measles morbillivirus\\
  NIOSH & \makecell[tl]{National Institute for Occupational Safety\\and Health}\\
  NSD & number size distribution\\
  N95 & \makecell[tl]{filtering facepiece respirator, filtering $\leq$95\,\% of\\airborne particles}\\
  O1 & oral mode 1\\
  O2 & oral mode 2\\
  OPC & optical particle counter\\
  OPS & optical particle sizer\\
  PSD & particle size distribution \\
  RH & relative humidity \\
  RNA & ribonucleic acid\\
  RTLF & respiratory tract lining fluid \\
  RV & residual volume\\
  SARS-CoV & \makecell[tl]{severe acute respiratory syndrome\\coronavirus}\\
  SARS-CoV-2 & \makecell[tl]{severe acute respiratory syndrome\\coronavirus 2}\\
  SSA & sea spray aerosol \\
  SMPS & scanning mobility particle sizer\\
  T & tidal breathing\\
  TB & Tuberculosis\\
  TLC & total lung capacity\\
  URT & upper respiratory tract\\
  UV-APS & ultra-violet aerodynamic particle sizer\\
  UVB & ultraviolet B radiation\\
  VC & vital capacity\\
  VSD & volume size distribution \\
  VZV & varicella-zoster virus\\
  ZSR & Zdanovskii--Stokes--Robinson\\
\end{tabular}
\end{ruledtabular}
\end{table}

\begin{table*}[!ht]
\caption{\label{tab:symbol}List of symbols.}
\begin{ruledtabular}
\begin{tabular}{llp{12.0cm}}   % {llp{5.0cm}}
Symbol&
Unit&
Quantity\\[2pt]
\hline & \\[-1.7ex]
     $A_\text{i}$ & cm$^{-3}$ & {Parameter in lognormal fit function, number concentration at $D_i$}\\
     $a_\text{w}$ & & Water activity in the aqueous solution\\
     $b$ & & {Empirical factor representing relationship between $C_\text{N}$ and $V_\text{B}/V_\text{VC}$ %representing shape of the exponential relationship between particle emission and ventilation ratio $V_\text{B}/V_\text{VC}$,
     } \\
     $c_\text{g}$ & mol cm$^{-3}$ & Molar concentration of water vapor in the gas phase \\
     $c_\text{gs}$ & mol cm$^{-3}$ & Molar concentration of water vapor in the near-surface gas phase \\
     $C_\text{H}$ & cm$^{-3}$ & Mode number concentration in Heintzenberg formula\\
     $C_\text{N}$ & cm$^{-3}$ & Number concentration\\
     $C_\textsf{T}$ & cm$^{-3}$ & Number concentration during tidal breathe\\
     $C_\text{v}$ & \micro m$^{3}$ cm$^{-3}$ & Volume concentration\\
     $C_\text{c}$ &  & Cunningham slip correction\\
     $C_\text{D}$ &  & Drag coefficient\\
     $D$ & \micro m & Diameter (unit in eqs. \ref{eq:1} to \ref{eq:3} is m)\\
     $D_\text{dry}$ & \micro m & Dry diameter \\
     $D_\text{wet}$ & \micro m & wet diameter for a supersaturation \\
     $D_\text{eq}$ & \micro m & Equilibrium diameter after shrinking in dry atmosphere \\
     $D_\text{exh}$ & \micro m & Initial diameter after exhalation \\   
     $D_\text{g}$ & cm$^2$ s$^{-1}$ & Gas diffusion coefficient of water vapor \\   
     $D_\text{i}$ & \micro m & Mode mean geometric diameter \\
     d$C_\text{N}/$dlog$D$ & cm$^{-3}$ & Bin normalized number concentration\\
     d$C_\text{V}/$dlog$D$ & \micro m$^{3}$ cm$^{-3}$ & Bin normalized volume concentration\\
     d$Q_\text{N}/$dlog$D$ & h$^{-1}$ & Bin normalized number emission rate\\
     d$Q_\text{V}/$dlog$D$ & \micro m$^{3}$ h$^{-1}$ & Bin normalized volume emission rate\\
     $f$ & h$^{-1}$ & frequency of respiratory activity\\
     $g$ & m s$^{-2}$ & gravitational acceleration \\
     $g_\text{m}$ & & Mass equivalent hygroscopic growth factor\\
     $g_\text{d}$ & & Diameter equivalent hygroscopic growth factor\\
     $i_\text{s}$ & & Van't Hoff factor of the solute\\
     %$Ke$ & & Kelvin term\\
     $M$ & kg mol$^{-1}$& Molar weight (indices: w\,=\,water, s\,=\,dry solute)%of water
     \\
     %$M_\text{s}$ & kg mol$^{-1}$& Molar weight of the dry solute\\
     $n$ & & Number of moles\\
     $Q_\text{N}$ & h$^{-1}$ & Number emission rate\\
     $Q_\text{V}$ & \micro m$^{3}$ h$^{-1}$ & Volume emission rate\\
     $q$ & cm$^{-3}$ & Volume exhaled by respiratory activity\\
     $RH$ & $\%$ & Relative humidity\\
     $R^2$ & & Coefficient of determination\\
     $Re$ & & Reynolds number\\
     $R$ & kg m$^2$ s$^{-2}$ mol$^{-1}$ K$^{-1}$ & Universal gas constant\\
     $R_0$ & & basic reproduction rate for infectious diseases\\
     $s$ &  & Water vapor saturation\\
     $S$ & $\%$ & Water vapor supersaturation\\
     $T$ & {K} & Absolute Temperature \\
     %$t$ & s & Time of resperatory event\\
     $t_e$ & s & Evaporation time \\
     $t_\text{s}$ & s & Sedimentation time\\
     $v_\text{s}$ & m s$^{-1}$ & Relative velocity between air and particle\\
     $V_\text{s}$ & cm$^{3}$ & Volume of the dry solute\\
     $V_\text{w}$ & cm$^{3}$ & Volume of pure water within the droplet\\
     $V$ & cm$^{3}$ & Volume\\
     $\dot{V}$ & cm$^3\,\text{h}^{-1}$ & Time-averaged air emission rate (in Table \ref{tab:RespParam}, unit is $\text{L}\,\text{h}^{-1}$)\\
     $V_\text{B}$ & cm$^{3}$ & Breathed volume\\
     $V_\text{VC}$ & cm$^{3}$ & Vital capacity\\
     $V_\text{T}$ & cm$^{3}$ & Tidal volume\\
     $\Delta t$ & s & Duration of respiratory activities/events\\
     $\lambda$ & \micro m & Gas mean-free-path\\
     $\kappa$ &  & Hygroscopicity parameter\\
     %$\rho$ & kg m$^{-3}$ & density\\
    $\rho$ & kg m$^{-3}$ & Density (indices: g\,=\,gas, p\,=\,particle, s\,=\,solute, w\,=\,water)%e.g., air) 
    \\
    %$\rho_\text{p}$ & kg m$^{-3}$ & Particle density \\
    %$\rho_\text{s}$ & kg m$^{-3}$ & Solute density \\
    %$\rho_\text{w}$ & kg m$^{-3}$ & Water density \\
    $\eta$ & kg m$^{-1}$ s$^{-1}$ & Gas dynamic viscosity \\
     $\theta$ & {\degree\,C} & Temperature \\
     $\sigma_\text{i}$ & & Modal geometric standard deviation for log normal fit \\
     $\sigma_\text{H}$ & & {Modal geometric standard deviation in Heintzenberg formulation} \\
     $\sigma_\text{s}$ & kg s$^{-2}$ & {solution surface tension} \\
     $\nu_s$ & &  Stoichiometric  dissociation  number\\
     $\Phi_s$ & &  Molar osmotic coefficient in aqueous solution\\
     %\textit{subscripts}\\
     %g&gas (e.g. air) \\
     %p&particle\\
     %s&solute\\
     %w&water\\
\end{tabular}
\end{ruledtabular}
\end{table*}
\vspace{-2.15ex}
\begin{acknowledgments}\vspace{-1.2ex}
This work has been supported by the Max Planck Society (MPG). MB, EB and SS acknowledge support by the German Federal Ministry of Education and Research (BMBF) as part of the B-FAST (Bundesweites Forschungsnetz Angewandte Surveillance und Teststrategie) project (01KX2021) within the NUM (Netzwerk Universitätsmedizin) and the MPG. The funders had no role in study design, data collection and analysis, decision to publish, or preparation of the manuscript. JAH thanks the University of Denver for intramural funding for faculty support. The work on the section on hygroscopic properties of the respiratory particles was support by the Russian Science Foundation (grant agreement no. 18-17-00076). The authors thank Luiz Machado, Thomas Klimach, Bruna Holanda, Uwe Kuhn, Benjamin Bandowe, Cornelius Zetzsch, Oliver Schlenczek, Prasad Kasibhatla, and James Davies for support and inspiring discussions. We appreciate the helpful support by Andreas Zimmer in the literature research and the creative support by Dom Jack in figure layout. MP and CP are particularly grateful for the support by Gerhard Heun, Gabriele Kr\"uger, and Rita Heun during writing of this manuscript. 
\end{acknowledgments}

%apsrmp4-2.bst 2018-12-27 (MD) hand-edited version of apsrmp4-1.bst
%Control: author (75) reversed first initials jnrlst

\end{document}